 \documentclass
 {llncs}
\usepackage{enumitem}
\usepackage[stable]{footmisc}
\usepackage[retainorgcmds]{IEEEtrantools}
\usepackage[T1]{fontenc}
\usepackage[utf8]{inputenc}
\usepackage[english]{babel}
\usepackage{hyperref}
\usepackage{tikz}
\usepackage{lipsum}
\usepackage{dutchcal}     
\usepackage[leqno]{amsmath}
\usepackage{fancybox}
\usepackage{graphicx}
\usepackage{wrapfig}
\usepackage{lineno}
\usepackage[misc,geometry]{ifsym}
\usepackage{graphicx}
\usepackage{color}
\usepackage{xcolor}
\definecolor{applegreen}{rgb}{0.55, 0.71, 0.0}
\definecolor{caribbeangreen}{rgb}{0.0, 0.8, 0.6}
\usepackage[normalem]{ulem}
\usepackage{ucs}
\usepackage{url}
\usepackage{makeidx}
\usepackage{amssymb}

\usepackage[all]{xy}
\usepackage{xspace}
\usepackage{subfig}

\usepackage{algorithm}
\usepackage{verbatim}
\usepackage{multirow}

\usepackage{rotating}
\usepackage{lscape}
\usepackage{epsfig}
\usepackage{color}
\usepackage{float}
\usepackage{amsbsy,wasysym}
\usepackage{amsfonts}
\usepackage{amstext}
\usepackage{amssymb}
\usepackage{mathrsfs}
\usepackage{array}
\usepackage{latexsym}
\usepackage{xspace}

\newcommand\xob[1]{\enma{|#1|}}
\renewcommand\xob[1]{\enma{#1_0}}

\newcommand\Arr{\enma{\mathsf{Arr}}}
\newcommand\obj[1]{\enma{\left|#1\right|}}
\newcommand\otherobj[1]{\enma{#1_0}}
\renewcommand\obj[1]{\enma{#1_0}}
\renewcommand\otherobj[1]{\enma{\left|#1\right|}}
\newcommand\objbra[1]{\enma{\left|#1\right|}}
\renewcommand\objbra[1]{\enma{(#1)_0}}

\newcommand\id[1]{\opname{id}_{{#1}}}
\renewcommand\id{\opname{id}}

\newcommand\stri{\nmf{\pmb{String}}}
\newcommand\inte{\nmf{\pmb{Integer}}}
\newcommand\cat{category}
\newcommand\cats{categories}

\newcommand\sqarr[2]{\enma{#1\!\rightsquigarrow\!#2}}
\newcommand\ovrx[2]{\enma{\ovr{#1}}_{#2}}
\newenvironment{myeq}[1]%
{\begin{equation}\label{eq:#1}}
{\end{equation}}     
\newcommand\bfem[1]{{\bf\em #1}}
\newcounter{defCounter}
\newenvironment{defin}[1][]{
	\refstepcounter{defCounter}
	\vspace{-1.ex}
	\begin{trivlist}
		\item[\hskip \labelsep {\bfseries Definition \arabic{defCounter} (#1) }]
	}%
	{\end{trivlist}}

\newcommand\tcat{\catname{I\!\!I}}
\renewcommand\tcat{\enma{\mathbf{\large 1}}}

\DeclareMathAlphabet{\mathantt}{OT1}{antt}{li}{it}
\DeclareMathAlphabet{\mathpzc}{OT1}{pzc}{m}{it}

\newcommand\opname[1]{\ensuremath{\mathsf{#1}}}

\newcommand\niceletter[1]{\ensuremath{{\cal #1}}}
\newcommand\nicex[1]{\niceletter{#1}}
\newcommand\nia{\nicex{A}}
\newcommand\nib{\nicex{B}}

\newcommand\nil {\nicex{L}}
\newcommand\nim{\nicex{M}}

\newcommand\nix{\nicex{X}}

\newcommand\so{\ensuremath{\mathsf{src}}}
\newcommand\ta{\ensuremath{\mathsf{trg}}}
\renewcommand\so{\ensuremath{\mathsf{so}}} 
\renewcommand\ta{\ensuremath{\mathsf{ta}}}  

\renewcommand\id{\opname{id}}
\newcommand\Id[1]{\opname{Id}_{{#1}}}
\renewcommand\Id{\opname{Id}}

\newcommand\figref[1]{Fig.~\ref{fig:#1}}
\newcommand\exref[1]{Example~\ref{ex:#1}}
\newcommand\defref[1]{Def.~\ref{def:#1}}
\newcommand\conref[1]{Constr.~\ref{con:#1}}

\newcommand\sectref[1]{Sect.~\ref{sec:#1}}

\renewcommand\eqref[1]{(\ref{eq:#1})}
\newcommand\remaref[1]{Remark \ref{rema:#1}}

\newcommand\eqdef{\ensuremath{\stackrel{\mathrm{def}}{=}}}
\newcommand\eqnat{\ensuremath{\stackrel{\mathrm{nat}}{=}}}
\newcommand\respace{\!}
\newcommand\lsemm{\left[ \respace\left[ }
\newcommand\rsemm{\right]\respace\right]}
\newcommand\semm[1]{\mbox{$\lsemm\, \mathit{#1} \, \rsemm$}}

\newcommand\timm{{\times}}
\newcommand\inn{{\in}}

\newcommand\superset{\supset}

\newcommand\self{\ensuremath{\underline{\mathrm{self}}}}

\newcommand\ovr[1]{\ensuremath{\overline{#1}}}

\newcommand\isoind{\opname{iso}}
\newcommand\iso{\isoind}

\newcommand\comprehension[2]{\ensuremath{\left\{{#1}\left|\;{#2}\right.\right\}}}
\renewcommand\comprehension[2]{\ensuremath{\left\{{#1}| \;{#2}\right\}}}  
\newcommand\comprehensionColon[2]{\ensuremath{\left\{{#1}{:} \;{#2}\right\}}}  
\newcommand\compr[2]{\comprehension{#1}{#2}}
\newcommand\comprCol[2]{\comprehensionColon{#1}{#2}}

\newcommand\noteq{\ne}

\newcommand\catname[1]{\mbox{\bfseries\itshape #1}}
\renewcommand\catname[1]{\enma{\pmb{#1}}}

\newcommand\graphcat{\catname{Graph}}
\newcommand\setcat{\catname{Set}}

\newcommand\relcat{\catname{Rel}}
\newcommand\catcat{\catname{Cat}}

\makeatletter
  \def\fps@figure{tp}
\makeatother

\newcommand\oid{o-id}

\newcommand\mt{multitarget}

\newcommand\eg{e.g.}
\newcommand\ie{i.e.}
\newcommand\wrt{w.r.t.}
\newcommand\etc{{\em etc}}
\newcommand\etal{{\em et al}}

\newcommand\respace{\!}
\newcommand\lsemm{\left[ \respace\left[ }
\newcommand\rsemm{\right]\respace\right]}
\newcommand\semm[1]{\enma{$\lsemm\, \mathit{#1} \, \rsemm$}}

\newcommand\semmx[2]{\enma{$\lsemm\, \mathit{#2} \, \rsemm^{#1}$}}
\newcommand\semmA[2]{\enma{$\lsemm\, \mathit{#2} \, \rsemm^{A}$}}

\newcommand\dolan
{\mbox{$\Delta $}}

\newcommand\T{\mbox{$\LogicalTime$}}

\newcommand\same[2]{\mbox{$\IntersttMap^{#1}_{#2}$}}

\newcommand\mathsymbol[1]{\mbox{$#1$}}
\newcommand\world
{\mathsymbol{\cal W}}
\newcommand\cmter
{\mathsymbol{\cal C}}

\newcommand\relcat{\mbox{$\categoryname{Rel}$}}

\newcommand\formalObj[1]{\mbox{\sffamily #1}}

\newcommand\informalObj[1]{\mbox{\textit #1}}

\newcommand\FormalInterpret[1]{\mbox{$\formalObj{i}_f$}}
\newcommand\SubstantInterpret[1]{\mbox{$\informalObj{i}_s$}}

\renewcommand\eg{e.g.}

\newcommand\x{\diag{}}   

\renewcommand\same[2]{\mbox{${\mathit{same}}^{#1}_{\mathit{#2}}$}}

\newcommand\categoryname[1]{\mathbf{#1}}

\newcommand\setcat{\mbox{$\setcatmath$}}
\renewcommand\setcat{\mbox{$\categoryname{Sets}$}}

\newcommand\catcat{\mbox{$\categoryname{Cats}$}}

\newcommand\corrce{correspondence}

\newcommand\ovr[1]{\mbox{$\overline{\mathstrut#1}$}}

\newcommand\mb[1]{\mbox{$#1$}}
\newcommand\nin{\mb{\cal N}}
\newcommand\nix{\mb{\cal X}}
\newcommand\nil{\mb{\cal L}}

\renewcommand\x{\mb{X}}

\newcommand\id{\textsl{id}}
\renewcommand\id{\mbox{$\iota$}}

\renewcommand\mt[1]{\mbox{$T_{#1}$}}

\newcommand\req{\newrevem{q}}

\newlength{\cellH}
\newlength{\cellW}

\newcommand{\mynote}[2]{\mycolorednote{#1}{#2}{red}}
\newcommand{\mycolorednote}[3]{
    \fbox{\bfseries\sffamily\scriptsize#1}
    {\small$\Rightarrow$
    \textcolor{#3}{\textsf{\emph{#2}}}
    $\Leftarrow$}}    
\newcommand\zd[1]{\mynote{ZD}{#1}}
\newcommand\zdpar[1]{\par\mynote{ZD}{#1}\par}

\newcommand\hg[2]{\hgmargin{#1}\textcolor{blue}{#2}}

\newcommand\mymarginnew[1]{%
                     \marginpar{\vspace{-1em}\flushleft\textit{\footnotesize\textcolor{blue} {#1}}}} 
    \newcommand\mymarginNew[2]{%
    	\marginpar{\vspace{#2}\flushleft\textit{\footnotesize\textcolor{blue} {#1}}}}

\newcommand\zdmargin[1]{\mymarginnew{ZD:~#1}}

\newcommand\zdnewwno[3]{}

\def\modiffyNew#1#2#3#4#5#6{%
	{\small\underline{\sf{#1}}}
	{{#2}}
	{
		\mymarginNew{#3}{#6}}
	{#4 }
	{{#5}}
}%

\newcommand{\modifyNew}[4]{
	\modiffyNew%
	{}
	{\textcolor{red}{\sout{#1}}}
	{\textcolor{blue}{#3}}
	{}
	{\textcolor{blue}{#2}}
	{#4} 
}

\newcommand\zdmodd[4]{\modifyNew{#1}{#2}{ZD:~#3}{#4}}

\newcommand\zdmodok[3]{\modifyok{#1}{#2}{ZD:#3}}

\newcommand\zdmoddno[4]{\modiffyno{#1}{#2}{ZD:#3}{#4}}

\newcommand{\modifyok}[3]{{#2}}

\newcommand{\modiffyno}[4]{{#1}}

\newcommand\newno[2]{}

\newcommand\corring{corresponding}

\newcommand\corrce{correspondence}

\newcommand\eg{e.g.}
\newcommand\ie{i.e.}
\newcommand\wrt{w.r.t.}
\newcommand\etc{etc.}
\newcommand\etal{{\em et al}}

\newcommand\trafon{transformation}

\newcommand\syncon{synchronization}

\newcommand\syncer{synchronizer}

\newcommand\fwk{framework}

\newlength{\lawgap}

\newcommand\frar[3]{\ensuremath{#1{:}\;#2\rightarrow #3}}
\newcommand\fprar[4]{\ensuremath{#1{:}\;#2\stackrel{#4}{\rightarrow} #3}}
\renewcommand\fprar[4]{\ensuremath{#1{:}\;#2\rTo^{#4} #3}}
 
\newcommand\wrar[3]{\ensuremath{#1{:}\;#2\rightsquigarrow #3}} 

\newcommand\drar[3]{\mbox{$#1\!:#2\Rightarrow #3$}} 

\newcommand\relrar[3]{\mbox{$#1\!:#2\nrightarrow#3$}}

\newcommand\flar[3]{\ensuremath{#1\!:#2\leftarrow #3}} 

\usepackage{diagrams}

\newarrowhead{bt}\blacktriangleright\blacktriangleright\blacktriangleright\blacktriangleright
\newarrowtail{bt}\blacktriangleleft\blacktriangleleft\blacktriangleleft\blacktriangleleft
\newarrowmiddle{bar}\rtbar\ltbar\dtbar\utbar
\newarrowmiddle{|}\vert\vert--
\newarrowmiddle{||}\Vert\Vert==
\newarrowmiddle{=}==\Vert\Vert
\newarrowtail{=}==\Vert\Vert
\newarrowhead{l}\langle\langle\langle\langle 
\newarrowhead{r}\rangle\rangle\rangle\rangle 
\newarrowfiller{=}==\Vert\Vert
\newarrowfiller{d}\cdot\cdot\cdot\cdot
\newarrowmiddle{x}****
\newarrowmiddle{b}\bullet\bullet\bullet\bullet
\newarrowtail{b}\bullet\bullet\bullet\bullet
\newarrowmiddle{3}\equiv\equiv\vfthree\vfthree
\newarrowfiller{3}\equiv\equiv\vfthree\vfthree
\newarrowtail{3}\equiv\equiv\vfthree\vfthree
\newarrowtail{<=}\Leftarrow\Rightarrow{\@cmex7E}{\@cmex7F}
\newarrowfiller{bold}{\bf-}{\bf-}{\bf|}{\bf|}  
\newarrowfiller{o}{\hho}{\circ}{\circ}{\circ}  
\newarrowmiddle{>}\rtla\ltla\dtla\utla
\newarrowmiddle{d}\cdot\cdot\cdot\cdot

\newarrow{To}----{>}
\newarrow{Toang}----{->}
\newarrow{BMapsto}b---{>}
\newarrow{Mapsto}{|}---{>}
\newarrow{Mapsto}{b}---{>}
\newarrow{Dermapsto}{|}{dash}{}{dash}{>}
\newarrow{Dermapsto}{b}{dash}{}{dash}{>}
\newarrow{Dashmapsto}{b}{dash}{dash}{dash}{>}
\newarrow{Dotmapsto}{b}...{>}
\newarrow{Maps}b----
\newarrow{ID}33333

\newarrow{Dashes}{}{dash}{}{dash}{}
\newarrow{Dots}.....

\newarrow{EQ}=====
\newarrow{IDto}3333r  
\newarrow{NRelto}--+-{->}
\newarrow{Relto}--b-{->}
\newarrow{BSpanto}--b-{->} 
\newarrow{Embed}>---{>}
\newarrow{Embed}{blacktriangle}---{>}
\newarrow{EmbedDashed}{blacktriangle}{dash}{dash}{dash}{>}
\newarrow{Embdash}{blacktriangle}{dash}{dash}{dash}{>}
\newarrow{Embdot}{blacktriangle}...{>}
\newarrow{FEmb}{blacktriangle}---{>}
\newarrow{WEmb}{blacktriangle}---{>}
\newarrow{WEmbDot}{blacktriangle}...{>}
\newarrow{FEmbDot}{blacktriangle}...{>}
\newarrow{FEmb}{blacktriangle}---{blacktriangle}
\newarrow{WEmb}{blacktriangle}---{blacktriangle}
\newarrow{WEmbDot}{blacktriangle}...{blacktriangle}
\newarrow{FEmbDot}{blacktriangle}...{blacktriangle}
\newarrow{WEmbDash}{blacktriangle}{dash}{dash}{dash}{blacktriangle}
\newarrow{FEmbDash}{blacktriangle}{dash}{dash}{dash}{blacktriangle}

\newarrow{Dashto}{}{dash}{}{dash}{>} 
\newarrow{Dertoda}{}{dash}{}{dash}{>} 
\newarrow{Dotto}....{>} 
\newarrow{Dertodo}....{>} 

\newarrow{RDiagto}3333{r}
\newarrow{RDiagderto}{}{3}{}{3}{r}
\newarrow{LDiagto}3333{l}
\newarrow{LDiagderto}{}{3}{}{3}{l}
\newarrow{Mapsderto}{b}{dash}{}{dash}{>}
\newarrow{Into}C---{>}
\newarrow{Indashto}C{dash}{}{dash}{>}
\newarrow{Derinto}C{dash}{}{dash}{>}
\newarrow{Congruent}33333
\newarrow{Cover}----{blacktriangle}
\newarrow{Monic}{>}--->
\newarrow{Isoto}{>}---{triangle}
\newarrow{ISA}===={=>}
\newarrow{Isato}===={=>}
\newarrow{Doubleto}===={=>}
\newarrow{ClassicMapsto}{|}--->
\newarrow{Entail}{|}----
\newarrow{PArrow}{o}--->
\newarrow{MArrow}----{>>}
\newarrow{MPArrow}{o}---{>>}
\newarrow{Ogogoto}oooo{->}
\newarrow{Metato}--3->
\newarrow{MetaMapto}{|}-3->
\newarrow{OneToMany}+---{o}
\newarrow{HalfDashTo}{}{dash}{}-{>>}
\newarrow{DLine}=====
\newarrow{Line}-----
\newarrow{Tline}33333
\newarrow{Dashline}{}{dash}{}{dash}{}
\newarrow{Dotline}{}{o}{}{o}{}
\newarrow{Curlyto}{curlyvee}--->           
\newarrow{Bito}<--->
\newarrow{Bito}{<}---{>}
\newarrow{Bidito}{<}==={>}
\newarrow{Bitrito}{bt}333{bt}
\newarrow{Corrto}<--->
\newarrow{Dercorrto}{<}{dash}{}{dash}{>}

\newarrow{Instofto}{curlyvee}...{>}
\newarrow{Instofderto}{curlyvee}{dash}{}{dash}{->}
\newarrow{Upd}===={=>}
\newarrow{Derupdto}{}{dash}{}{dash}{>}
\newarrow{Mchto}----{>}
\newarrow{Dermchto}{}{dash}{}{dash}{>}
\newarrow{Viewto}----{>->}
\newarrow{Derviewto}{}{dash}{}{dash}{>>}
\newarrow{Viewto}===={=>}
\newarrow{Hetmchto}===={=>}
\newarrow{Derviewto}{=}{}{=}{}{=>}
\newarrow{Idleto}===={=>}
\newarrow{Deridleto}{=}{}{=}{}{=>}

\newlength{\StatArrBody}
\newlength{\NodeFrameThickness}
\newcommand\putell[2]{\putUL{\ell.#1}{#2}}
\newcommand\putekk[2]{\putUL{\ekk.#1}{#2}}
\newcommand\putemm[2]{\putUL{\emm.#1}{#2}}
\newcommand\putekel[2]{\putUL{\klbr.#1}{#2}}
\newcommand\putelem[2]{\putUL{\lmbr.#1}{#2}}

\newcommand\getUL[2]{\enma{\get^{#1}_{#2}}}
\newcommand\getekk[1]{\getUL{\ekk}{#1}}
\newcommand\getell[1]{\getUL{\ell}{#1}}
\newcommand\getelem[1]{\getUL{\ell\emm}{#1}}
\newcommand\getemm[1]{\getUL{\emm}{#1}}

\newcommand\emm{\enma{\mathpzc{m}}}
\renewcommand\emm{\enma{\mu}}

\newcommand\ekk{\enma{\mathpzc{k}}}

\newcommand\idl{\enma{\mathfrak{id}}}
\renewcommand\idl{\enma{\mathpzc{id}}}

\newcommand{\kl}{\ekk\timm \ell}

\newarrow{Updto}----{>}
\newarrow{Derupdto}{}{dash}{}{dash}{>}
\newarrow{Mchto}----{>}
\newarrow{Dermchto}{}{dash}{}{dash}{>}
\newarrow{Uto}----{>}
\newarrow{Uderto}{}{dash}{}{dash}{>}
\newarrow{Mto}----{>}
\newarrow{Mderto}{}{dash}{}{dash}{>}

\newcommand\xx{\!\!}
\newcommand\xxx{\!\!\!}
\newcommand\xxxx{\xxx\xx\,}

\newcommand\swssseTilearrowABx[2]{
	\begin{array}{c}
		#1 
		\\
		\xxx\swarrow\xxx\xx\swarrow%
		\downarrow\!\!\downarrow%
			\searrow\xxx\xx\searrow
	    \\ [-2.5pt]
	      \!\! #2
		\end{array}
}
\newcommand\swseTilearrowABx[1]{
	\begin{array}{c}
		#1 
		\\
		\xxx\swarrow\xxx\xx\swarrow%
		\searrow\xxx\xx\searrow
		\\ [-2.5pt]
	\end{array}
}

\newcommand\swssseTilearrowAB[2]{\stackrel{#1}{
		{\xxx\swarrow\xxx\xx\swarrow%
			\downarrow\!\!\downarrow%
			\searrow\xxx\xx\searrow}
}}

\newcommand\seTilearrow[1]{
 \lefteqn{~~~{:}#1}{\searrow\xxx\xx\searrow}
 }%
\newcommand\swTilearrow[1]{%
 \lefteqn{~~{:}#1}{\swarrow\xxx\xx\swarrow}
}%
\renewcommand\seTilearrow[1]{
 \lefteqn{~~~{:}#1}{\searrow\xxxx\xxx\searrow}
 }%
\renewcommand\swTilearrow[1]{%
 \lefteqn{~~{:}#1}{\swarrow\xxxx\xxx\swarrow}
}%

\newcommand\spacenam[1]{\ensuremath{\mathbf{#1}}}
\newcommand\spX[1]{\ensuremath{\spacenam{#1}}}
\newcommand\spA{\spX{A}}
\newcommand\spB{\spX{B}}
\newcommand\spC{\spX{C}}
\newcommand\spD{\spX{D}}
\newcommand\spG{\spX{G}}

\newcommand\spP{\spX{P}}
\newcommand\spQ{\spX{Q}}
\newcommand\spR{\spX{R}}
\newcommand\spS{\spX{S}}
\newcommand\spT{\spX{T}}
\newcommand\spXX{\spX{X}}

\newcommand\deltas[1]{\ensuremath{\Delta_{#1}}}

\newcommand\vecfont[1]{\ensuremath{\mathbf{#1}}}
\renewcommand\vecfont[1]{\ensuremath{{\cal #1}}}

\renewcommand\nia{\vecfont{A}}
\renewcommand\nim{\vecfont{M}}

\newcommand\T{\classname{T}}

\newcommand\dif{\enma{\diagopername{dif}}}
\renewcommand\deltas{\enma{\diagopername{deltas}}}

\newcommand\diagopername[1]{\ensuremath{\mathsf{#1}}}

\newcommand\get[1]{\mbox{\diagopername{get}$_{#1}$}} %
\newcommand\Get[1]{\mbox{\diagopername{Get}$_{#1}$}} %
\newcommand\putl[1]{\mbox{\diagopername{put}$_{#1}$}} 
\newcommand\putlback[1]{\ensuremath{\overleftarrow{\putl[1]}}}
\newcommand\Put[1]{\mbox{\diagopername{Put}$_{#1}$}} 
\renewcommand\get{\enma{\diagopername{get}}} %

\renewcommand\Get{\mbox{\diagopername{Get}}} %

\renewcommand\putl{\mbox{\diagopername{put}}} 
\renewcommand\Put{\mbox{\diagopername{Put}}} 
\newcommand\Putl{\mbox{\diagopername{Put}}} 
\renewcommand\putlback{\ensuremath{\overleftarrow{\putl}}}
\renewcommand\putlback{\ensuremath{\putl^\circlearrowleft}}

\newcommand\fppg{\mbox{\diagopername{fPpg}}}
\newcommand\bppg{\mbox{\diagopername{bPpg}}}

\newcommand\ide[2]{\ensuremath{\mathsf{id}_{{}_{#1}}^{#2}}}
\renewcommand\ide[2]{\ensuremath{\mathsf{id}_{{}_{#1}}{\!#2}}}

\newcommand\falt{\ensuremath{\diagopername{fAln}}}
\newcommand\balt{\ensuremath{\diagopername{bAln}}}

\newcommand\x{\ensuremath{\diagopername{x}}}

\newcommand\lawname[1]{\textsf{{#1}}}
\newcommand\lawnamebr[1]{\textsf{{(#1)}}}
\renewcommand\lawnamebr[2]{\ensuremath{\mathsf{{(#1)}}_{#2}}}
\newcommand\putget{\lawname{Putget}}

\newcommand\putgetcodi{\lawnamebr{Putget}{0}}

\newcommand\idppglaw{\lawname{IdPpg}}

\newcommand\fbfppglaw{\lawname{fbfPpg}}

\newcommand\bfbppglaw{\lawname{bfbPpg}}

	\newcommand\getio{\enma{\get1_0}}
\newcommand\getiio{\enma{\get2_0}}

\renewcommand\putl{\enma{\opname{put}}}

\newcommand\lowlowindex[2]{\enma{#1_{\!_{#2}}}}

\newcommand\minusrr{\enma{{-}\rrind}}
\newcommand\minusmtl{\enma{{-}\mtl}}

\newcommand\minusput{\enma{{-}\putl}}

\newcommand\plusx[1]{\lowlowindex{+}{#1}}

\newcommand\plusdelx[1]{\enma{\plusx{#1}\deltas}}

\newcommand\wbcodex{{\em Wb Codex}}
	\newcommand\spba{\enma{\spA\spB}}
	\newcommand\spca{\enma{\spA\spC}}
	\newcommand\spcb{\enma{\spB\spC}}
\newcommand\spbb{\enma{\spB\spB}}
\newcommand\spcc{\enma{\spC\spC}}

\newcommand\tlens{\enma{\pmb{\mathcal{t}}}}
\newcommand*{\newmathcal}[1]{%
	\textit{\fontfamily{qzc}\selectfont#1}%
}
\renewcommand\tlens{\enma{\pmb{\newmathcal{1}}}}
\newcommand\tlenscat{\catname{\{ }\!\tlens\catname{ \} }  }
\renewcommand\tlenscat{\catname{\tlens}}

\newcommand\quit{\nmf{quit}}
\newcommand\trans{\nmf{trans}}
\renewcommand\quit{\nmf{qt}}
\renewcommand\trans{\nmf{tr}}

\newcommand\john{\enma{\mathrm{John}}}
\newcommand\jon{\enma{\mathrm{Jon}}}
\newcommand\mary{\enma{\mathrm{Mary}}}
\newcommand\ann{\enma{\mathrm{Ann}}}
\newcommand\expr{\nmf{Expr.}}
\newcommand\dep{\nmf{Depart.}}
\newcommand\name{\nmf{Name}}
\newcommand\oid{\nmf{OID}}

\newcommand\isolensfun{\enma{\catname{isolens}}}

\newcommand\paramed{parameterized}
\newcommand\paramen{parameterization}
\newcommand\paramon{parameterization}
\newcommand\lfunee{\enma{L_{\eps,\err}}}

\newcommand\emno{} 

\newcommand\pairing{\enma{\diagopername{pairing}}}

\newcommand\grt[1]{\textcolor{caribbeangreen}{#1}}

\newcommand\nivv{\enma{\mathpzc{v}}}

\newcommand\xyz{\enma{\mathsf{xxx}}}
\newcommand\bx{Bx}
\newcommand\bxmde{\enma{\mbox{\bx}_{_{\mathrm{MDE}} }}}
\renewcommand\bxmde{Bx}
\newcommand\mtl{\enma{{\mathcal{mtl}}}}
\renewcommand\mtl{\enma{{\mathcal{b\;\!\!x \!L}}}}
\renewcommand\mtl{\enma{{\mathcal{L\! b\;\!\!x}}}}
\renewcommand\mtl{\enma{{\mathcal{L}_{\mathsf{bx}}}}}
\newcommand\mtlwb{\enma{{\mathcal{b\;\!\!x \!L\!wb}}}}
\renewcommand\mtlwb{\enma{{\mathcal{b\;\!\!x \!L}_\wbind}}}

\newcommand\codilabel[1]{\enma{#1^{\codiind}}}
\newcommand\codiind{\enma{\pmb \circledast}}
\renewcommand\codiind{\enma{\pmb \ast}}
\newcommand\codiindd{\enma{\pmb{\ast\ast}}}

\newcommand\wblabel[1]{\enma{#1_{\wbind}}}
\newcommand\wbind{\enma{\mathsf{wb}}}

\newcommand\alalenscatwbind{\enma{\alalenscat
		_{\mathsf{wb}} 
}}

\newcommand\getcat{\catname{Get}}

\newcommand\policyupd[1]{\enma{\mathsf{pol}^\upd_{#1}}}
	\newcommand\polupdX{\policyupd{\nix}}
\newcommand\Q{\enma{Q}}
\newcommand\QQ{\enma{\pmb Q}}
\newcommand\modcat{\catname{Mod}}

\newcommand\semmx[2]{\enma{\lsemm\, \mathit{#2}\,\rsemm^{#1}}}
\newcommand\semmA[1]{\semmx{A}{#1}}
\newcommand\vdefcat{\catname{VDef}}
\newcommand\pre{\enma{\mathsf{pre}}}

\newcommand\semmX[1]{\semmx{X}{#1}}
\newcommand\semmAp[1]{\semmx{A'}{#1}}
\newcommand\semmBp[1]{\semmx{B'}{#1}}
\newcommand\semmXp[1]{\semmx{X'}{#1}}
\newcommand\semmQA[1]{\semmx{\QQ(A)}{#1}}

\newcommand\Bx{Bx}

\newcommand\DDel{\enma{{\pmb \Delta}}}

\newcommand\del{\enma{\delta}}

\newcommand\della{\enma{\hat{\del}}}
\newcommand\dellp{\enma{\check{\del}}}

\newcommand\jacobx[1]{\enma{\mathbf{J}#1}}
\newcommand\jacobf{\jacobx{f}}

\newcommand\fu{\nmf{f}}
\newcommand\wi{\nmf{w}}

\newcommand\rcat{\enma{\mathbf{R}}}

\newcommand\alalens{ala-lens}
\newcommand\Alalens{Ala-lens}

\newcommand\warr{\enma{\rightsquigarrow}}
\newcommand\warrxy[2]{\enma{#1\warr#2}}
 
\newcommand\warrbb{\warrxy{b}{b'}}

\newcommand\byx[1]{\mbox{\quad #1}}

\newcommand\byeqna[2]{\mbox{\quad #1 \eqref{eqna-#2}}}
\newcommand\byeqnaa[3]{\mbox{\quad #1 \eqref{eqna-#2}~#3  }}

\newcommand\elll{\enma{\ell}}
\newcommand\vii{\enma{v^{}}}
\newcommand\uiii{\enma{u^{}}}

\newcommand\niget{\enma{\mathpzc{get}}}
\newcommand\nifo[1]{\enma{\mathcal{#1}}}
\newcommand\nin{\nifo{N}}
\newcommand\defcat{\enma{\mathpzc{Def}}}

\newcommand\smm{symmetric monoidal}

\newcommand\ome{\enma{\omega}}
\newcommand\isop{\enma{\iota}}
\newcommand\hell{\enma{\hat\ell}}
\newcommand\hp{{\enma{\hat p}}}
\renewcommand\hp{{\enma{\isop(p)}}}
\newcommand\hget{\enma{\widehat{\get}}}
\newcommand\hputl{\enma{\widehat{\putl}}}

\newcommand\enma[1]{{\ensuremath{#1}}}

\newcommand\rcatx[1]{\enma{\rcat^{{#1}}}}
\newcommand\realsx[1]{\enma{\mathbb{R}^{#1}}}

\newcommand\reals{\enma{\mathbb{R}}}

\newcommand\pol{\nmf{pol}}

\renewcommand\pol{\enma{\mathit{pol}}}

\newcommand\err{\nmf{err}}
\newcommand\rre{\nmf{r\!r\!e}}

\newcommand\eps{\enma{\varepsilon}}

\newcommand\apalenscat{\catname{aLaLens}}
\newcommand\allenscat{\catname{aLLens}}
\newcommand\alalenscat{\catname{aLaLens}}
\newcommand\aalenscat{\catname{aaLens}}

\newcommand\lenscat{\catname{Lens}}

\newcommand\alalenscatcodi{\codilabel{\alalenscat}}

\newcommand\alenscatcodi{\codilabel{\alenscat}}
\newcommand\allenscatcodicodi{\codilabel{\catname{aL}}\!\codilabel{\lenscat}}

\newcommand\putlcodi{\codilabel{\putl}}

\newcommand\alalenscatwb{\wblabel{\alalenscat}}

\newcommand\allenscatwb{\wblabel{\allenscat}}

\newcommand\alalensdifcat{\alalenscat\catname{Dif}}

\newcommand\codiscr{\diagopername{codiscr}}
\renewcommand\codiscr{\catname{codiscr}}
\newcommand\codifun{\codiscr}
\newcommand\pgetcat{\catname{pGet}}

\newcommand\icorr{\ensuremath{\mathsf{indCorr}}}
\newcommand\lner{learner}

\newcommand\learncat{\catname{Learn}}

\newcommand\paarr{\enma{\;\paar\;}} 
\newcommand\lm{\enma{\ell;\emm}}
\newcommand\lmbr{\enma{(\ell;\emm)}}
\newcommand\paarrx[1]{\paarr} 

\newcommand\paar{\enma{\!\,|\!|}}
\renewcommand\ll{\enma{\ell_1\!|\!|\ell_2}}
\newcommand\pp{{p_1\paar p_2}}
\renewcommand\ss{{A_1\paar A_2}}

\renewcommand\ll{\enma{\ell_1\paar\ell_2}}
\newcommand\llbr{\enma{(\ll)}}
\newcommand\putpsi{\putxyz{(1)}{p_1}{S_1}}
\newcommand\putpsii{\putxyz{(2)}{p_2}{S_2}}

\newcommand\catcattimm{\enma{(\catcat,\!\timm)}}
\newcommand\setcattimm{\enma{(\setcat,\!\timm)}}

\newcommand\pcatcat{\catname{pCat}}

\newcommand\psetcat{\catname{pSet}}
\newcommand\paracat{\catname{Para}}
\newcommand\rrind{[\reals^m,\reals^n]}
\renewcommand\rrind{\reals}
\newcommand\paracatrr{\enma{\catname{Para}_{_{\rrind}} }}
\newcommand\paracatre{\catname{Para_\reals}}
\newcommand\paracatrcat{\catname{Para_\rcat}}
  
\newcommand\leefun{\enma{L_{\eps,\err}}}
\newcommand\leefunovr{\ovr{\leefun}}

\newcommand\lnercat{\learncat}
\newcommand\slenscat{\catname{sLens}}
\newcommand\alenscat{\catname{aLens}}

\newcommand\ltol{\catname{l2l}}
\newcommand\pq{\enma{{pq}}}

\newcommand\klbr{\enma{(\kl)}}

\newcommand\citell{\cite{ll-bx19}[ll@bx19]}

\newcommand\figri{} 
\newcommand\figrii{}
\newcommand\figriii{}

\newcommand\putgeto{\enma{\putget_0}}

\newcommand\putxyz[3]{\ensuremath{\putl^{#1}_{{#2},{#3}}}}
\newcommand\putUL[2]{\ensuremath{\putl^{#1}_{#2}}}
\newcommand\getxy[2]{\ensuremath{\get^{{#1}}_{{#2}} }}

\newcommand\putxy[2]{\ensuremath{\putl_{(#1,#2)}}}
\renewcommand\putxy[2]{\ensuremath{\putl_{#1,#2}}}

\newcommand\putps{\putxy{p}{S}}

\newcommand\upd{\ensuremath{\mathsf{upd}}}
\newcommand\req{\ensuremath{\mathsf{req}}}
\renewcommand\self{\ensuremath{\mathsf{self}}}

\renewcommand\x{\enma{\mathbf{x}}}

\newcommand\Corr{\ensuremath{\mathsf{Corr}}}

\newcommand\amdx[1]{\ensuremath{#1^{@}}}
\newcommand\amex[1]{\amdx{#1}}

\newcommand\ppgB{\ppgrx{B}}

\newcommand\xconfig[1]{\ensuremath{\mathsf{#1}}}
\newcommand\Span{\xconfig{Span}}

\newcommand\compUpd[2]{\ensuremath{\mathsf{#1}^{{#2}}}}

\newcommand\Kcorr{\compUpd{K}{\bigstar}}

\newcommand\prt{\ensuremath{\partial}}

\renewcommand\ta{\ensuremath{\mathsf{ta}}}
\renewcommand\so{\ensuremath{\mathsf{so}}}
\renewcommand\ta{\ensuremath{\mathsf{cod}}}
\renewcommand\so{\ensuremath{\mathsf{dom}}}
\cornersize{.5} 
\newcommand\dbox[1]{\fbox{$#1$}}
\renewcommand\dbox[1]{\ovalbox{$#1$}}

\newcommand\namefont[1]{\ensuremath{\mathsf{#1}}}
\newcommand\nmf[1]{\namefont{#1}}

\newcommand\hgg[1]{}

\newcommand\citefst{}
\newcommand\citefjbx{}
\renewcommand\id{\opname{id}}

\title{
	{
		General Supervised Learning as Change Propagation with Delta Lenses}
\thanks{An extended version of paper with the same title published at FOSSACS 2020. Unfortunately, both the paper and the previous version of the extended version uploaded to arxiv on Feb 26, 2020, had bad typos in Definition 4 and Fig.4, which are now fixed.}
}
{ 
\author{
Zinovy Diskin
\raisebox{0.75ex}{\scriptsize{(\Letter)}} 
}
\authorrunning{Z. Diskin} 
 \institute{
 		{McMaster University, Hamilton, Canada}
	\\
    \email{diskinz@mcmaster.ca}
}
}
\newcommand\dend{\end{document}}

\newcommand\shortlong[2]{#1}  
\renewcommand\shortlong[2]{#2} 
\newcommand\hippoOutin[2]{#1} 

\renewcommand\zdpar[1]{}

\newlength{\figwidth}
\newlength{\onemodwid}
\newlength{\twomodwid}
\newlength{\mygap}
\begin{document}
	\maketitle
	\begin{abstract}
	Delta lenses are an established mathematical framework for modelling and designing bidirectional model transformations (Bx). Following the recent observations by Fong et al, the paper extends the delta lens framework with a a new ingredient: learning over a parameterized space of model transformations seen as functors. We will define a notion of an asymmetric learning delta lens with amendment (ala-lens), and show how ala-lenses can be organized into a symmetric monoidal (sm) category. We also show that sequential and parallel composition of well-behaved (wb) ala-lenses are also wb so that wb ala-lenses constitute a full sm-subcategory of ala-lenses. 
\end{abstract}

	\pagestyle{plain}
\shortlong{
	
	\newcommand\figro{\figref{policyfun-ml1}}
	\newcommand\figroo{\figref{policyfun-ml2}}
	\renewcommand\figrii{\figref{policyfun-brief}}
\section{Introduction}\label{sec:intro}

The goal of the paper is to develop a formal model of {\em supervised learning} in a very general context of 
{\em bidirectional model \trafon\  {\em or} Bx},  
\ie, \syncon\ of two arbitrary complex structures (called {\em models}) related by a transformation.%
\footnote{Term {\em Bx} 
	refers to a wide area including 
	file \syncon, data exchange in databases, and model \syncon\ in Model-Driven software Engineering (MDE), 	see \cite{bx-cross} for a survey. 
	In the present paper,  Bx will mainly refer to Bx in the MDE context.}  
Rather than learning \paramed\ functions between Euclidean spaces as is typical for machine learning (ML), we will consider learning mappings between model spaces and formalize them as parameterized functors between categories, \frar{f}{P\timm\spA}{\spB}, with $P$ being a parameter space. The basic ML-notion of  a {\em training pair} $(A,B')\in\obj{\spA}\times\obj{\spB}$ 
will be considered as an inconsistency between models caused by a change ({\em delta}) \frar{v}{B}{B'} of the target model $B=f(p,A)$ that was first consistent with $A$ \wrt\ the \trafon\ (functor) $f(p,\_)$. An inconsistency is repaired by an appropriate change of the source structure, \frar{u}{A}{A'}, changing the parameter to $p'$, and an {\em amendment} of the target structure \frar{v^@}{B'}{B^@} so that $f(p',A')=B^@$ is a consistent state of the \paramed\ two-model system. %

The setting above without \paramen\ and learning (\ie, $p'=p$ always holds), and without amendment ($v^@=\id_{B"}$ always holds), is well known in the 
Bx literature under the name of {\em delta lenses}--- mathematical structures, in which consistency restoration via change propagation is modelled by functorial-like algebraic operations over categories \cite{me-jot11,bryce-act19}. There are several types of delta lenses tailored for modelling different \syncon\ tasks and scenarios, particularly, symmetric and asymmetric; below we will often omit the adjective 'delta'. Despite their extra-generality, (delta) lenses have been proved useful in the design and implementation of practical model \syncon\ systems with triple graph grammars (TGG) \cite{tony-viewtgg-ecmfa14,tony-bxbook1}; 
enriching lenses with amendment is a recent extension of the \fwk\ motivated and formalized in \cite{me-faoc19}. A major advantage of the lens \fwk\ for \syncon\ is its compositionality: a lens satisfying several equational laws specifying basic \syncon\ requirements is called {\em well-behaved (wb)}, and basic lens theorems state that sequential and parallel composition of wb lenses is again wb. In practical applications, it allows the designer of a complex \syncer\ to avoid integration testing: if elementary \syncer s are tested and proved to be wb, their composition is automatically wb as well. 

The present paper makes the following contributions to the delta lens \fwk\ for \Bx. It i) motivates model \syncon\ enriched with learning and, moreover, with {\em categorical} learning, in which the parameter space is a category rather than a set (\sectref{learn4bx}), ii) introduces the notion of a {\em wb asymmetric learning (delta) lens {\em with} amendment} (a {\em wb ala-lens} in shot), 
and iii) proves compositionality of wb ala-lenses and shows how their universe can be organized into a symmetric monoidal (sm) category: see Theorems 1-3 on pages \pageref{th:seqwb}-\pageref{th:alalens2smcat}. (All proofs (rather straightforward but notationally laborious) can be found in the long version of the paper  \citelongver{Appendices}). One more compositional result is a definition of a {\em compositional} bidirectional \trafon\ language (\defref{mtl} on p.\pageref{def:mtl}) that formalizes an important requirement to model \syncon\ tools, which (surprisingly) is missing from the Bx literature (and it seems also from the practice of MDE tooling).

	\medskip\par\noindent  
\shortlong{{\em Notation.}}{{\em About notation used in the paper.}
In a general context, an application of function $f$ to argument $x$ will be denoted by $f(x)$. But many formulas in the paper will specify terms built from two operations going in the opposite directions (this is in the nature of the lens formalism): in our diagrams,  operation \get\ maps from the left to the right while operation \putl\ maps in the opposite direction. To minimize the number of brackets,  and relate a formula to its supporting diagram, we will also use the dot notation in the following way.  
If $x$ is an argument in the domain of \get, we tend to write formula $x'=\putl(\get(x))$ as $x'=\putl(x.\get)$ while if $y$ is an argument in the domain of \putl, we tend to write the formula $y'=\get(\putl(y))$ as  $(\putl.y).\get=y'$ or $(\putl.y)\get=y'$.
Unfortunately, this discipline is not always well aligned with the in-fix notation for sequential (;) and parallel/monoidal (||) composition of functions, 
so that some notational mix remained. 
}

Given a category \spA, its objects are denoted by capital letters $A$, $A'$, etc. to recall that in MDE applications, objects are complex structures, 
which themselves have elements $a, a',....$; the collection of all objects of category \spA\ is denoted by \obj{\spA}. An arrow with domain $A\in\obj{\spA}$ is written as \frar{u}{A}{\_} or $u\in\spA(A,\_)$; we also write $\so(u)=A$ (and sometimes $u^\so=A$ to shorten formulas). 
Similarly, formula \frar{u}{\_}{A'} denotes an arrow 
with codomain $u.\ta=A'$. 
A subcategory $\spB\subset\spA$ is called {\em wide} if it has the same objects. Given a functor \frar{f}{\spA}{\spB}, its object function is denoted by \frar{\obj{f}}{\obj{\spA}}{\obj{\spB}} (sometimes \otherobj{f}). 




	\renewcommand\figri{\figref{bxmlPut}}
	\renewcommand\figrii{\figref{policyfun-brief}}
\section{Background: Update propagation, policies, and delta lenses}\label{sec:backgr}

We will consider a simple example demonstrating main concepts and ideas of \bxmde. Although \Bx\ ideas work well only in domains conforming to the slogan {\em any implementation satisfying the specification is good enough} such as code generation and (in some contexts) model refinement (see \cite{me-jss15} for discussion), and have rather limited applications in databases (only so called updatable views can be treated in the \bx-way), we will employ a simple database example: it allows demonstrating the core ideas without any special domain knowledge required by typical \bx-amenable areas.  The presentation will be semi-formal as our goal is to motivate the delta lens formalism that abstracts the details away rather than formalize the example as such. 


\subsection{Why deltas}
\Bx-lenses first appeared in the work on file \syncon, and if we have two sets of strings, say, $B=\{\john, \mary\}$ and $B'=\{\jon, \mary\}$, we can readily see the difference: $\john\noteq\jon$ but $\mary=\mary$. We thus have a structure in-between $B$ and $B'$ (which maybe rather complex if $B$ and $B'$ are big files), but this structure can be recovered by string matching and thus updates can be identified with pairs. 
The situation dramatically changes if $B$ and $B'$ are object structures, \eg, $B=\{o_1, o_2\}$ with $\name(o_1)=\john$, $\name(o_2)=\mary$ and similarly  $B'=\{o'_1, o'_2\}$ with $\name(o'_1)=\jon$, $\name(o'_2)=\mary$. Now string matching does not say too much: it may happen that $o_1$ and $o'_1$ are the same object (think of a typo in the dataset), while $o_2$ and $o'_2$ are different (although equally named) objects. Of course, for better matching we could use full names or ID numbers or something similar (called, in the database parlance, primary keys), but absolutely reliable keys are rare, and typos and bugs can compromise them anyway. Thus, for object structures that \bxmde\ needs to keep in sync, deltas between models  need to be independently specified, \eg, by specifying a {\em sameness relation} $u\subset B\timm B'$ between models. For example, $u=\{o_1, o_1'\}$ says that $\john@B$ and $\jon@B'$ are the same person while $\mary@B$ and $\mary@B'$ are not. Hence, model spaces in \bxmde\ are categories (objects are models and arrows are update/delta specifications) rather than sets (codiscrete categories). 

\subsection{Consistency restoration via update propagation:
	An Example}\label{sec:ex-updprop}
\renewcommand\figri{\figref{ex-whyDL}}

Figure~\ref{fig:ex-whyDL} presents a simple example of delta propagation for consistency restoration. Models consist of objects (in the sense of OO programming)  with attributes (a.k.a. labelled records), \eg, the source model $A$ consists of three objects identified by their oids \#A, \#J, \#M (think about employees of some company) with attribute values as shown in the table (attribute \expr\ refers to Experience measured by a number of years, and \dep\ is the column of department names). The schema of the table, \ie, the triple $S_\spA$ of attribute names (\name, \expr, \dep) with their domains of values \stri, \inte, \stri\ resp., determines a model space \spA.%
\shortlong{}{
\footnote{
Formally, schema $S_\spA$ is a graph consisting of three arrows named \name, \expr, \dep, having the common source named \oid\ and the targets \stri, \inte, \stri\ resp. This graph freely generates a category (just add four identity arrows) that we denote by $S_\spA$ again.  We assume that a general model of such a schema is a functor \frar{X}{S_\spA}{\relcat} that maps arrows to relations. If we need some of these relations to be functions, we label the arrows in the schema with a special constraint symbol, say, [fun], so that schema becomes a generalized sketch in the sense of Makkai (see \cite{makkai-ske,me-entcs08}). In $S_\spA$, all three arrows are labelled by [fun] so that a legal model must map them to functions. For example,  model $A$ in the figure is given by functor \frar{{\_}^A}{S_\spA}{\relcat} with the following values:  $\oid^A=\{ \#A, \#J, \#M \}$, sets $\stri^A$ and $\inte^A$ actually do not depend on $A$---they are the predefined sets of strings and integers resp., and  $\name^A(\#A)=\ann$, $\name^A(\#J)=\john$, $\expr^A(\#A)=10$, etc.   
} 
} 
The target model space \spB\ is given by a similar schema $S_\spB$ consisting of two attribute names. For any model $X \in \spA$, we can compute its \spB-view $\get(X)$ by selecting those oids $\#O\in \oid^X$ for which $\dep^X(\#O)\in \{\mathrm{Testing, ML, DB}\}$; we will refer to departments, whose names are in \{Testing, ML \} as to {\em IT-departments} and the view $\get(X)$ as the {\em IT-view} of $X$. For example, the upper part of the figure shows the IT-view $B$ of model $A$. We assume that all column names in schemas $S_\spA$,  and $S_\spB$ are qualified by schema names, \eg, $\oid@S_\spA$, $\oid@S_\spB$ etc, so that schemas are disjoint except elementary domain names like \stri\ and \inte. Also disjoint are \oid-values, \eg,  \#J@$A$ and \#J@$B$ are different elements, but, of course, constants like John and Mary are elements of set \stri\ shared by both schemas. To shorten long expressions in the diagrams, we will often omit qualifiers and write $\#J=\#J$ meaning $\#J@A=\#J@B$ or $\#J@B=\#J@B'$ depending on the context given by the diagram; often we will also write $\#J$ and $\#J'$ for such OIDs. Also, when we write $\#J=\#J$ inside block arrows denoting updates, we actually mean a pair, \eg, $(\#J@B, \#J@B')$. 
   
Given two models over the same schema, say, $B$ and $B'$ over  $S_\spB$, an update \frar{v}{B}{B'} is a relation $v\subset \oid^B\timm\oid^{B'}$; if the schema were containing more nodes, an update should provide such a relation $v_N$ for each node $N$ in the schema. However, we do not require naturality: in the update $v_2$ specified in the figure, for object $\#J\in\oid^B$, we have $\#J.v_2.\name^{B'} \noteq \#J.\name^B$ but it is a legal update that modifies the value of the attribute. 

Note an essential difference between the two parallel updates \frar{v_1, v_2}{B}{B'} specified in the figure.  Update $v_1$ says that John's name was changed to Jon (\eg, by fixing a typo), and the experience data for Mary were also corrected (either because of a typo or, \eg, because the department started to use a new ML method for which Mary has a longer experience). Update $v_2$ specifies the same story for John but a new story for Mary: it says that Mary@$B$ left the IT-view and Mary@$B'$ is a new employee in one of IT-departments.

\setlength{\twomodwid}{\textwidth}

\begin{figure}
    \vspace{-0.25cm}
       \centering
   \shortlong{
    \includegraphics[width=0.9\twomodwid]%
    {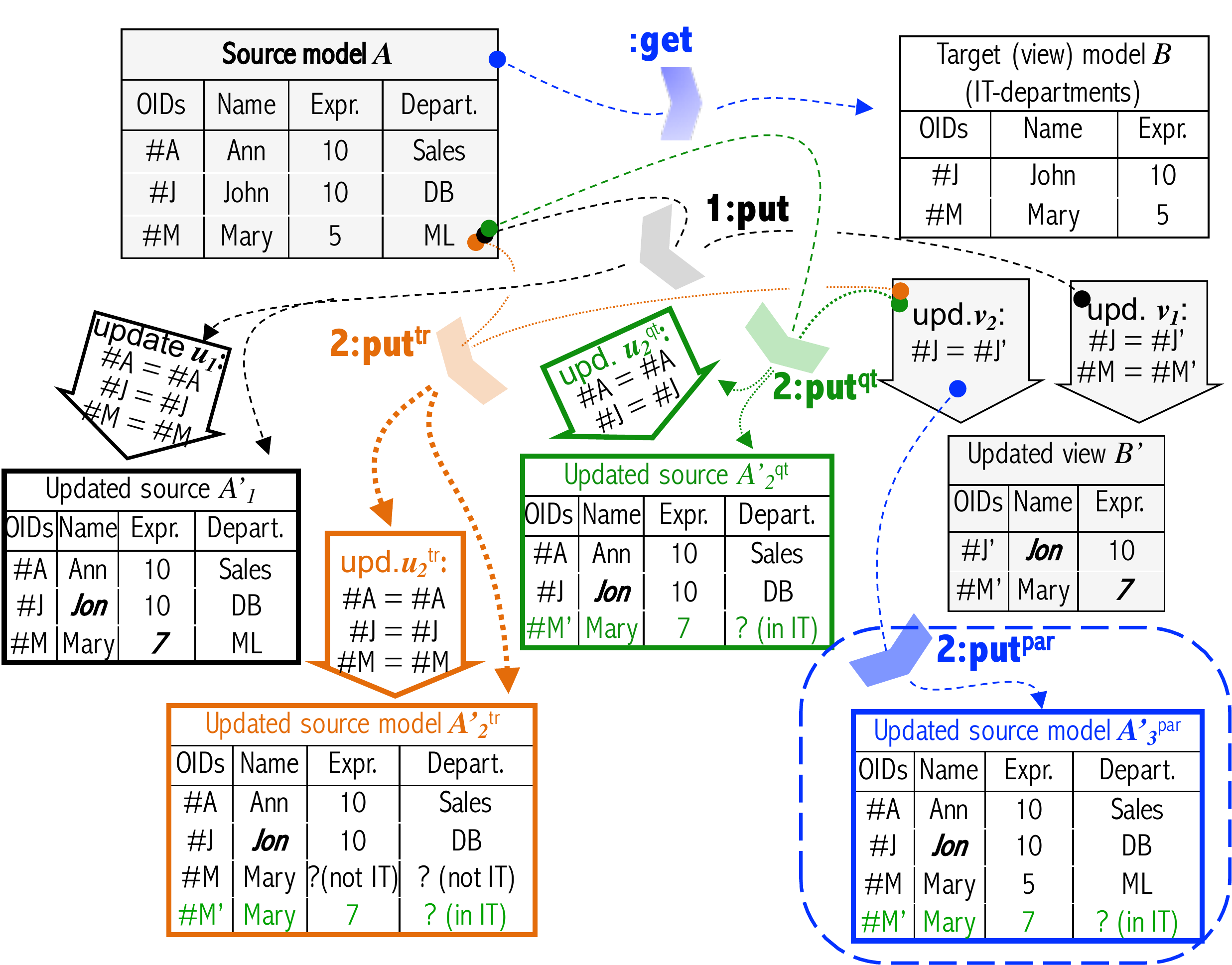}
    }{
   \includegraphics[width=1.\twomodwid]%
{ll-figures/pdfs/ex-whyDLbig-pic.pdf}
}
    \caption{Example of update propagation}
    \label{fig:ex-whyDL}
   \vspace{-0.5cm}
\end{figure}

\subsection{Update propagation and update policies}
The updated view $B'$ is inconsistent with the source $S$ and the latter is to be updated accordingly --- we say that update $v$ is to be propagated (put back) to $A$. Propagation of $v_1$ is easy: we just update accordingly the values of the corresponding attributes according to update \frar{u_1}{A}{A'_1} specified in the figure inside the black block-arrow $u_1$. Importantly, propagation needs two pieces of data: the view update $v_1$ and the original state $A$ of the source as shown in the figure by two data-flow lines into the chevron 1:\putl\ denoting invocation of the backward propagation operation \putl\ (read ``put view update back to the source'').  The quadruple $1=(v_1,A,~u_1, A')$ is an {\em instance} of operation \putl, hence the notation $1{:}\putl$ (borrowed from the UML). Note that the updated source model $A'$ is actually derivable from $u_1$ as its target, but we included it explicitly into \putl's output to make the meaning of the figure more immediate.   

Propagation of update $v_2$ is more challenging: Mary can disappear from the IT-view because a) she quit the company, b) she transitioned to a non-IT department, and c) the view definition has changed, \eg, the view now only shows employee with experience more than 5 years (and for more complex views, the number of possibilities is much bigger).  Choosing between these possibilities is often called choosing an {\em (update) policy}. We will consider the case of changing the view (conceptually, the most radical one) in \sectref{learn4bx}, and below discuss policies a) and b). 

For policy a) (further referred to as {\em quiting} and briefly denoted by \quit), the result of update propagation is shown in the figure with green colour: notice the update (block) arrow $u_2^\quit$ and its result, model $A_{2}^{'\quit}$, produced by invoking operation $\putl^\quit$. Note that while we know the new employee Mary works in one of IT departments, we do not know in which one. This is specified with a special value '?' (a.k.a. labelled null in the database parlance). %

For policy b) (further referred to as {\em transition} and denoted \trans), the result of update propagation is shown in the figure with orange colour: notice update arrow $u_{2}^\trans$ and its result, model $A_{2}^{'\trans}$ produced by $\putl^\trans$. Mary \#M is the old employee who transitioned to a new non-IT department, for which her expertize is unknown. Mary \#M' is the new employee in one of IT-departments (recall that the set of departments is not exhausted by those appearing in a particular state $A\in\spA$). There are also updates whose backward propagation is uniquely defined and does not need a policy, \eg, update $v_1$ is such. 

An important property of update propagations we considered (ignore the blue propagation in the figure that shows policy c)) is that they restore consistency: the view of the updated source equals to the updated view initiated the update: $\get^0(A')=B'$. Moreover, this equality extends for update arrows: $\get(u_i)=v_i$, $i=1,2$, where \get\ is an extension of the view mapping $\get_0$ for update arrows. Such extensions can be derived from view definitions if the latter are determined by so called monotonic queries (which encompass a wide class of practically useful queries including Select-Project-Join queries); for views defined by non-monotonic queries, in order to obtain \get's action on source updates \frar{u}{A}{A'}, a suitable policy is to be added to the view definition (see \cite{viewmain-vldb98,viewmain-elke06,me-jot11} for a discussion). Moreover, normally \get\ preserves identity updates, $\get(\id_A)=\id_{\get(A)}$, and update composition: for any \frar{u}{A}{A'} and \frar{u'}{A'}{A''}, equality $\get(u;u')=\get(u);\get(u')$ holds. 

\subsection{Delta lenses and their composition}
Our discussion of the example can be summarized in the following algebraic terms. We have two categories of {\em models} and {\em updates}, \spA\ and \spB, and a functor \frar{\get}{\spA}{\spB} incrementally computing \spB-views of \spA-models (we will often write $A.\get$ for $\get(A)$). We also suppose that for a chosen update policy, we have worked out precise procedures for how to propagate any view update backwards. This gives us 
a family of operations 
\flar{\putl_A}{\spA(A,\_)}{\spB(A.\get, \_)}
indexed by \spA-objects, $A\in\obj{\spA}$, for which we write $\putl_A.v$ or $\putl_A(v)$ interchangeably. 
{
\renewcommand\spS{\spA}  
\renewcommand\spT{\spB}
\renewcommand\S{{A}}
\renewcommand\T{{B}}
\begin{defin}[Delta Lenses (\cite{me-jot11})]\label{def:alens} 
	Let \spS, \spT\ be two categories. An {\em (asymmetric delta) lens} from \spS\ to \spT\ is a pair $\ell=(\get, \putl)$, where \frar{\get}{\spS}{\spT} is a functor and $\putl$ is a family of operations 
	\flar{\putl_A}{\spA(A,\_)}{\spB(A.\get, \_)}
	indexed by objects of \spS, $\S\in\obj{\spS}$. Given $\S$, operation $\putl_\S$ maps any arrow \frar{v}{\S.\get}{\T'} 
	to an arrow \frar{u}{\S}{\S'} such that $\S'.\get=B'$. The last condition is called (co)discrete Putget law:
		\\ [5pt]
	\begin{tabular}{l@{\quad}l}
\lawnamebr{Putget}{0}
		& 
		\mbox{ 
			$(\putl_\S.v).\ta.\get_0  = v.\ta$ for all $\S\in|\spS|$ and $v\in\spT(\S.\get,\_)$
		} 
	\end{tabular}
\\ [5pt]	
where $\get_0$ denotes the object function of functor \get.	We will write a lens as an arrow \frar{\ell}{\spS}{\spT} going in the direction of \get.
\end{defin}
Note that family \putl\ corresponds to a chosen update policy, \eg, in terms of the example above, for the same view functor \get, we have two families of operations \putl, $\putl^\quit$ and $\putl^\trans$, corresponding to the two updated policies we discussed. 
These two policies determine two lenses $\ell^\quit=(\get,\putl^\quit)$ and $\ell^\trans=(\get,\putl^\trans)$ sharing the same \get. %
%

\begin{defin}[\shortlong
	{Well-behavedness}{Well-behavedness}]\label{def:wbalens}{\emno 
		A {\em (lens) equational law} is an equation to hold for all values of two variables: 
		$\S\in\xob{\spS}$ and \frar{v}{\S.\get}{T'}} (as in the laws below).	
	A lens is called {\em well-behaved (wb)} if the following two equational laws hold:
	\\[1ex]
	\lawgap=1 ex
	\noindent \begin{tabular}{l@{\quad}l}
		\lawnamebr{Stability}{
		}	& \mbox{ 
			$\id_\S = \putl_\S.\id_{\S.\get}$ for all $\S\in|\spS|$ }
		\\ [\lawgap] \lawnamebr{Putget}{} 
		& 
		\mbox{ 
			$(\putl_\S.v).\get  = v$ for all $\S\in|\spS|$ and all $v\in\spT(\S.\get,\_)$
		} 
	\end{tabular}  
\end{defin} 
} 
\begin{remark}[On lens laws]\label{rema:laws1}{\emno 
		a) Stability says that the lens does nothing if nothing happens on the target side (no trigger--no action, hence, the name of the law) 
		
		b) 	Putget requires the goal of update propagation to be achieved after the propagation act is finished (see examples in \sectref{ex-updprop}). Note the distinction between the Putget$_0$ condition included into the very definition of a lens, and the full Putget law required for the wb specialization of lenses. It is needed to ensure smooth tiling of \putl-squares (\ie, arrow squares describing application of \putl\ to a view update and its result) both horizontally and vertically (not considered in this paper). Also, if we want to accurately define operations \putl\ independently of the functor \get, we still need a function \frar{\get_0=|\get|}{|\spA|}{|\spB|} and the codiscrete Putget law to ensure smooth tiling (cf.\cite{bryce-act19}).

        c) A natural requirement for the family \putl\ would be its  compatibility with update composition: 
        for any \frar{v}{A.\get}{B'}, \frar{v'}{B'}{\_} the following is to hold:
        
        \medskip
        \noindent \begin{tabular}{l@{\quad}l}
        \lawnamebr{Putput}{} 
        	& 
        	\mbox{ 
        		$\putl_A.(v;v') = (\putl_A.v);(\putl_{A'}.v')$ where $A'=(\putl_A.v)^\ta$
        	} 
        \\ [5pt]
        \end{tabular} 

\noindent (note that $A'.\get=B'$ due to \putgetcodi\ law).   However, this law does not hold in typical \Bx applications (see \cite{me-bx17} for examples and discussion) and thus is excluded from the  \wbcodex. 
    } 
\end{remark}

\medskip
\newcommand\figalenscomp{\figref{alens-compose}}
\begin{figure}
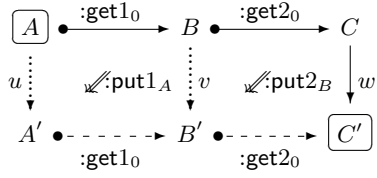

		\centering
{ 
\begin{tabular}{c}
\begin{diagram}[w=1.1\cellW,h=0.9\cellH] 
\dbox{A} 
& \rMapsto^{{:}\getio}  & {B} 
& \rMapsto^{{:}\getiio} &  {C}%
\\
\dDotto<u & \swTilearrow{\putl{1}_A}~~~ & \dDotto>{v} %
                 & \swTilearrow{\putl2_B} ~~~& \dTo>{w} 
\\
{A'} & \rDermapsto_{{:}\getio}  & %
 B'            & \rDermapsto_{{:}\getiio} & \dbox{C'} %
\end{diagram}%
\end{tabular}
} 
\caption{Lens composition} %
\label{fig:alenscompose}%
\end{figure}

Asymmetric lenses are sequentially associatively composable. 
Having two lenses \frar{\ell1=(\get1,\putl1)}{\spA}{\spB} and \frar{\ell2=(\get2,\putl2)}{\spB}{\spC}, we build a lens \frar{\ell=(\get,\putl)}{\spA}{\spC} with $\get = \get1;\get2$ and $\putl$ being the family defined by composition as shown in \figref{alenscompose} (where objects produced by functors \get s are non-framed, arrows are dashed, and arrows produced by \putl s are dotted): for $A\in|\spA|$ and \frar{w}{A.\get}{C'}, $\putl_A.w=\putl1_A.\putl2_B.w$.  The identity lens is given by identity mappings, and we thus have a category \alenscat\ of asymmetric delta lenses \cite{me-jot11,jr-unified}. It's easy to see that sequential  composition preserves well-behavedness; we thus have an embedding $\alenscat_\wbind\subset\alenscat$.

Next we will briefly outline the notion of an asymmetric lens with amendment (aa-lens): a detailed discussion and motivation can be found in \cite{me-faoc19}

\medskip
\noindent
\begin{minipage}{1.\textwidth}
{
	\renewcommand\spS{\spA}  
	\renewcommand\spT{\spB}
	\renewcommand\S{{A}}
	\renewcommand\T{{B}}
	\begin{defin}[Lenses with amendment ]\label{def:aalens} 
		Let \spS, \spT\ be two categories. An {\em (asymmetric delta) lens with amendment (aa-lens)} from \spS\ to \spT\ is a triple $\ell=(\get, \putl^\spba,\putl^\spbb)$, where \frar{\get}{\spS}{\spT} is a functor, $\putl^\spba$ is a family of operations 
		\flar{\putl^\spba_A}{\spA(A,\_)}{\spB(A.\get, \_)}
		indexed by objects of \spS, $\S\in|\spS|$ exactly like in \defref{alens} and 
		$\putl^\spbb$ is a family of operations 
		\frar{\putl^\spbb_A}{\spB(A.\get, \_)}{\spB(A.\get.\ta,\_)}
		also indexed by objects of \spS\ but now mapping an arrow \frar{v}{A.\get}{B'} to an arrow \frar{v^@}{B'}{B'^@} called an {\em amendment} to $v$. We require for all $\S\in|\spS|$,  $v\in\spT(\S.\get,\_)$: 
				\\ [5pt]
		\begin{tabular}{l@{\quad}l}
			\lawnamebr{Putget}{0}
			& 
			\mbox{ 
				$(\putl^\spba_\S.v).\ta.\get_0  = (v;\putl^\spbb_A).\ta$ 
			} 
		\end{tabular}
		\\ [5pt]
		An aa-lens is called {\em well-behaved (wb)} if the following two equational laws hold:
		\\[1ex]
		\lawgap=1 ex
		\noindent \begin{tabular}{l@{\quad}l}
			\lawnamebr{Stability}{
			}	& \mbox{ 
				$\putl^\spba_\S(\id_{\S.\get})=\id_\S $ and}
				\mbox{$\putl_\S^\spbb(\id_{\S.\get})=\id_{\S.\get}$ 
				}
			\\ [5pt] \lawnamebr{Putget}{} 
			& 
			\mbox{ 
				$(\putl^\spba_\S.v).\get  = v;(\putl^\spbb_A.v)$}
		\end{tabular}  
	\\[5pt] 
	for all $\S\in|\spS|$ and all $v\in\spT(\S.\get,\_)$.
	We will sometimes refer to the Putget law for aa-lenses as the {\em amended} Putget.
	\end{defin} 
}  
\end{minipage}
\begin{remark}[On lens laws]\label{rema:laws2}
\shortlong{}{b)}	
Besides Stability and Putget, the MDE context for ala-lenses suggests other laws discussed in \cite{me-faoc19}, which were (somewhat recklessly) included into the \wbcodex. To avoid confusion, Johnson and Rosebrugh aptly proposed to call lenses satisfying Stability and Putget {\em SPg lenses} so that the name unambiguously conveys the meaning. In the present paper, wb will mean exactly SPg.   
\shortlong{}{
	c) Hippocraticness prohibits amending delta $v$ if consistency can be achieved for $v$ without amendment (hence, the name).  Note that neither equality $\putUL{\upd}{p,S}(v)=p$ nor $\putUL{\req}{p,S}(v)=u$ are required: the lens can improve consistency by changing $p$ and $u$ but changing $v$ is disallowed.  
} 
\end{remark}
\begin{remark}[Putget for codiscrete lenses]
	\label{rema:putget4codi}
In the codiscrete setting, the codiscrete Putget law, \putgetcodi, actually determines amendment in a unique way. The other way round, a codiscrete lens without the requirement to satisfy Putget, is a wb codiscrete lens with amendment (which satisfies the amended Putget). 
\end{remark}



\begin{wrapfigure}{R}{0.45\textwidth}
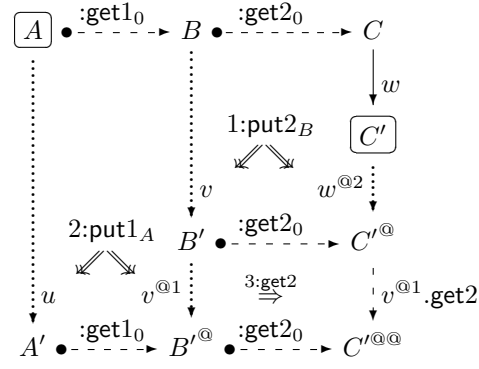

	\centering
\begin{tabular}{l}
$
\begin{diagram}[h=2.em]
\dbox{A}&\rDermapsto^{{:}\get1_{0}}
&B&\rDermapsto^{{:}\get2_{0}}
&C
\\ 
&&&& \dTo>{w}
\\ 
&& 
&
\swseTilearrowABx{1{:}\putl2_{B}}{}
&\dbox{C'}
\\ 
&&
\dDotto>{v}
&&\dDotto<{w^{@2}}
\\ 
&
\swseTilearrowABx{2{:}\putl1_{A} }{}
&B'
&\rDashmapsto^{{:}\get2_{0}}&{C'}^@
\\ 
\dDotto>{u} 
&&\dDotto<{v^{@1}}&\stackrel{3{:}\get2}{\Rightarrow}&\dDashto>{v^{@1}.\get2}
\\ 
A'&\rDashmapsto^{{:}\get1_{0}}&{B'}^@&\rDashmapsto^{{:}\get2_{0}}
&C'^{@@}
\end{diagram}
$
\end{tabular}
\caption{Composition of aa-lenses}
\label{fig:aalenscompose}
\vspace{-0.75cm}
\end{wrapfigure}
\noindent Composition of aa-lenses, say, \frar{\ell1}{\spA}{\spB} and \frar{\ell2}{\spB}{\spC} is specified by diagram in \figref{aalenscompose} (objects produced by functors \get s are non-framed, arrows are dashed, and arrows produced by \putl s are dotted).  Operation invocation numbers 1-3 show the order of applying operations to produce the composed \putl: 
$\putl^\spca_A.w=\putl^\spba_A.\putl^\spcb_B.w$ and 
the composed lens amendment $\putl^{\spcc}=w^@$ is defined by composition $w^{@2};(v^{@1}.\get2)$.  In 
\sectref{seqlensass-proof}, 
we will see that composition of aa-lenses is associative (in the more general setting of ala-lenses, \ie, aa-lenses with learning) and they form a category \aalenscat\ with a subcategory of wb aa-lenses $\aalenscat_\wbind\subset\aalenscat$. Also, an ordinary a-lens can be seen as a special aa-lens, for which all amendments are identities, and aa-lens composition specified in \figref{aalenscompose} coincides with a-lens composition in \figref{alenscompose}; moreover, the aa-lens wb conditions become a-lens wb conditions. Thus, we have embeddings $\alenscat\subset\aalenscat$ and $\alenscat_\wbind\subset\aalenscat_\wbind$.

          \renewcommand\figriii{\figref{policyfun}}
 \section{Asymmetric Learning Lenses with Amendments
}\label{sec:learn4bx}

\zdmoddno{We will begin}{Theywill}{efjjwekjw}{-3em} with a brief motivating discussion, and then proceed with formal definitions 

\subsection{Does \Bx\ need categorical learning?}
Enriching delta lenses with learning capabilities has a clear practical sense for Bx. 
%
Having a lens \frar{(\get, \putl)}{\spA}{\spB} and inconsistency 
$A.\get\noteq B'$, the idea of learning extends the notion of the search space and allows us to update the transformation itself so that the final consistency is achieved for a new \trafon\ $\get'$: $A.\get'=B'$.  For example, in the case shown in \figref{ex-whyDL}, 
disappearance of Mary \#M in the updated view $B'$ can be caused by changing the view definition, which now requires to show only those employees whose experience is more than 5 years and hence Mary \#M is to be removed from the view, while  Mary \#M' is a new IT-employee whose experience satisfies the new definition. Then the update $v_2$ can be propagated as shown in the bottom right corner of \figref{ex-whyDL}, where index $\mathsf{par}$ indicates a new update policy allowing for view definition (parameter) change.

To  manage the extended search possibilities, we parameterize the space of transformations as a family of mappings \frar{\get_p}{\spA}{\spB} indexed over some parameter space $p\in\spP$. For example, we may define the IT-view to be parameterized by the experience of employees shown in the view (including {\em any} experience as a special parameter value). 
 Then we have two interrelated propagation  operations that map an update \sqarr{B}{B'} to a parameter update \sqarr{p}{p'}  and a source update \sqarr{A}{A'}. 
 Thus, the extended search space allows for new update policies that look for updating the parameter as an update propagation possibility. The possibility to update the transformation appears to be very natural in at least two important \Bx\ scenarios: a) model transformation design and b) model \trafon\ evolution (cf. \citemtbe), which necessitates the enrichment of the delta lens \fwk\ with \paramen\ and learning. 
   Note that all transformations $\get_p$, $p\in\spP$ are to be elements of the same lens, and operations \putl\ are {\em not} indexed by $p$, hence, formalization of learning by considering a family of ordinary lenses would {\em not} do the job. 
 
 \vspace{-2ex}
  \subsubsection{Categorical vs. codiscrete learning}\label{sec:codicodi}
 Suppose that the parameter $p$ is itself a set, \eg, the set of departments forming a view can vary depending on some context. 
 Then an update from $p$ to $p'$ has a relational structure as discussed above, \ie, \frar{e}{p}{p'} is a relation $e\subset p\timm p'$ 
 specifying which departments disappeared from the view and which are freshly added. This is a general phenomenon: as soon as parameters are structures (sets of objects or graphs of objects and attributes), a parameter change becomes a structured delta and the space of parameters gives rise to a category \spP. The search/propagation procedure returns an arrow \frar{e}{p}{p'} in this category, which updates the parameter value from $p$ to $p'$. 
 Hence, a general model of supervised learning should assume \spP\ to be a category (and we say that learning is {\em categorical}). The case of the parameter space being a set is captured by considering a codiscrete category \spP\ whose only arrows are pairs of its objects; we call such learning {\em codiscrete}. 

	\newcommand\figari{\figref{apa-arity}} 
	\newcommand\defri{\defref{def:aalens-arity}}
	\renewcommand\ppgB{\ensuremath{\putl^B}}
\renewcommand\spS{\spA}
\renewcommand\spT{\spB}
\subsection{Ala-lenses
}
\label{sec:alalens}

\renewcommand\putps{\putxy{p}{A}}
\renewcommand\putpsi{\putxyz{(1)}{p_1}{A_1}}
\renewcommand\putpsii{\putxyz{(2)}{p_2}{A_2}}


The notion of a {\em \paramed\ functor (p-functor)} is fundamental for ala-lenses, but is not a lens notion per se and is thus placed into Appendix \sectref{pcatcat}. We will work with its exponential (rather than its product based equivalent) formulation but will do uncurrying and currying back if necessary, and often using the same symbol for an arrow $f$ and its uncurried version $\check f$.
\begin{defin}[\alalens es]
	Let \spS\ and \spT\ be categories. An {\em \alalens}
	from \spS\ (the {\em source} of the lens) to \spT\ (the {\em target}) is 
	a pair $\ell=(\get, \putl)$ whose first component is a p-functor \fprar{\get}{\spS}{\spT}{\spP} and the second component is a triple of (families of) operations 
	$$\putl=(\putps^\upd,\putps^\req,\putps^\self)$$
	indexed by pairs $p\in\obj{\spP}$, $A\in\obj{\spS}$; arities of the operations are  specified below after we introduce some notation. Names \req\ and \upd\ are chosen to match the terminology in \citefst. 
	
Categories \spS, \spT\ are called {\em model spaces}, their objects are called {\em models} and their arrows are {\em (model) updates} or {\em deltas}.
Objects of \spP\ are called {\em parameters} and are denoted by small letters $p,p',..$ rather than capital ones to avoid confusion with \citefst, in which capital $P$ is used for the entire parameter set.  Arrows of \spP\ are called {\em parameter deltas}.
For a parameter  $p\in\obj{\spP}$, we write 
$\get_p$ 
for the functor \frar{\get(p)}{\spS}{\spT} (read ``get \spT-views of \spS''), and if $A\in\obj{\spS}$ is a source model, its $\get_p$-view is denoted by $\get_p(A)$ or $A.\get_p$ or even $A_p$ (so that ${\_}_p$ becomes yet another notation for functor $\get_p$). 

Given a parameter delta \frar{e}{p}{p'} and a source model $A\in\obj{\spS}$, the model delta \frar{\get(e)}{\get_p(A)}{\get_{p'}(A)} will be denoted by $\get_e(A)$ or $e_A$ (rather than $A_e$ as we would like to keep capital letters for objects only). In the uncurried version, $\get_e(A)$ is nothing but $\check \get(e,\id_A)$

Since $\get_e$ is a natural transformation, for any delta \frar{u}{A}{A'} we have a commutative square $e_A;u_{p'}=u_p;e_{A'}$ (as shown by the right face of the prism in \figari). 
We will denote the diagonal of this square by $u.\get_e$ or \frar{u_e}{A_p}{A'_{p'}}.  Thus, we use notation
     \begin{myeq}{getnotation}
\begin{array}{l}
	A_p\eqdef A.\get_p\eqdef \get_p(A)\eqdef\get(p)(A)
	\\	
	\frar{u_e\eqdef u.\get_e \eqdef\get_e(u)\eqdef\get(e)(u)\eqdef e_A;u_{p'}\eqnat u_p;e_{A'}}{A_p}{A'_{p'}}
\end{array}
\end{myeq}
	\newcommand\psobpr{\enma{\obj{\spP}\times \obj{\spS}}}
Now we describe operations \putl. They all have the same indexing set \psobpr,
and the same domain $\spB(A_p,\_)$, \ie, for any index $(p,A)$ (we will omit brackets below) and any model delta 
\frar{v}{A_p}{B'} in \spT,
three  values $\putps^\x(v)$, $\x\in\{\req,\upd,\self\}$ are uniquely defined:
\begin{myeq}{putarity}
\begin{array}{l}
\mbox{$\putps^\upd(v)\in\spP(p,\_)$ is a parameter delta from $p$}
\\ [5pt]
\mbox{$\putps^\req(v)\in\spA(A,\_)$ is a model delta from $A$}
\\ [5pt]
\mbox{$\putps^\self(v)\in\spB(B',\_)$ is a model delta from $B'$ called an {\em amendment}}
\end{array}
\end{myeq}

Note that the definition of $\putl^\self$ involves an equational dependency between all three operations: for all $A\in\obj{\spS}$,  $v\in\spT(A.\get,\_)$, we require 
				\\ [5pt]
\begin{tabular}{l@{\quad}l}
	\lawnamebr{Putget}{0}
	& 
	\mbox{ 
		$(\putl^\req_A.v).\ta.\get_{p'}  = (v;\putl^\self_A).\ta$ 
		where $p'=(\putl^\upd_A.v).\ta$
	} 
\end{tabular}
\\ [5pt]
as demonstrated by the diagram in \figari\ (which will be explained in detail in a moment).

We will write an ala-lens as an arrow \fprar{\ell=(\get,\putl)}{\spS}{\spT}{\spP}.

\medskip
	A lens is called {\em (twice) codiscrete} if categories \spS, \spT, \spP\ are codiscrete and thus \fprar{\get}{\spS}{\spT}{\spP} is a parameterized function. If only \spP\ is codiscrete, we call \elll\ a {\em codiscretely learning} delta lens, while if only model spaces are codiscrete, we call \elll\ a {\em categorically learning} codiscrete lens.  
	\end{defin}


\shortlong{
\noindent
\begin{minipage}{1.\textwidth}
\begin{wrapfigure}{R}{0.5\textwidth} \vspace{-0.6cm}
		\centering
	\centering
	\shortlong{
	   \includegraphics[width=0.45\textwidth]%
	{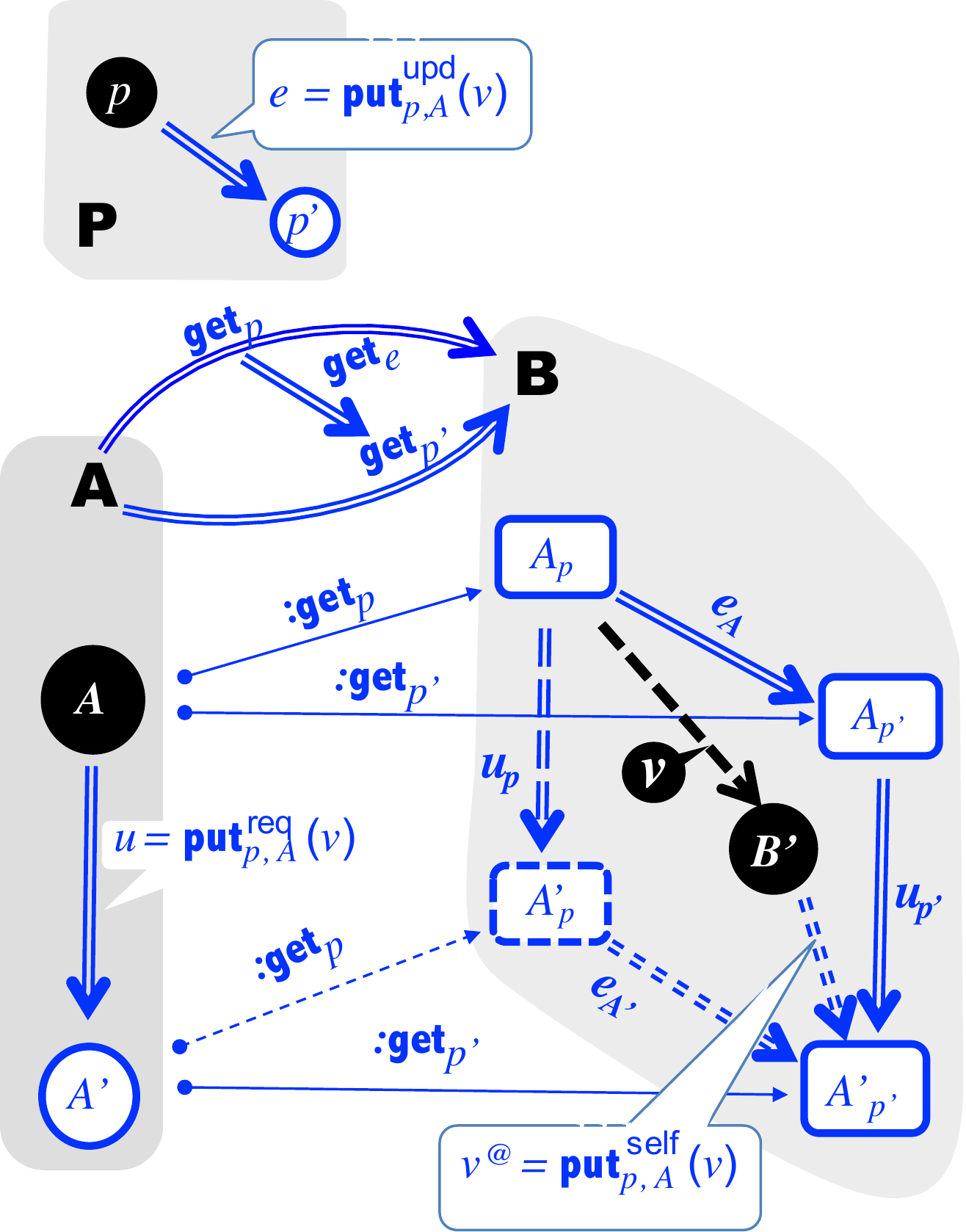}
}
{
	   \includegraphics[width=0.45\textwidth]%
{ll-figures/pdfs/alalensArity-pic-S2A.pdf}
}	
	\caption{Ala-lens operations}
	\label{fig:apa-arity}
\vspace{-0.6cm} \end{wrapfigure}
Diagram in \figari\ shows how a lens' operations are interrelated. The upper part shows an arrow \frar{e}{p}{p'} in category \spP\ and two corresponding functors from \spS\ to \spT. The lower part is to be seen as a 3D-prism with visible front face $AA_{p'}A'_{p'}A'$ and visible upper face $AA_pA_{p'}$, the bottom and  two back faces are invisible, and the \corring\ arrows are dashed. The prism denotes an algebraic term: given elements are shown with black fill and white font while derived elements are blue (recalls being mechanically computed) and blank (double-body arrows are considered as ``blank''). The two pairs of arrows originating from $A$ and $A'$ are not blank because they denote pairs of nodes (the UML says {\em links}) rather than mappings/deltas between nodes. 
\end{minipage}
Equational definitions of deltas $e,u,v^@$ are written up in the three callouts near them. 
The right back face of the prism is formed by the two vertical derived deltas 
$u_p=u.\get_p$ and $u_{p'}=u.\get_{p'}$, and the two matching them horizontal derived deltas $e_S=\get_e(A)$ and  $e_{A'}=\get_e(A')$; together they form a commutative square due to the naturality of $\get(e)$ as explained earlier.
}  
{
\begin{figure}
		\centering
	\centering
	\shortlong{
	   \includegraphics[width=0.45\textwidth]%
	{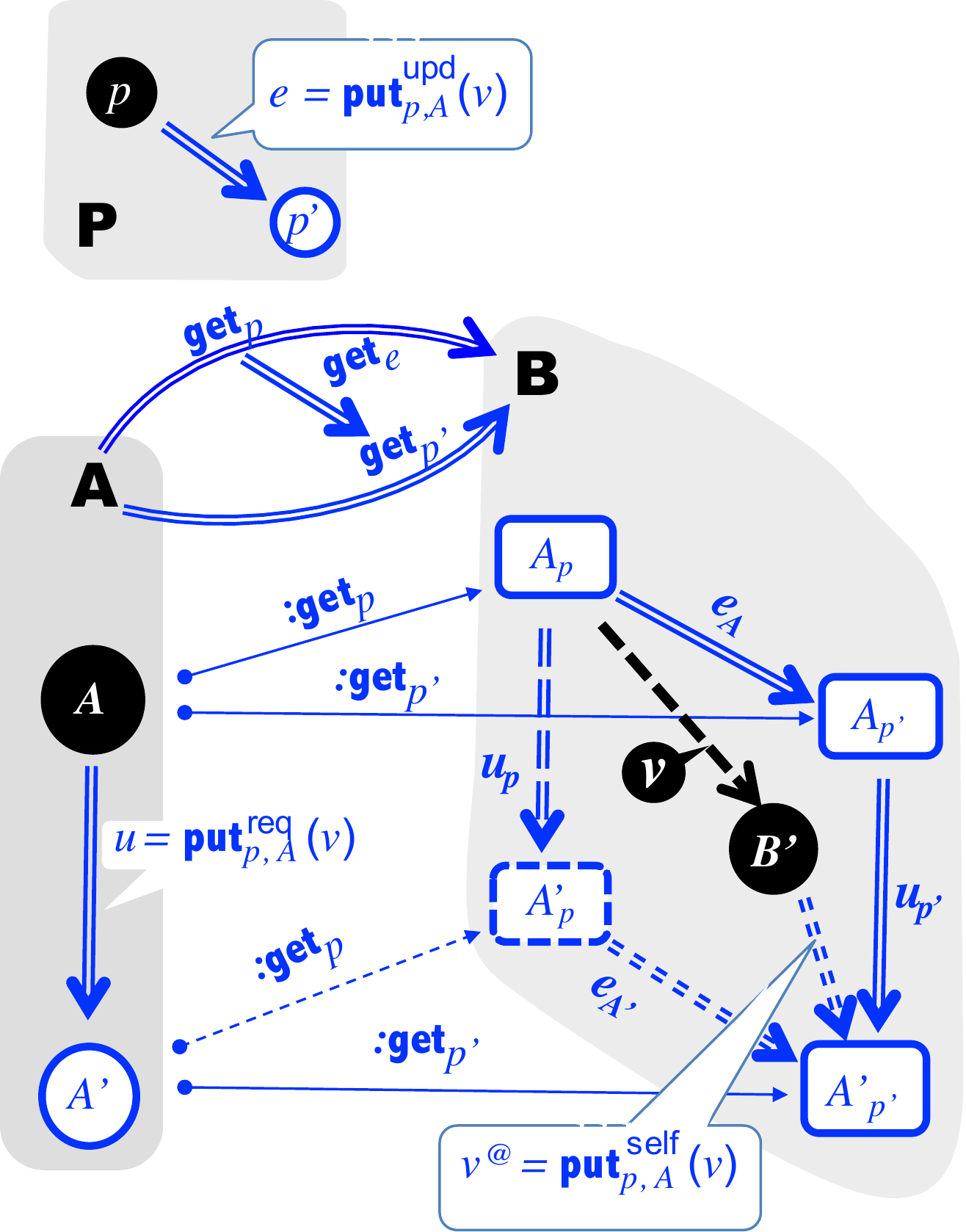}
}
{
	   \includegraphics[width=0.55\textwidth]%
{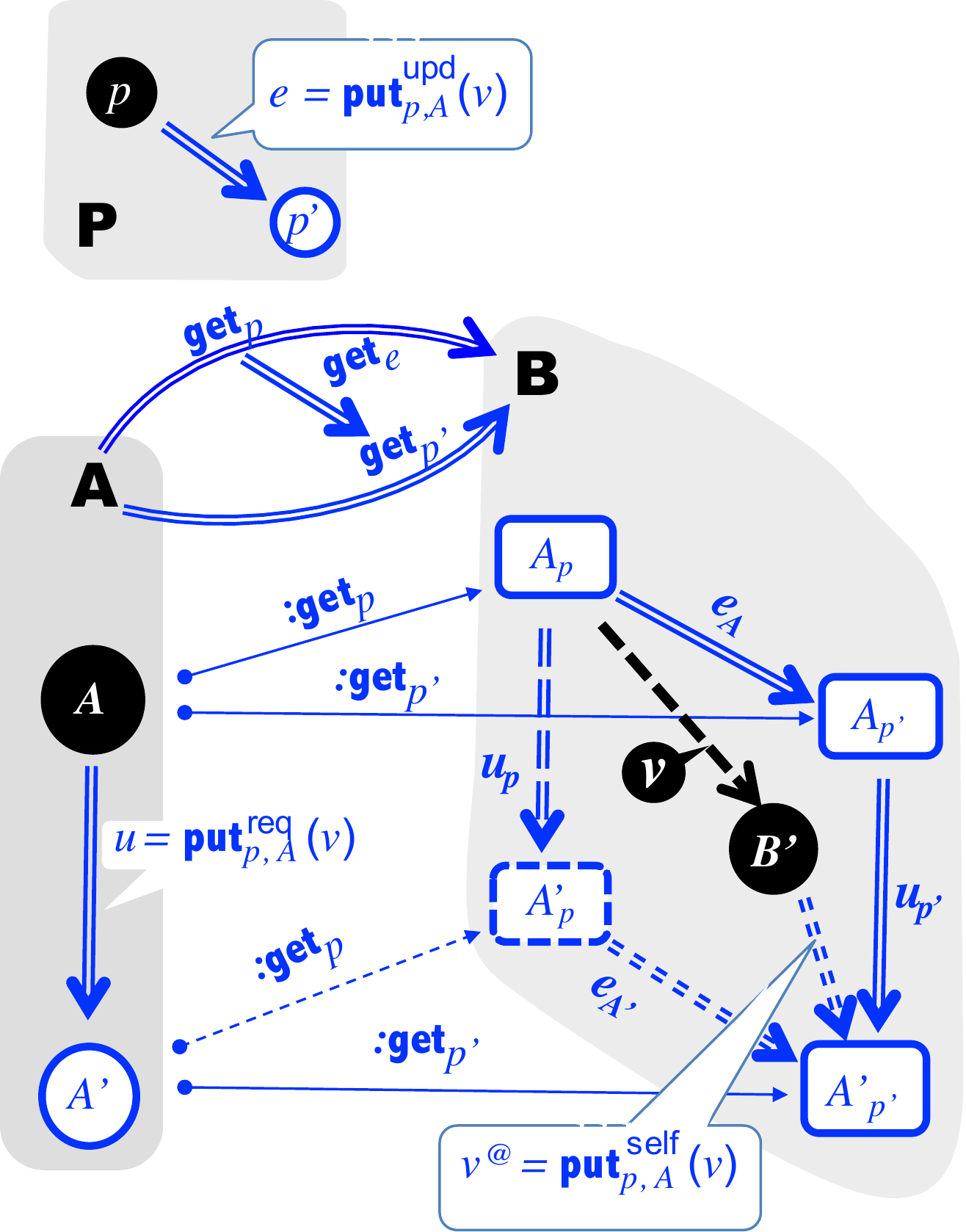}
}	
	\caption{Ala-lens operations}
	\label{fig:apa-arity}
\end{figure}
Diagram in \figari\ shows how a lens' operations are interrelated. The upper part shows an arrow \frar{e}{p}{p'} in category \spP\ and two corresponding functors from \spS\ to \spT. The lower part is to be seen as a 3D-prism with visible front face $AA_{p'}A'_{p'}A'$ and visible upper face $AA_pA_{p'}$, the bottom and  two back faces are invisible, and the \corring\ arrows are dashed. The prism denotes an algebraic term: given elements are shown with black fill and white font while derived elements are blue (recalls being mechanically computed) and blank (double-body arrows are considered as ``blank''). The two pairs of arrows originating from $A$ and $A'$ are not blank because they denote pairs of nodes (the UML says {\em links}) rather than mappings/deltas between nodes. Equational definitions of deltas $e,u,v^@$ are written up in the three callouts near them. 
The right back face of the prism is formed by the two vertical derived deltas 
$u_p=u.\get_p$ and $u_{p'}=u.\get_{p'}$, and the two matching them horizontal derived deltas $e_S=\get_e(A)$ and  $e_{A'}=\get_e(A')$; together they form a commutative square due to the naturality of $\get(e)$ as explained earlier.
}

\begin{defin}[\shortlong
	{Well-behavedness}{Well-behavedness}]\label{def:wbalens}{\emno 
%
	An \alalens\ is called {\em well-behaved (wb)} if the following two laws hold 
	for all $p\in \xob{\spP}$, $A\in\xob{\spS}$ and \frar{v}{A_p}{B'}}:
	\\[1ex]
	\lawgap=1 ex
	\noindent \begin{tabular}{l@{\quad}l}
		\lawnamebr{Stability}{
		}	& \mbox{
		if $v=\id_{A_p}$ 
		then all three propagated updates 
		$e,u,\amex{v}$ are identities:} 
		\\
		& $\putps^\upd(\id_{A_p})=\id_p$, \quad%
		$\putps^\req(\id_{A_p}) = \id_S$, \quad%
		$\putps^\self(\id_{A_p})=\id_{A_p}$
		\\ [\lawgap] \lawnamebr{Putget}{
		} & 
	\mbox{ 
		 $(\putps^\req.v).\get_{e} = v; v^@$ 
		 where $e=\putps^\upd(v)$ and $v^@= \putps^\self(v)$
	 } 
 \hippoOutin{}{
	 	\\ [\lawgap]
				\lawnamebr{Hippocr}{
		}	& \mbox{\em 
			if $\get_p(u)=v$ for some \frar{u}{A}{A'}, 
			then $v^@=\id_{B'}$
	                       }
                       } 
                	\end{tabular}
\end{defin}  
\begin{remark}[Lens laws continued]
	\label{rema:lenslaw3}
Note that \remaref{putget4codi} about the Putget law is again applicable. 
\hippoOutin{}{
}  
\end{remark}
 
\setlength{\mygap}{-3ex}
\begin{example}[Identity lenses]\label{ex:idlens} 
	Any 
	category \spA\ gives rise to an \alalens\ $\idl_\spA$ with the following components. The source and target spaces are equal to \spA, and the parameter space is \tcat. 
	Functor $\get$ is the identity functor and all \putl s are identities. Obviously, this lens is wb.
\end{example}
\vspace{\mygap}
\begin{example}[Iso-lenses]\label{ex:isolens}
	Let \frar{\iota}{\spA}{\spB} be an isomorphism between model spaces. It gives rise to a wb \alalens\ \frar{\ell(\iota)}{\spA}{\spB} with $\spP^{\ell(\iota)}=\tcat=\{*\}$ as follows. Given any $A$ in \spA\ and \frar{v}{\iota(A)}{B'} in \spB, we define $\putl^{\ell(\iota).\req}_{*,A}(v) = \iota^{-1}(v)$ while the two other put operations map $v$ to identities. 
\end{example}
\vspace{\mygap}
\begin{example}[Bx lenses]\label{ex:bxlens} Examples of wb aa-lenses modelling a Bx can be found in \cite{me-faoc19}: they all can be considered as \alalens es with a trivial parameter space \tcat. 
\end{example}
\vspace{\mygap}
\begin{example}[Learners] \label{ex:learner}
	Learners defined in \citefst\ are codiscretely learning codiscrete lenses with amendment, and as such satisfy (amended) Putget (see \remaref{putget4codi}). 
\end{example}

	   \renewcommand\kl{\ensuremath{\ekk;\ell}} 
		\renewcommand\figrii{\figref{kl-lens}}

\section{
	Compositionality of ala-lenses 
} \label{sec:alalenscat}

This section explores the compositional structure of the universe of ala-lenses; especially interesting is their sequential composition. We will begin with a small example demonstrating sequential composition of ordinary lenses and showing that the notion of update policy transcends individual lenses. Then we define sequential and parallel composition of ala-lenses (the former much more involved than for ordinary lenses) and show that wb ala-lenses can be organized into an sm-category. Finally, we formalize the notion of a compositional update policy via the notion of a compositional bidirectional language.  

\subsection{Compositionality of  update policies: An example }\label{sec:ex-policyfun}
\renewcommand\figrii{\figref{ex-policyfun}}
\newcommand\figrsch{\figref{ex-sixlenses}}

\figrii\  extends  the example in \figri\ with a new model space \spC\ whose schema consists of the only attribute \name, and a view of the IT-view, in which only employees of the ML department are to be shown. Thus, we  now have two functors, \frar{\get1}{\spA}{\spB} and \frar{\get2}{\spB}{\spC}, and their composition \frar{\Get}{\spA}{\spC} (referred to as the {\em long} get). The top part of \figrii\ shows how it works for model $A$ considered above. 

\setlength{\twomodwid}{\textwidth}

\begin{figure}
    \vspace{-0.25cm}
       \centering
   \shortlong{
    \includegraphics[width=0.9\twomodwid]%
    {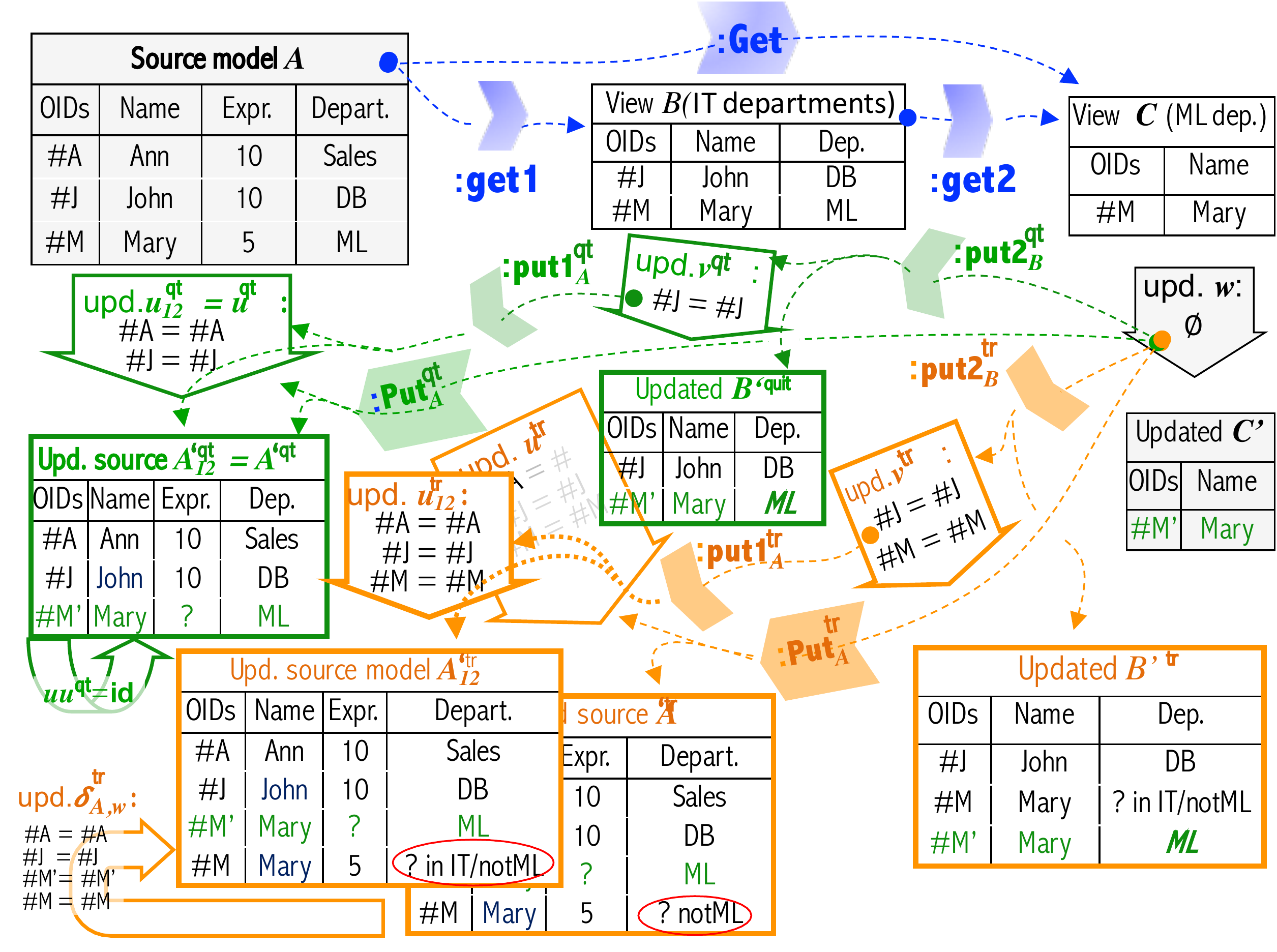}
}{
    \includegraphics[width=1.\twomodwid]%
{ll-figures/pdfs/ex-policyfun-pic.pdf}
}
    \caption{Example cont'd: functoriality of update policies}
    \label{fig:ex-policyfun}
   \vspace{-0.5cm}
\end{figure}

\begin{wrapfigure}{R}{0.33\columnwidth} \vspace{-0.25cm}
\centering
    \includegraphics[
                    width=0.33\columnwidth%
                 ]%
                 {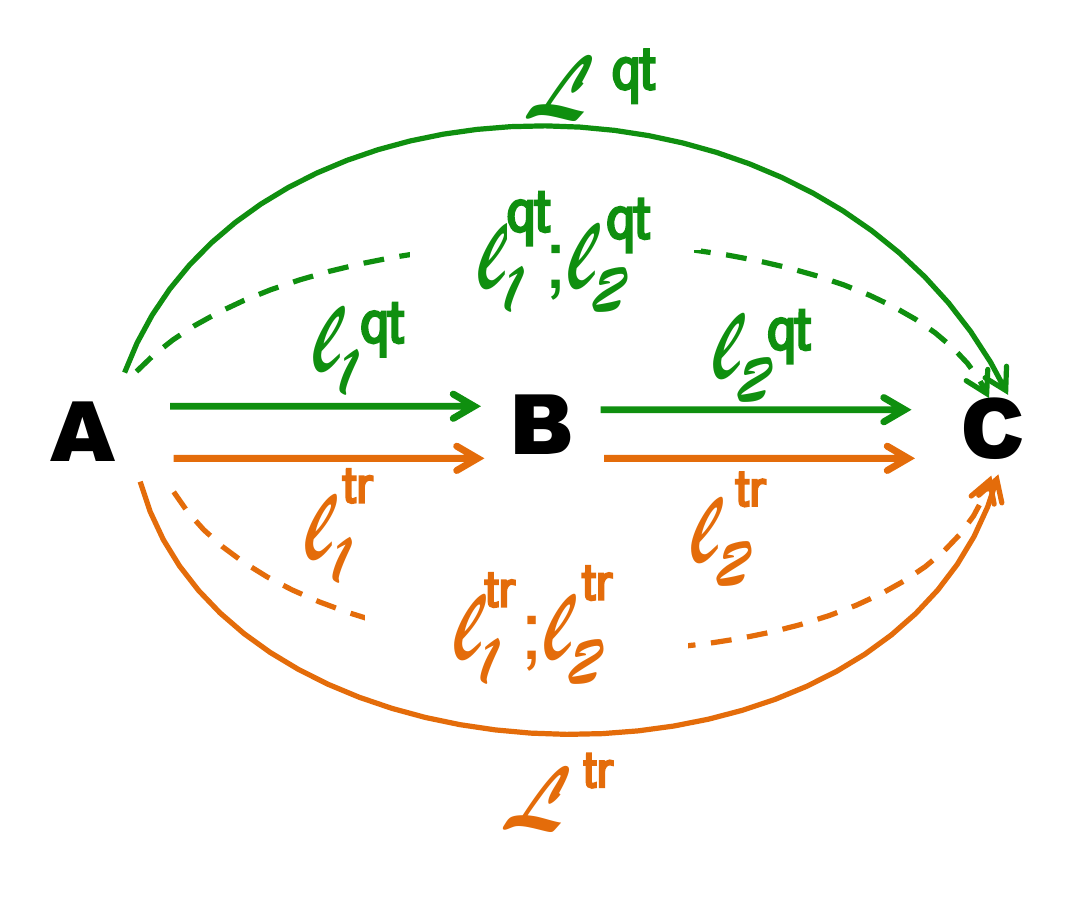}
                 \vspace{-20pt}
\caption{Lens combination schemas for \figref{ex-policyfun}}
\label{fig:ex-sixlenses}
\vspace{-17.5pt}
\end{wrapfigure}
Each of the two policies,  
policy \quit\ (green) and policy \trans\  (orange), in which person's disappearance from the view are interpreted, resp., as quiting the company and transitioning to a department not included into the view, is applicable to the new view mappings $\get2$ and $\Get$, thus giving us six lenses shown in \figrsch\ with solid arrows; amongst them, lenses, $\nil^{\quit}$ and  $\nil^\trans$  are obtained by applying policy \pol\ to the (long) functor \Get;, and we will refer to them {\em long} lenses. In addition, we can compose lenses of the same colour 
as shown in \figrsch\ by dashed arrows (and we can also compose lenses of different colours ($\ell_1^\quit$ with $\ell_2^\trans$ and $\ell_1^\trans$ with $\ell_2^\quit$) but we do not need them). Now an important question is how composed lenses and long lenses are related: whether $\nil^\pol$ and $\ell_1^\pol;\ell_2^\pol$ for
$\pol\in\{\quit,\trans\}$, are equal (perhaps up to some equivalence) or different?  

\figrii\ demonstrates how the mechanisms work with a simple example. 
We begin with an update $w$ of the view $C$ that says that Mary $\#M$ left the ML department, and a new Mary $\#M'$ was hired for ML. Policy \quit\ interprets Mary's disappearance as quiting the company, and hence this Mary doesn't appear in view $B'^\quit$ produced by $\putl2^\quit$ nor in view $A'^\quit_{12}$ produced from $B'^\quit$ by $\putl1^\quit$, and updates $v^\quit$ and $u^\quit_{12}$ are written accordingly. Obviously, Mary also does not appear in view $A^{'\quit}$ produced by the long lens's $\Putl^\quit$. Thus, $\putl1_A^\quit(\putl2_A^\quit(w))=\Putl_A^\quit(w)$, and it is easy to understand that such equality will hold for any source model $A$ and any update \frar{w}{C}{C'} due to the nature of our two views $\get1$ and $\get2$. Hence, 
    $\nil^\quit=\ell_1^\quit ; \ell_2^\quit$ where $\nil^\quit=(\Get, \Putl^\quit)$ and $\ell_i^\quit=(\get i, \putl{i}^\quit)$. 
    
    The situation with policy \trans\ is more interesting. Model $A^{'\trans}_{12}$ produced by the composed lens $\ell_1^\trans ; \ell_2^\trans$, and model  $A^{'\trans}$ produced by the long lens $\nil^\trans=(\Get, \Putl^\trans)$ are different as shown in the figure  (notice the two different values for Mary's department framed with red ovals in the models). Indeed, the composed lens has more information about the old employee Mary---it knows that Mary was in the IT view, and hence can propagate the update more accurately. The comparison update \frar{\delta_{A,w}^\trans}{A'^\trans}{A'^\trans_{12}} adds this missing information so that equality $u^\trans ; \delta^\trans_{A,w}=u^\trans_{12}$ holds.  This is a general phenomenon: functor composition looses information and, in general, functor $\Get=\get1;\get2$ knows less than the pair $(\get1,\get2)$. Hence, operation $\Putl$ back-propagating updates over \Get\ (we will also say  {\em inverting} \Get) will, in general, result in less certain models than composition $\putl1\circ\putl2$ that inverts the composition $\get1;\get2$ (a discussion and examples of this phenomenon in the context of vertical composition of updates can be found in \cite{me-bx17}). Hence, comparison updates such as $\delta^\trans_{A,w}$ should exist for any $A$ and any \frar{w}{A.\Get}{C'}, and together they should give rise to something like a natural transformation between lenses, 
    \drar{\delta^\trans_{\spA,\spB,\spC} }%
    {\nil^\trans}%
    {\ell^\trans_1;\ell^\trans_2}. To make this notion precise, we need a notion of natural transformation between ``functors'' \putl, which we leave for future work. In the present paper, we will consider policies like \quit, for which strict equality holds.

\subsection{Sequential composition of \alalens es}\label{con:apa-seq}  

	Let \frar{\ekk}{\spA}{\spB} and \frar{\ell}{\spB}{\spC} be two \alalens es with parameterized functors \frar{\get^\ekk}{\spP}{[\spA,\spB]} and \frar{\get^\ell}{\spQ}{[\spB,\spC]} resp. 
	Their {\em composition}  is the following \alalens\ \kl.
	Its parameter space is the product $\spP\times\spQ$, and the \get-family is defined as follows. For any pair of parameters $(p,q)$ (we will write \pq), 
	\frar{\get^{\kl}_{pq}=\get^\ekk_p;\get^\ell_q}{\spA}{\spC}.
	Given a pair of parameter deltas, \frar{e}{p}{p'} in \spP\ and \frar{h}{q}{q'} in \spQ, their $\get^{\kl}$-image is the Godement product $*$ of natural transformations, $\get^{\kl}(eh)=\get^\ekk(e)*\get^\ell(h)$ ( we will also write $\get^\ekk_e \,{||} \,\get^\ell_h$)

	\begin{figure}
	\centering
\shortlong{
	\includegraphics[width=0.9\textwidth]%
	{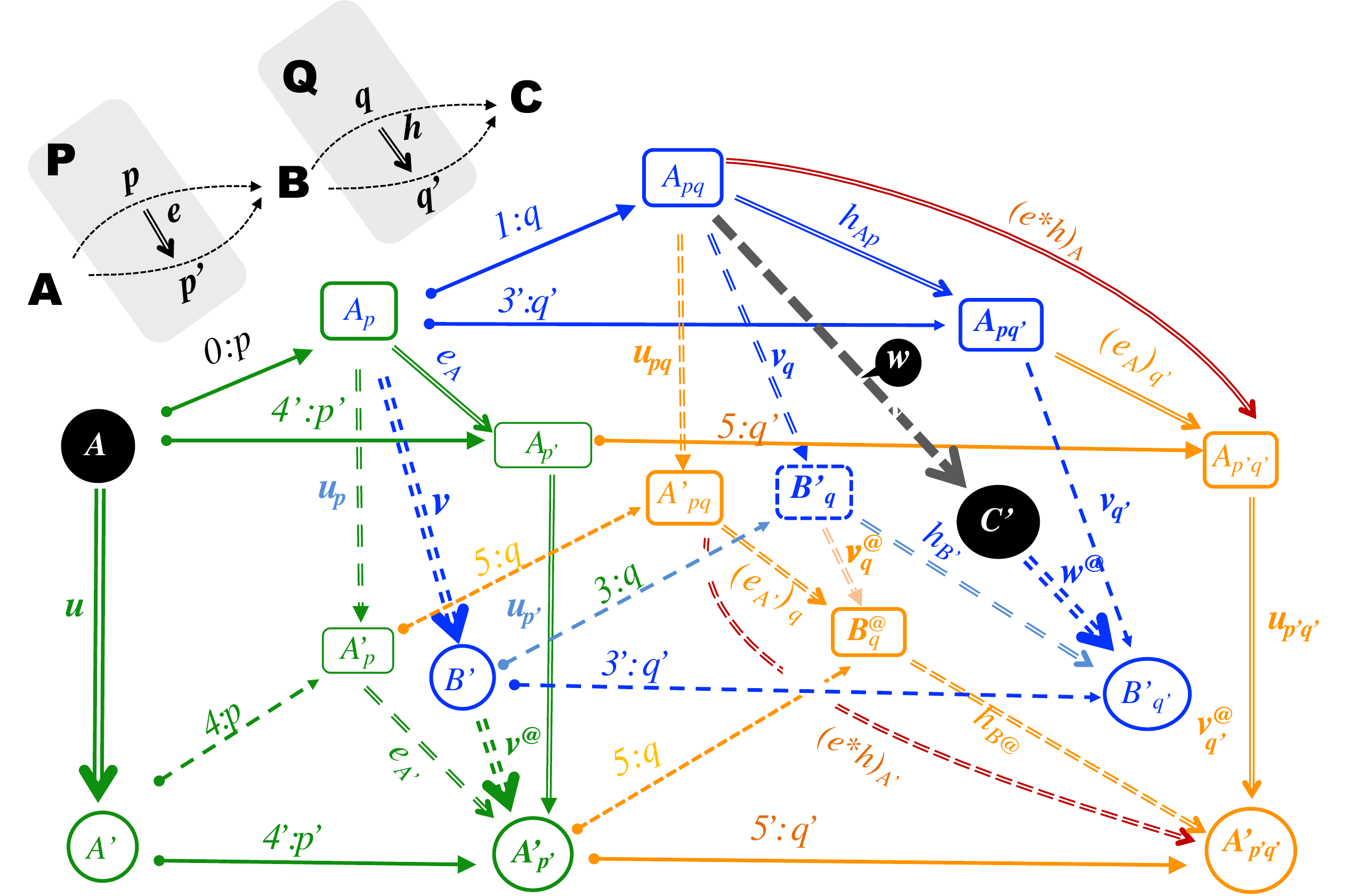}
}{
	\includegraphics[width=1.\textwidth]%
{ll-figures/pdfs/aalens-hComp-pic}
}
	\caption{Sequential composition of apa-lenses}
	\label{fig:kl-lens}
\end{figure}

	Now we define \kl's propagation operations \putl s. Let $(A, pq, A_{pq})$ with $A\in\xob{\spA}$, $\pq\in \objbra{\spP\timm\spQ}$, $A.\get^\ekk_p.\get^\ell_q=A_{pq}\in\obj{\spC}$ be a \zdmodok{state}{state}{} of lens \kl, and  \frar{w}{A_\pq}{C'} is a target update as shown in \figrii. 
	For the first propagation step, we run lens $\ell$ as shown in \figrii\ with the blue colour for derived elements: this is just an instantiation of the pattern of \figari\ with the source object being $A_p=A.\get_p$ and parameter $q$. The results are deltas 
	\begin{equation}\label{eq:kldef1}
	\frar{h=\putxy{q}{A_p}^{\ell.\upd}(w)}{q}{q'}, 
	\frar{v=\putxy{q}{A_p}^{\ell.\req}(w)}{A_p}{B'}, 
	\frar{\amex{w}=\putxy{q}{A_p}^{\ell.\self}(w)}{C'}{B'_{q'}}.
	\end{equation}
	%
	Next we run lens $\ekk$ at state $(p,A)$ and the target update $v$ produced by lens $\ell$; it is yet another instantiation of pattern in \figari\ (this time with the green colour for derived elements), which produces three deltas
	\begin{equation}\label{eq:kldef2}
	\frar{e=\putxy{p}{A}^{\ekk.\upd}(v)}{p}{p'},
	\frar{u=\putxy{p}{A}^{\ekk.\req}(v)}{A}{A'}, 
	\frar{\amex{v}=\putxy{p}{A}^{\ekk.\self}(v)}{B'}{A'_{p'}}.
\end{equation}
These data specify the green prism adjoint to the blue prism: the edge $v$ of the latter is the ``first half'' of the right back face diagonal $A_pA'_{p'}$ of the former. In order to make an instance of the pattern in \figari\ for lens \kl, we need to extend the blue-green diagram to a triangle prism by filling-in the corresponding ``empty space''.
These filling-in arrows are provided by functors $\get^\ell$ and $\get^\ekk$ and shown in orange  (where we have chosen one of the two equivalent ways of forming the Godement product -- note two curve brown arrows).
In this way we obtain yet another instantiation of the pattern in \figari\ denoted by \kl:
\begin{equation}\label{eq:kldef3}
\putxy{A}{\pq}^{\klbr\upd}(w) = (e,h), \quad
\putxy{A}{\pq}^{\klbr\req}(w) = u, \quad
\putxy{A}{\pq}^{\klbr\self}(w) = w^@; v^@_{q'}  
\end{equation}
where 
$v^@_{q'}$ denotes $v^@.\get_{q'}$. 
Thus, we built an \alalens\ \kl, which satisfies equation \putgeto\ 
by construction.  

\hippoOutin{}{ 
{\newcommand\xell{\enma{}}
\begin{defin}[Get and Put images of a lens]
	Given a lens from space \spS\ to space \spT, define 
	$$
	\begin{array}{l}
	\Get^\xell(\spS)=\comprCol{u.\get_p\, }{u\in\Arr\spS, p\in|\spP|}\subset \Arr\spT
	\\ 
	\Putl^\xell(\spT)=\comprCol{\putl^\xell_{p,S}.v\,}{p\in|\spP|, S\in|\spS|, \frar{v}{\get_p(S)}{\_}\in\Arr\spT}\subset\Arr\spS
	\end{array}
	$$
	
	Two consecutive lenses \frar{\ekk}{\spA}{\spB}, \frar{\ell}{\spB}{\spC} are called {\em well-matched} if 
	$
\Get^\ekk(\spA)\superset\Put^\ell(\spC).
	$
\end{defin}
} 
} 
\vspace{0.25\mygap}
\begin{theorem} [Sequential composition and lens laws] \label{th:seqwb}
	Given \alalens es \frar{\ekk}{\spA}{\spB} and \frar{\ell}{\spB}{\spC}, let lens \frar{\kl}{\spA}{\spC} be their sequential composition as defined above. 
	Then the  lens \kl\ is wb\ as soon as lenses $\ekk$ and $\ell$ are such.  
\hippoOutin{}{ 
	Moreover, if lenses \ekk\ and \elll\ are wb and additionally {well-matched} 
then lens \kl\ is wb too.  
} 
\end{theorem}
\shortlong{See  \citelongver{Appendix A.3} for a proof.}{The proof is in Appendix \ref{sec:seqlenswb-proof}.}
\subsection{Parallel composition of \alalens es}\label{con:apa-par}
Let \frar{\ell_i}{\spA_i}{\spB_i}, $i=1,2$ be two \alalens es with parameter spaces $\spP_i$. The lens 
\frar{\ll}{\spA_1\timm\spA_2}{\spB_1\timm\spB_2} is defined as follows. 
Parameter space $\ll.\spP=\spP_1\times\spP_2$. For any pair $p_1\paar p_2\in \objbra{\spP_1\timm\spP_2}$,  define $\getxy{\ll}{\pp}=\getxy{\ell_1}{p_1}\times\getxy{\ell_2}{p_2}$ (we denote pairs of parameters by $p_1\paar p_2$ rather than $p_1\otimes p_2$ 
to shorten long formulas going beyond the page width).
Further, for any pair of models $ 
\ss \in \objbra{\spA_1 \times \spA_2}
$ and deltas 
\frar{v_1\paar v_2}%
       {(\ss).\getxy{\ll}{\pp}} 
       {B'_1\paar B'_2},
       we define componentwise
    $$
       \begin{array}{l}
       \frar{e=\putxyz{\llbr\upd}{\pp}{\ss}(v_1\paar v_2)}{\pp}{p'_1\paar p'_2} 
        \end{array}
$$
by setting $e=e_1\paar e_2$ where %
	$e_i=\putxyz{{\ell_i}}{p_i}{S_i}(v_i)$, $i=1,2$  
and similarly for $\putxyz{\llbr\req}{\pp}{\ss}$ and \putxyz{\llbr\self}{\pp}{\ss}
The following result is obvious. 
\begin{theorem}[Parallel composition and lens laws]\label{th:parwb}
Lens $\ell_1\paar\ell_2$ is wb 
as soon as lenses $\ell_1$ and $\ell_2$ are such.
\end{theorem}
\subsection{Symmetric monoidal structure over \alalens es }
Our goal is to organize \alalens es into an sm-category. To make sequential   composition of \alalens es associative, we need to consider them up to some equivalence (indeed, Cartesian product is not strictly associative).
\begin{defin}[\Alalens\ Equivalence]
	Two parallel \alalens es \frar{\ell, \hell}{\spS}{\spT} are called {\em equivalent} if their parameter spaces are isomorphic via a functor \frar{\isop}{\spP}{\hat\spP} such that for any $A\in\xob{\spS}$,  $\frar{e}{p}{p'} \in\spP$ and \frar{v}{(A.\get_p)}{T'} the following holds (for $\x\inn\{\req,\self \}$):
	$$
	A.\get_e=A.\hget_{\isop(e)}, 
	\isop(\putps^\upd(v))=\hputl_{\hp,A}(v), 
		\mbox{ and }
	\putps^{\x}(v)=\hputl^\x_{\isop(p),A}(v) 
	$$
\end{defin}
\begin{remark}It would be more categorical to require delta isomorphisms (\ie, commutative squares whose horizontal edges are isomorphisms) rather than equalities as above. However, model spaces appearing in \bx-practice are skeletal categories (and even stronger than skeletal in the sense that all isos, including iso loops, are identities), for which isos become equalities so that the generality would degenerate into equality anyway. 
\end{remark}
\mygap=-1ex
\vspace{\mygap}
	It is easy to see that operations of lens' sequential and parallel composition are compatible with lens' equivalence and hence are well-defined for equivalence classes. 
Below we identify lenses with their equivalence classes by default. 
\vspace{\mygap}
\begin{theorem}[\Alalens es form an sm-category]\label{th:alalens2smcat}
	Operations of sequential and parallel composition of \alalens es defined above give rise to an sm-category \apalenscat, whose objects are  model spaces (= categories) 
	and arrows are (equivalence classes of) \alalens es.  
\end{theorem}
\shortlong{
See \citelongver{p.17 and Appendix A.4} for the construction.
}{
{\em Proof.} It is easy to check that identity lenses $\idl_\spA$ defined in \exref{idlens} are the units of the sequential lens composition defined above. The proof of associativity is rather ``intertwined'' and is placed into Appendix \ref{sec:seqlensass-proof}. Thus, \apalenscat\ is a category. 
Next we define a monoidal structure over this category. The monoidal product of objects is Cartesian product of categories, 
and the monoidal product of arrows is lens' parallel composition defined above. The monoidal unit is the terminal category \tcat. Associators, left and right unitors, and braiding are iso-lenses generated by the respective isomorphism functors (\exref{isolens}). Moreover, it is easy to see that the iso-lens construction from \exref{isolens} is actually a functor \frar{\catname{isolens}}{\catcat_{\iso}}{\apalenscat}. Then  as a) \catcat\ is \smm\ and fulfils all necessary monoidal equations, and b) \isolensfun\ is a functor, these equations hold for the \alalens images of $\catcat_\iso$-arrows, and \apalenscat\ is \smm\ too (cf. a similar proof in \citefst\ with $(\setcat,\times)$ instead of $(\catcat,\times)$).  
}  


\subsection{Functoriality of learning in the {\em delta} lens setting}\label{sec:ddlens-1}
\newcommand\IAoutin[2]{#1} 


%


As example in \sectref{ex-policyfun} shows, the notion of update policy transcends individual lenses. Hence, its proper formalization needs considering the entire category of ala-lenses and functoriality of a suitable mapping.

	\renewcommand\grt[1]{#1}
\noindent	
\begin{minipage}[c]{0.61\linewidth}
	\vspace{2ex}
		\begin{defin}[Bx-transformation language] \label{def:mtl}
			A {\em compositional bidirectional model \trafon\ language} \mtl\ is given  by 
			{\bf (i)} an  sm-category $\pgetcat(\mtl)$ of {\em \mtl-model spaces {\em and} \mtl-transformations} 
			supplied with forgetful functor into \pcatcat,  
			and 
			{\bf (ii)} an sm-functor 
									\frar{L_\mtl}{\pgetcat(\mtl)}{\alalenscat} 
				such that the lower triangle in the inset diagram  commutes. (Forgetful functors in this diagram are named ``$-X$'' with $X$ referring to the structure to be forgotten.)
					\end{defin}
	\end{minipage}\quad	
	\begin{minipage}[c]{0.3\linewidth}
	\begin{tabular}{c}
\begin{diagram}[small]
	&&&& 
	\alalenscatwbind
	\\ 
	&&& 
	\ruTo(4,2) 
	&
	\dEmbed(2,2)>{{-}\wbind}
	\\ 
	\grt{\pgetcat(\mtl)}
	&&\rTo^{L_\mtl}
	&&\alalenscat
	\\ 
	&
	\rdTo(2,2)<{{-}\mtl}
	&&
	\ldTo(2,2)_{\minusput} 
	& 
	\\ 
	&&\pcatcat&& 
\end{diagram}
\end{tabular}
\end{minipage}%
				
				An \mtl-language is {\em well-behaved (wb)} if functor $L_\mtl$ factorizes as shown by the upper triangle of the diagram. 
\par\medskip

\noindent {\em Example.} A major compositionality result of Fong \etal\  \citefst\ states the existence of an sm-functor from the category of Euclidean spaces and parameterized differentiable functions (pd-functions) \paracat\ into the category \learncat\ of learning algorithms ({\em learners}) as shown by the inset commutative diagram.  (The functor 

\begin{wrapfigure}{R}{0.29\textwidth}
\vspace{-1.25cm}
	\centering
\[
\begin{diagram}[small]
\\ 
\grt{\paracat} 
&&\rTo^{\grt{L_{\eps,\err}}} 
&&\learncat 
\\ 
&\rdTo<{\grt{{-}\reals}}
&
&\ldTo>{{-}\putl}
& 
\\ 
&&\psetcat
&&
\end{diagram}
\]
\vspace{-0.75cm}
\label{fig:policyfun-ml1}
\end{wrapfigure}
\noindent is itself parameterized by a {\em step size} $0<\eps\in\reals$ and an {\em error function} \err\ 
needed to specify 
the gradient descent procedure.) However, learners are nothing but codiscrete ala-lenses as shown in \sectref{classPlane}, and thus the inset diagram is a codiscrete  specialization of the diagram in \defref{mtl} above. That is, the category of Euclidean spaces and pd-functions, and the gradient descent method for back propagation, give rise to a (codiscrete) compositional bx-\trafon\ language. 

Finding a specifically \bx\ instance of \defref{mtl} (\eg, checking whether it holds for concrete languages and tools such as {\sc eMoflon} \cite{tony-emoflon-bx19} or {\sc groundTram} \cite{groundtram1}) is laborious and left for future work.

\hippoOutin{}{\vspace{6ex}}
\section{Related work}\label{sec:relwork}

\begin{wrapfigure}{R}{0.31\columnwidth}
\vspace{-0.75cm}
\centering
    \includegraphics[
                    width=0.31\columnwidth%
                 ]%
                 {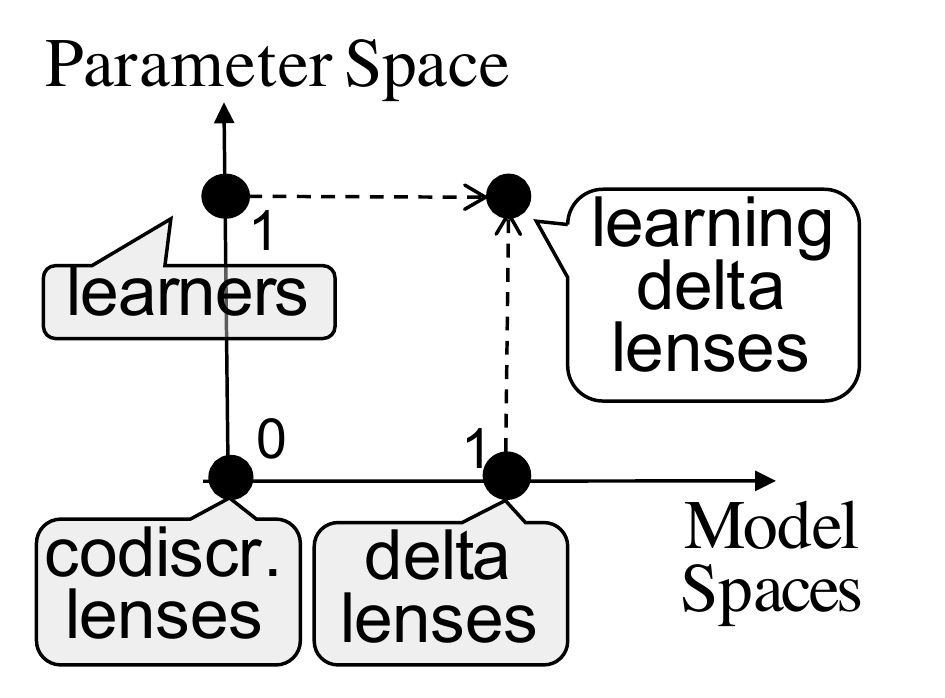}
                 \vspace{-17pt}
\caption{}
\label{fig:mlbx-plane}
\vspace{-15pt}
\end{wrapfigure}
Figure \ref{fig:mlbx-plane} on the right  is a simplified version of \figref{classPlane} convenient for our discussion here:  immediate related work should be found in areas located at points (0,1) (codiscrete learning lenses) and (1,0) (delta lenses) of the plane. For the point (0,1), the paper \citefst\ by Fong, Spivak and Tuy\'eras is fundamental:  they defined the notion of a codiscrete learning lens (called a learner), proved a fundamental results about sm-functoriality of the gradient descent approach to ML, and thus laid a foundation for the compositional approach to change propagation with learning. One follow-up of that work is  paper \citefj\ by Fong and Johnson, in which they build an sm-functor 
$\learncat\rightarrow\slenscat$
which maps learners to so called symmetric lenses. That paper is probably the first one where the terms 'lens' and 'learner' are met, but an initial observation that a learner whose parameter set is a singleton is actually a lens is due to Jules Hedges, see \citefj. 

There are conceptual and technical distinctions between \citefj\ and the present paper. On the conceptual level, by encoding learners as symmetric lenses, they ``hide'' learning inside the lens \fwk\ and make it a technical rather than conceptual idea. In contrast, we consider parameterization and supervised learning as a fundamental idea and a first-class citizen for the lens \fwk, which grants creation of a new species of lenses. Moreover, while an ordinary lens is a way to invert a functor, a learning lens is a way to invert a \paramed\ functor so that learning lenses appear as an extension of the \paramon\ idea from functors to lenses.  (This approach can probably be specified formally by treating \paramon\ as a suitably defined functorial construction.) Besides technical advantages (working with asymmetric lenses is simpler), our asymmetric model seems more adequate to the fact that we deal with functions rather than relations. On the technical level, the lens \fwk\ we develop in the paper is much more general than in \citefj: we categorificated both the parameter space and model spaces, and we work with lenses with amendment. 

As for the delta lens roots (the point (1,0) in the figure), delta lenses were motivated and formally defined in \cite{me-jot11} (the asymmetric case) and \cite{me-models11} (the symmetric one). Categorical foundations for the delta lens theory were developed by Johnson and Rosebrugh in a series of papers (see \citejrunified\ for references); this line is continued in Clarke's work \cite{bryce-act19}. 
The notion of a delta lens with amendments (in both asymmetric and symmetric variants) was defined in \cite{me-faoc19}, and several composition results were proved. 
Another extensive body or work within the delta-based area is modelling and implementing model \trafon s with triple-graph grammars (TGG) \cite{andy-20years,tony-emoflon-bx19}. TGG provide an implementation \fwk\ for delta lenses as is shown and discussed in \cite{tony-viewtgg-ecmfa14,frank-sosym15,tony-bxbook1}, and thus inevitably consider change propagation on a much more concrete level than lenses.  The author is not aware of any work of discussing functoriality of update policies developed within the TGG \fwk.  
The present paper is probably the first one at the intersection (1,1) of the plane. The preliminary results have recently been reported at ACT'19 in Oxford to a representative lens community, and no references besides \citefst, \citefj\ mentioned above were provided.
\section{Conclusion}\label{sec:concl}
\label{sec:future}

The perspective on \bx\ presented in the paper is an example of a fruitful interaction between two domains---ML and \bx. In order to be ported to \bxmde, the compositional approach to ML developed in \citefst\ is to be categorificated as shown in \figref{classPlane} on p.~\pageref{fig:classPlane}. This opens a whole new program for \bx: checking that currently existing \bx\ languages and tools are compositional (and well-behaved) in the sense of \defref{mtl}  p.~\pageref{def:mtl}.  The wb compositionality is an important practical requirement as it allows for modular design and testing of bidirectional transformations.   
Surprisingly, but this important requirement has been missing from the agenda of the \Bx\ community, \eg, the recent endeavour of developing an effective benchmark for \bx-tools \cite{tony-benchmarx1} does not discuss it.

In a wider context, the main message of the paper is that the learning idea transcends its applications in ML: it is applicable and usable in many domains in which lenses are applicable such as model transformations, data migration, and open games \cite{jules-corr19}. Moreover, the categorificated learning may perhaps find useful applications in ML itself. In the current ML setting, the object to be learnt is a function \frar{f}{\reals^m}{\reals^n} that, in the OO class modelling perspective, is a very simple structure: it can be seen as one object with a (huge) amount of attributes, or, perhaps, a predefined set of objects, which is not allowed to be changed during the search 
--- only attribute values may be changed. In the delta lens view, such changes constitute a rather narrow class of updates and thus unjustifiably narrow the search space.  Learning with the possibility to change dimensions $m,n$ may be an appropriate option in several contexts. On the other hand, while categorification of model spaces extends the search space, categorification of the parameter space would narrow the search space as we are allowed to replace a parameter $p$ by parameter $p'$ only if there is a suitable arrow \frar{e}{p}{p'} in category \spP. This narrowing may, perhaps, improve performance. All in all, the interaction between ML and \Bx\ could be bidirectional!


\appendix
\section{Appendices}
\vspace{-1ex}
\subsection{Category of parameterized functors \pcatcat\ 
}\label{sec:pcatcat}

\newcommand\hf{\enma{\hat f}}
\renewcommand\hg{\enma{\hat g}}
Category \pcatcat\ has all small categories as objects. \pcatcat-arrows $\spA\rightarrow\spB$ are 
{\em parameterized} functors ({\em p-functors}) \ie, functors \frar{f}{\spP}{[\spA, \spB]} 
with \spP\ a small category of {\em parameters} and 
$[\spA, \spB] $ 
the category of functors from \spA\ to \spB\ and their natural transformations. For an object $p$ and an arrow \frar{e}{p}{p'} in $\spP$, we write $f_p$ for the functor \frar{f(p)}{\spA}{\spB} and $f_e$ for the natural transformation  \drar{f(e)}{f_p}{f_{p'}}. 
We will write p-functors as labelled arrows \fprar{f}{\spA}{\spB}{\spP}. 
As \catcat\ is Cartesian closed, we have a natural isomorphism between $\catcat(\spP, [\spA,\spB])$ and $\catcat(\spP\timm\spA, \spB)$ and can reformulate the above definition in an equivalent way with functors $\spP\timm\spA\to\spB$. We prefer the former formulation as it corresponds to the notation  \fprar{f}{\spA}{\spB}{\spP} visualizing \spP\  as a hidden state of the transformation, which seems adequate to the intuition of \paramed\ in our context. (If some technicalities may perhaps be easier to see with the product formulation, we will switch to the product view thus doing currying and uncurrying without special mentioning.)
Sequential composition of 
of \fprar{f}{\spA}{\spB}{\spP} and \fprar{g}{\spB}{\spC}{\spQ}
is 
\fprar{f.g}{\spA}{\spC}{\spP\timm\spQ}
given by $(f.g)_{pq}\eqdef f_p.g_q$ for objects, \ie, pairs $p\inn\spP$, $q\inn\spQ$, and by the Godement product of natural transformations for arrows in $\spP\timm\spQ$. That is, given a pair  \frar{e}{p}{p'} in \spP\ and \frar{h}{q}{q'} in \spQ, we define the transformation \drar{(f.g)_{eh}}{f_p.g_q}{f_{p'}.g_{q'}} to be the Godement product $f_e * g_h$. 

Any category \spA\ gives rise to a p-functor \fprar{\Id_\spA}{\spA}{\spA}{\tcat}, whose parameter space is a singleton category \tcat\ with the only object $*$, $\Id_\spA(*)=\id_\spA$ and $\drar{\Id_A(\id_*)}{\id_\spA}{\id_\spA}$ is the identity transformation. It's easy to see that p-functors $\Id_\_$ are units of the sequential composition. 
To ensure associativity we need to consider p-functors up to an equivalence of their parameter spaces. Two parallel p-functors \fprar{f}{\spA}{\spB}{\spP}  and 
\fprar{\hf}{\spA}{\spB}{\hat\spP}, are {\em equivalent} if there is an isomorphism \frar{\alpha}{\spP}{\hat\spP} such that two parallel functors \frar{f}{\spP}{[\spA,\spB]} and \frar{\alpha;\hf}{\spP}{[\spA,\spB]} are naturally isomorphic;
then we write $f\approx_\alpha\hf$. It's easy to see that if \frar{f\approx_\alpha \hf}{\spA}{\spB} and \frar{g\approx_\beta\hg}{\spB}{\spC}, then \frar{f;g\approx_{\alpha\timm\beta} \hf;\hg}{\spA}{\spC}, \ie, sequential composition is stable under equivalence. Below we will identify p-functors and their equivalence classes.
%
Using a natural isomorphism $(\spP\timm\spQ)\timm\spR\cong \spP\timm(\spQ\timm\spR)$, strict associativity of the functor composition and strict associativity of the Godement product, we conclude that sequential composition of (equivalence classes of) p-functors is strictly associative. Hence, \pcatcat\ is a category.

 \renewcommand\wi{\enma{\mathsf{[w]}}}
\renewcommand\fu{\enma{\mathsf{[f]}}}
\renewcommand\wi{}
\renewcommand\fu{}

\begin{wrapfigure}{R}{4.cm}
		\vspace{-5ex}
	\centering
	\begin{diagram}[small]
	\pcatcat&\lMonic^\fu&\psetcat 
	\\  
	\uMonic>\wi&&\uMonic>\wi 
	\\ 
	\catcattimm&\lMonic^\fu&\setcattimm
  \end{diagram} 
\vspace{-5mm}
\caption{}
	\label{fig:paraplane}
	\vspace{-7mm}
\end{wrapfigure}

Our next goal is to supply it with a monoidal structure. We borrow the latter from the sm-category \catcattimm, whose tensor is given by the product. There is an identical on objects embedding $\catcattimm\rMonic\pcatcat$ that maps a functor \frar{f}{\spA}{\spB} to a p-functor \fprar{\bar f}{\spA}{\spB}{\tcat} whose parameter space is the singleton category \tcat. Moreover, as this embedding is a functor, the coherence equations for the associators and unitors that hold in \catcattimm\ hold in \pcatcat\ as well (this proof idea is borrowed from \citefst). In this way, \pcatcat\ becomes an sm-category. 
In a similar way, we define the sm-category \psetcat\ of small sets and parametrized functions between them --- the codiscrete version of \pcatcat. The diagram in \figref{paraplane} shows how these categories are related.

 \vspace{-0.5ex}
\subsection{Ala-lenses as categorification of ML-learners}\label{sec:classPlane}
Figure~\ref{fig:classPlane} shows a discrete two-dimensional plane with each axis having three points: a space is a singleton, a set, a category encoded by coordinates 0,1,2 resp. 
\begin{figure}
	\vspace{-0.25cm}
	\centering
	\includegraphics[width=0.7\textwidth]%
	{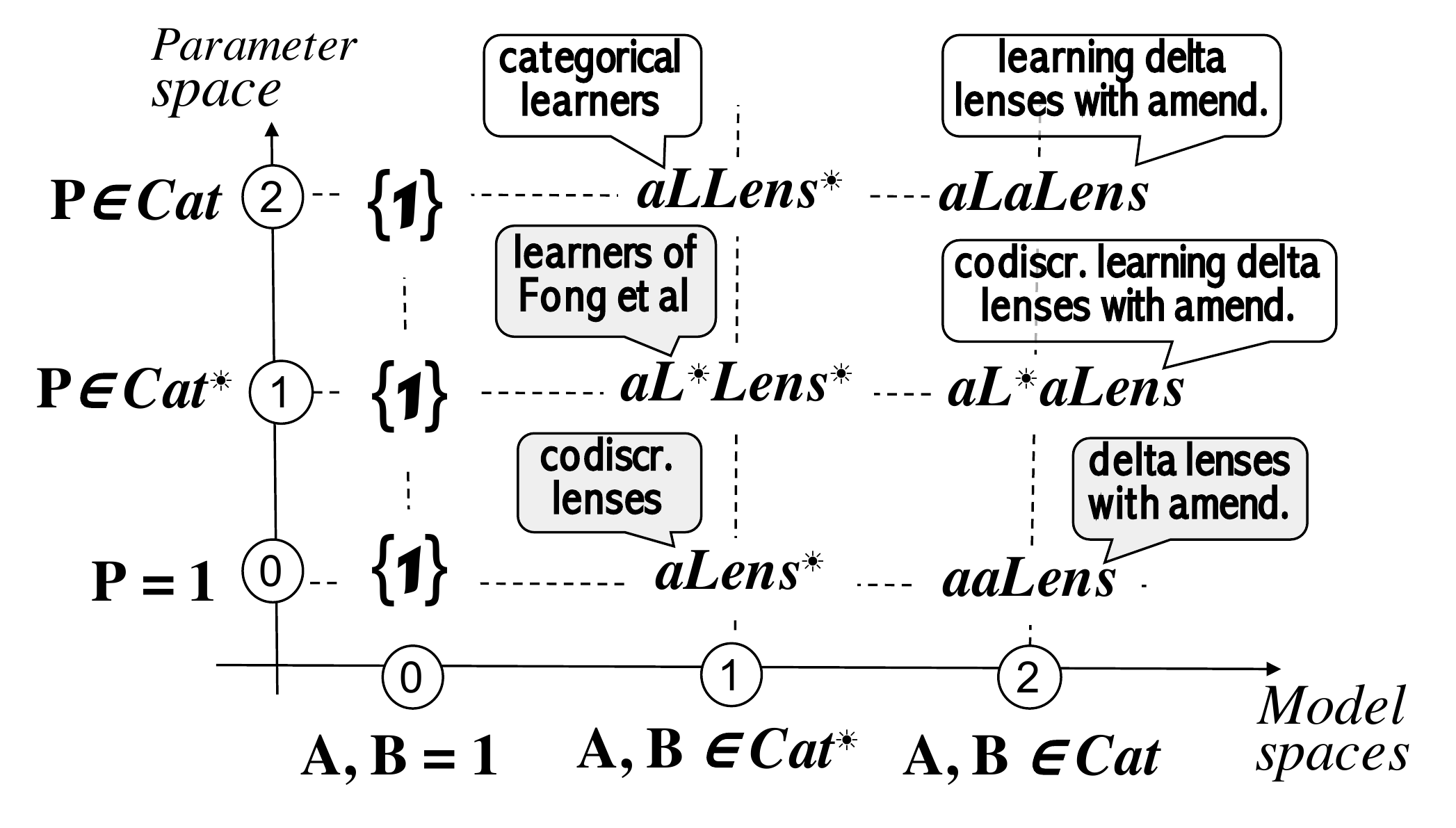}
	\caption{The universe of categories of learning delta lenses
	}
	\label{fig:classPlane}
	\vspace{-0.25cm}
\end{figure}
Each of the points $x_{ij}$ 
is then the location of a corresponding sm-category of (asymmetric) learning (delta) lenses. 
Category \{\tlenscat\} is a terminal category whose only arrow is the identity lens \frar{\tlenscat=(\id_\tcat, \id_\tcat)}{\tcat}{\tcat} propagating from a terminal category $\tcat$ to itself. 
Label $\codiind$ refers to the codiscrete specialization of the 
construct being labelled: $\catcat^\codiind$ is the category of small codiscrete categories, $\catname{L}^{\codiind}$ means codiscrete learning (\ie, the parameter space \spP\ is a set considered as a codiscrete category) and $\alenscat^\codiind$ refers to codiscrete model spaces. The category of learners defined in \citefst\ is located at point (1,1), and the category of learning delta lenses with amendments 
defined in the present paper 
is located at (2,2).  There are also two semi-categorificated species of learning lenses: categorical learners at point (1,2) and codiscretely learning delta lenses at (2,1), which are special cases of ala-lenses. 
\zdpar{A phrase about Mealy machines and Appendix \sectref{lensesasmm} is to be written here. See the commented piece in the source if needed} 
%
%


\clearpage
\bibliographystyle{splncs04} 
\bibliography{refsGrand17}


\vfill

{\small\medskip\noindent{\bf Open Access} This chapter is licensed under the terms of the Creative Commons\break Attribution 4.0 International License (\url{http://creativecommons.org/licenses/by/4.0/}), which permits use, sharing, adaptation, distribution and reproduction in any medium or format, as long as you give appropriate credit to the original author(s) and the source, provide a link to the Creative Commons license and indicate if changes were made.}

{\small \spaceskip .28em plus .1em minus .1em The images or other third party material in this chapter are included in the chapter's Creative Commons license, unless indicated otherwise in a credit line to the material.~If material is not included in the chapter's Creative Commons license and your intended\break use is not permitted by statutory regulation or exceeds the permitted use, you will need to obtain permission directly from the copyright holder.}

\medskip\noindent\includegraphics{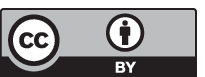}

\dend
\clearpage
\section{Old intro, grrr}

In a seminal paper \citefst, Fong, Spivak and Tuy\'eras showed how to compose supervised machine learning (ML) algorithms so that the latter form a symmetric monoidal (sm) category \learncat, and built an sm-functor 
	\begin{myeq}{lfunee}
		\frar{\lfunee}{\paracatrr}{\learncat}, 
	\end{myeq}
which maps a parameterized differentiable function 
\frar{f}{P\times\reals^m}{\reals^n} (with $P=\reals^k$ being the parameter space )
to a learning algorithm called a {\em learner}; 
the latter improves an initially  given function \frar{f(p, \_)}{\reals^m}{\reals^n} 
by learning from a set of training pairs $(a,b)\in \reals^m\timm \reals^n$. The functor is itself parameterized by a {\em step size} $0<\eps\in\reals$ and an {\em error function} \frar{\err}{\reals\timm\reals}{\reals} needed to specify the subject function of the gradient descent procedure.  
Recently, Fong and Johnson noticed in \citefjbx\ (quoting them directly) ``surprising links between two apparently disparate areas'': 
ML (treated compositionally as above) and bidirectional model \trafon s, \Bx (also treated compositionally in a \fwk\ of mathematical structures called {\em lenses} \cite{foster07}), whereas ``naively at least, there seemed to be little reason to expect them to be closely related mathematically''.%
\footnote{
	Term {\em \bx} abbreviates ``bidirectional something (or $x$)'' and refers to bidirectional change propagation in different contexts in different domains: file \syncon\ in versioning, data exchange in databases, model \syncon\ and model transformation in Model-Driven software Engineering (MDE), 	see \cite{bx-cross} for some of these contexts. \Bx also refers to a community working across those domains but self-integrated by using the conceptual and terminological \fwk\ provided by lenses. The latter lay a foundational common ground for a vast variety of \syncon\ tasks, and a whole zoo of lenses has been created to address different particular problems (cf. \cite{me-grand17}). 
	 Thus, within \Bx the application context can vary, sometimes significantly, but in the present paper, abbreviation \Bx will mainly refer to \Bx in the MDE context. Sometimes we will need both a general \Bx and a special \Bx for MDE, then the latter will be denoted by \bxmde.  
}   

The goal of the present paper is to show that incorporating the supervised learning idea into \bx\ is both practically useful and theoretically natural within the lens \fwk. We need a new species of lenses---lenses with learning capabilities or {\em learning} lenses---to be created and added to the lens library. In fact, learners by Fong, Spivak and Tuy\'eras can be seen as {\em codiscretely}  learning {\em codiscrete} lenses, and the category of classical (asymmetric) codiscrete lenses \alenscatcodi\  is a full subcategory of \learncat\ for which the parameter space is a singleton set.%
\footnote{According to \citefj, this was first noticed by Jules Hedges.}  Here  the attribute 'codiscrete' refers to the fact that spaces over which codiscrete lenses operate are sets (rather than categories but are) considered as codiscrete categories: every pair of elements $(x,y)$ is an arrow $x\rightarrow y$ and all arrows are such (the index \codilabel{} aims to recall this type of connectivity sometimes referred to as {\em chaotic}).  The prefix $\pmb a$ in the name \alenscat\ 
refers to so called {\em asymmetric} lenses (as opposed to symmetric ones); although the only lenses we will consider in the paper are asymmetric and, as a rule, we omit this attribute, it's still useful  to keep in the names of categories for future use. 

The main motivation for the present paper is that codiscrete lenses are inadequate to \bxmde\ (this is discussed in detail in \cite{me-jot11,me-models11} and briefly outlined below in \sectref{backgr}) and we thus need learning {\em delta} lenses that work over categories rather than sets. The
\noindent inset figure shows the story in a nutshell. We have two orthogonal ways of enriching codiscrete lenses: making the parameter space a category (perhaps, codiscrete) rather than being a singleton set, and making model spaces non-codiscrete categories,  and the goal of the paper is to integrate them into a new type of lenses located at point 11, and analyze some of its properties. 
\zdmargin{this diagram is inaccurate -- compare it with a more accurate description in \sectref{codicodi} \figref{classPlane}.}
We will define the notion of {\em a(n asymmetric) learning delta lens (al-lens)} and show that al-lenses can be organized into an sm-category \allenscat\ such that learners can be identified with twice codiscrete al-lenses: 
$
\learncat=\allenscatcodicodi\subset\allenscat
$. 
where the left \codiind\ means that the parameter space is codiscrete (we will phrase this as that ``learning is codiscrete'') and the right \codiind\ means 
that model spaces over which lenses operate are codiscrete (\sectref{codicodi} will provide a more accurate version of the classification plane). 
{
\renewcommand\minusrr{}
\renewcommand\minusput{}
\renewcommand\codiindd{{}}	
\renewcommand\minusmtl{}  

\begin{figure}
	\renewcommand\grt[1]{#1}
	\centering
	\begin{tabular}{c@{\qquad}c@{\qquad}c}
\begin{diagram}[small]
	\\ 
	\grt{\paracatrr}
	&&\rTo^{\grt{L_{\eps,\err}}}
	&&
	\grt{\begin{array}{c}
			\learncat=
			\\ [-2.5pt]
			~~\allenscatcodicodi
		\end{array}
	}
	\\ 
	&\rdEmbed(2,2)<{\minusrr} 
	&
	&
	\ldTo(2,2)_{\minusput^\codiindd} 
	&
	\\ 
	&&\psetcat&&
\end{diagram}
& 
\begin{diagram}[small]
	\\ 
	\grt{\pgetcat_\mtl}
	&&\rTo^{L_\mtl}
	&&\allenscat
	\\ 
	&
	\rdEmbed(2,2)<{\minusmtl} 
	&&
	\ldTo(2,2)_{\minusput} 
	& 
	\\ 
	&&\pcatcat&& 
\end{diagram}
&
\begin{diagram}[small]
	\\ 
	\grt{\pgetcat_\mtl}
	&&\rTo^{L_{\mtlwb}}
	&&\alalenscatwb
	\\ 
	&
	\rdEmbed(2,2)<{\minusmtl} 
	&&
	\ldTo(2,2)_{\minusput} 
	& 
	\\ 
	&&\pcatcat&& 
\end{diagram}

\\ [5em]
(a) ML setting
& 
	(b) \Bx\ setting 
	&
	(c) \Bx setting refined
\end{tabular}
\caption{
 Functoriality of learning algorithms and policies
}
\label{fig:policyfun-brief}
\end{figure}
} 
Thus, the first contribution of the paper is a categorification of ML's compositional learners. The latter are specified by commutative diagram (a) in \figrii\ (where nodes are sm-\cats\ and arrows are sm-functors, and two diagonal arrows are obvious forgetful functors), and their categorificated version is the notion of compositional update policies definable with some bidirectional \trafon\ language \mtl\ as specified by commutative diagram (b): in this diagram, $\pgetcat_\mtl$ is a  category of \mtl-defined model spaces as objects and \mtl-defined parameterized transformations as arrows 
(called {\em get}s to be read ``get the  transformation done''), and functor $L_\mtl$ builds a learning lens for a given parametrized functor  \get\ by using some policy of inverting that \get\ (we also say an {\em update policy} over \get) so that the triangle commutes. Note that while the world of learners is built over the sm-category of \paramed\ functions \psetcat, the world of learning lenses is built over the sm-category of \paramed\ functors \pcatcat.

The second contribution is a refinement of the first by considering learning lenses with equational laws.  
The point is that lenses appearing in \bx\ applications satisfy several equational laws assuring that update propagation restores consistency (or, at least, improves it); such lenses are (often loosely) called {\em well-behaved (wb)}. The laws can be either strict or laxed, in which case the equality is only achieved after a mediating update called an {\em amendment} is applied. This gives rise to a new species of {\em lenses with amendment, aa-lenses,} recently formally defined in  \cite{me-faoc19}. To follow our agenda in \figrplane, we will define {\em (asymmetric) learning lenses with amendment, ala-lenses}, and show that they form an sm-category \alalenscat; now lenses without amendment can be seen as ala-lenses with an identity amendment and inherit the sm-structure, $\allenscat\subset_{\mathrm{sm}}\alalenscat$.  Moreover, we show that sequential and parallel (tensor) composition of ala-lenses preserve two major equational laws of being well-behaved (Stability and Putget)  so that we have the following chain of embeddings: 
\begin{myeq}{wbembed}
\allenscatwb\subset\alalenscatwb\subset\alalenscat
\end{myeq}
whose middle member allows us to consider well-behavedness in a relaxed way more suitable for applications.
This results in a commutative diagram (c) refining diagram(b) with equational laws as described. 

The compositional nature of the gradient descent method in ML shown in \citefst, \ie, the existence of functor $L_{\eps, \err}$ specified in diagram (a), can be seen as a (codiscrete) instance of the construct specified by diagram (c) (see \exref{learner} on p.~\pageref{ex:learner} for a more accurate formulation). 
 Finding specifically \Bx non-codiscrete instances is a future work as checking compositionality of a given model transformation language is a laborious endeavour. The situation is somewhat paradoxical: although functoriality of the mapping $L_\mtlwb$ 
is an important requirements to a practical \bx-\trafon\ language as it is this requirement that allows for a compositional design of such languages, it is  missing from the current practice of \bx\ model transformation language design and evaluation. 
The necessity of this requirement, and its formal definition, is the main contribution of the paper to the practical side of \Bx. 

The paper is structured as follows.  Section~\ref{sec:backgr} briefly motivates delta lenses and explains the basic notions of the delta lens \fwk. Section~\ref{sec:learn4bx} discusses why the supervised learning idea is useful and natural for \bx, and compares codiscrete and categorical learning. The cornerstone of this analysis is the \cat\ \pcatcat\ of all (small) \cats\ and (equivalence classes of) \paramed\ functors between them, and its full subcategory \psetcat\ of all (small) sets and (equivalence classes of) \paramed\ functions between them. This material does not actually belong to lenses per se, and is placed in Appendix \sectref{pcatcat};  in addition, such an arrangement may help reading Sect.3, which we begin with codiscrete setting based on \psetcat\ formally defined in \sectref{pcatcat} as a special case of \pcatcat. 
Some details behind the scene in \sectref{learn4bx}  will be fully clear after ala-lenses, and their sequential and parallel compositions, are formally defined in Sect. 4 and 5 resp. The three main results of the paper are Theorems 1-2 (pages \pageref{th:seqwb}-\pageref{th:parwb}) stating that the two 
compositions preserve the two major ala-lens laws, and Theorem 3 showing how the universe of ala-lenses can be organized into an sm-category. All proofs (rather straightforward but notationally laborious) are placed into Appendices.  Section~\ref{sec:relwork} reviews the related work, and \sectref{concl}  concludes.  

\dend

\clearpage
\medskip
\noindent{\Large\sc Appendices} 
%

	\newcommand\ideny{identity}
	\renewcommand\lm{\enma{\ell\emm}}
	\renewcommand\klbr{\enma{(\ekk\ell)}}
	\renewcommand\lmbr{\enma{(\ell\emm)}}
	\renewcommand\putekel[2]{\putUL{\kl.#1}{#2}}
	\renewcommand\putelem[2]{\putUL{\lm.#1}{#2}}
\subsection{Sequential composition of \alalens es and lens laws: Proof of Theorem 1 on page \pageref{th:seqwb}}\label{sec:seqlenswb-proof}

\hippoOutin{Proof.}{Proof of a).}

Stability of \kl\ is obvious. To prove Putget for \kl, we need to prove that $	(\putekel{\req}{pq,A}.w).\getUL{\kl}{\putekel{\upd}{pq,A}}=	w;w^{\klbr.@}$ for any $A\in\obj{\spA}$, $p\in\obj{\spP}$, $q\in\obj{\spQ}$ and \frar{w}{A_{pq}}{C'}. 
Let $\putekel{\upd}{pq,A}$ be pair $(e,h)$ with some \frar{e}{p}{p'} and \frar{h}{q}{q'}. We compute:
\begin{eqnarray}
	(\putekel{\req}{pq,A}.w).\getUL{\kl}{eh}
	&=& 
	(\putekk{\req}{p,A}.\putell{\req}{q,A_p}.w).\getekk{e}.\getell{h}  \byx{by constr. of \kl}
	\\ &=&
	(v;v^@).\getell{h} \byx{Putget for \ekk\ (where $v=\putell{\req}{q,{A_p}}.w$)} \nonumber
		\\ &=&
	(v;v^@).\getell{q} \,;\, h_{B^@}\byx{def. of \getell{h}} \nonumber
		\\ &=&
	v_q\,;\,(v^@_q ; h_{B^@}) \byx{functoriality of \getell{q} \& assoc. of ;}\nonumber
			\\ &=&
	v_q\,;\,(h_{B'};v^@_{q'}) \byx{naturality of \getell{h}}\nonumber
		\\ &=&
	(w;w^@)\, ;\, v^@_{q'} \byx{Putget for \elll\ }\nonumber
	\\ &=&
	w;(w^@;v^@_{q'}) \byx{associativity of ;}\nonumber
	\\ &=& 
	w;w^{\klbr.@}\nonumber
\end{eqnarray}
%
\hippoOutin{}{
Proof of (b). For any given $A, p, q$ in the respective spaces, and delta \frar{w}{A_{pq}}{C'} such that $u.\get^{\kl}_{pq}=w$ for some \frar{u}{A}{\_}, we need to prove $w^{\klbr.@} = \id_{C'}$. We have seen that 
$$
w^{\klbr.@}=w^{\ell.@}{\bf \,;\;}(v^{\ekk.@}.\getell{q'}) 
\mbox{\quad where $v=\putl^{\ell.\req}_{A_p,q}.w$}
$$ We will prove that both components of ; are identities. The first one is identity due to $w=(u.\getekk{p}).\getell{q}$ and \elll\ is Hippocratic. For the second component, $v^{\ekk.@}$ is identity as $v=\putl^{\ell.\req}_{A_p,q}.w$, lenses are well-matched and lens \ekk\ is Hippocratic. Now the second component is identity as functor $\getell{q'}$ preserves identities. 
}
%
%
%

	
	\newcommand\figkl{\figref{kl-lens}}
	\renewcommand\figrii{\figkl} 
	\newcommand\defrii{\defref{def:aalens-hComp}}
	
	\renewcommand\kl{\ensuremath{\ekk;\ell}} 
	\renewcommand\lm{\enma{\ell;\emm}}
	\renewcommand\lmbr{\enma{(\ell;\emm)}}
	\renewcommand\kl{\enma{\ekk\ell}}
	\renewcommand\lm{\enma{\ell\emm}}
	\renewcommand\klbr{\enma{(\ekk\ell)}}
	\renewcommand\lmbr{\enma{(\ell\emm)}}
	\renewcommand\putekel[2]{\putUL{\kl.#1}{#2}}
	\renewcommand\putelem[2]{\putUL{\lm.#1}{#2}}
	\subsection{Sequential \alalens\ composition is associative}\label{sec:seqlensass-proof}   

Let  \frar{\ekk}{\spA}{\spB}, \frar{\ell}{\spB}{\spC}, \frar{\emm}{\spC}{\spD} be three consecutive lenses with parameter spaces \spP, \spQ, \spR\ resp. We will  denote their components by an upper script, \eg, $\get^\ekk_p$ or $\putl^{\ell.\upd}_{q,B}$, and lens composition by concatenation: \kl\ is $\ekk;\ell$ etc; $\putekel{\upd}{p,A}$ denotes $\putl^{(\kl).\upd}_{p,A}$

We need to prove $\klbr\emm=\ekk\lmbr$. We easily have associativity for the get part of the construction: $(\spP\timm\spQ)\timm\spR\cong\spP\timm(\spQ\timm\spR)$ (to be identified for equivalence classes), and $(\get^\ekk_p;\get^\ell_q);\get^\emm_r=\get^\ekk_p;(\get^\ell_q;\get^\emm_r)$, which means that $\get^{\klbr\emm}_{(pq)r}=\get^{\ekk\lmbr}_{p(qr)}$, where $p,q,r$ are parameters (objects) from $\obj{\spP}, \obj{\spQ}, \obj{\spR}$ resp., and pairing is denoted by concatenation.

Associativity of puts is more involved. 
Suppose that we extended the diagram in \figkl\ with lens \emm\ data on the right, \ie, with a triangle prism, whose right face is a square $D_{pqr}D_{r'}D^@D'_r$ with diagonal \frar{\ome;\ome^@}{D_{pqr}}{D^@} where $r\in\obj{\spR}$ is a parameter, $D_{pqr}=\get^\emm_r(C_{pq})$ and \frar{\ome}{D_{pqr}}{D'} is an arbitrary delta to be propagated to $\spP$ and \spA, and reflected with amendment $\ome^@=\putxyz{\emm.\self}{r}{C_{pq}}(\ome)$. Below we will omit parameter subindexes near $B$ and $C$.

We begin with term substitution in equations (\ref{eq:kldef1}-\ref{eq:kldef3}) in \conref{apa-seq}, which gives us equational definitions of all put operations (we use the function application notation $f.x$ as the most convenient): 
\begin{eqnarray}\label{eqna:kldef2}
	\putekel{\req}{pq,A}.w %
		&=& %
		(\putekk{\req}{p,A}.\putell{\req}{q,B}.w)
	{:}~A\rightarrow A', \label{eq:eqna-req}
	\\	
	\putekel{\upd}{pq,A}.w
	&= &%
	(\putekk{\upd}{p,A}.\putell{\req}{q,B}.w
	){:}~p\rightarrow p'
	\paarr 
	(\putell{\upd}{q,B}.w){:}~q\rightarrow q' \label{eq:eqna-upd}
	\\  
	\putekel{\self}{pq,A}.w %
		&=& (\putell{\self}{q,B}.w) \,;\, \getell{q'}
		(\putekk{\self}{p,A}.\putell{\req}{q,B}.w)    
	{:}~C'\rightarrow C^{@}\rightarrow C^{@@} \label{eq:eqna-self}
\end{eqnarray}%
(note the interplay between different puts in \eqref{eqna-upd} and \eqref{eqna-self}, and also their ``duality'': \eqref{eqna-upd} is a \paar-tem while \eqref{eqna-self} is a ;-term). 

Now we apply these definitions to the lens \klbr\emm\ and substitute. Checking $\putl^{\klbr\emm.\req}$ is straightforward similarly to associativity of gets, but we will present its inference to show how the notation works (recall that \frar{\ome}{D_{pqr}}{D'} is an arbitrary delta to be propagated)
\begin{equation}\label{eq:eqna-reqreq}
\begin{array}{lll}
	\putl^{\klbr\emm.\req}.\ome
	&=&\putekel{\req}{pq,A}.\putemm{\req}{r,C}.\ome  
	\byeqna{by}{req}
	\\ &=& \putekk{\req}{p,A}.(\putell{\req}{q,B}.\putemm{\req}{r,C}.\ome) \byeqna{by}{req}
	\\ &=& \putekk{\req}{p,A}.\putelem{\req}{qr,B}.\ome 
	\byeqna{by}{req}
	\\&=& \putl^{\ekk;\lmbr}_{p,A}.\ome 
	\byeqna{by}{req}
	\end{array}
\end{equation}
Computing of $\putl^{\klbr;\emm.\upd}$ is more involved (below a pair $(x,y)$ will be denoted as either $xy$ or $x\paar y$ depending on the context). 
\begin{equation}
\begin{array}{lll}   
\putl^{\klbr;\emm.\upd}_{(pq)r,A}.\ome
&=& 
\left(\putekel{\upd}{pq,A}.\putemm{\req}{r,C}.\ome{:}~~p\paar q\rightarrow p'\paar q'\right)
\paarrx{(a)}
\left (\putemm{\upd}{r,C}.\ome{:}~~r\rightarrow r'\right)
\byeqna{by}{upd}
\\ &=& 
\left(
\putekk{\upd}{p,A}.\putell{\req}{q,B}.\putemm{\req}{r,C}.\ome %
\; \paar\; 
\putell{\upd}{q,B}.\putemm{\req}{r,C}\ome
\right)
\paar \;
\putemm{\upd}{r,C}.\ome
\byeqna{by}{upd} \quad \paar \mbox{ same } 
\\ &=& 
\putekk{\upd}{p,A}.\putell{\req}{q,B}.\putemm{\req}{r,C}.\ome %
\; \paar\; 
\left(
\putell{\upd}{q,B}.\putemm{\req}{r,C}\ome
\paar \;
\putemm{\upd}{r,C}.\ome  
\right)
\byx{by assoc. of \paar}{} 
\\ &=& 
\putekk{\upd}{p,A}.\putelem{\req}{qr,B}.\ome
\paarr 
\putelem{\upd}{qr,B}.\ome
\mbox{~~~~~}\byeqna{by}{req} \;\paar \!\byeqna{by}{upd}
\\ &=& 
\putl^{\ekk;\lmbr.\upd}_{p(qr),A}.\ome
\byeqna{again by}{upd}
\end{array}
\end{equation}
Associativity of 
$\putl^{(\ekk;\ell;\emm).\self}_{pqr,A}$ can be proved in a similar manner using associativity of ; (see \eqref{eqna-self}) rather than associativity of \paar\ (see \eqref{eqna-upd}) used above. Below $w$ stands for $\putemm{\req}{r,C}.\ome$
\begin{equation}
\begin{array}{lll}
\putl^{\klbr.\mu.\self}_{(pq)r,A}.\ome
  &=&(\putemm{\self}{r,C}.\ome)\,;\,
\getemm{r'}\left(
\putekel{\self}{p,A}.w 
 \right) \byeqnaa{by}{self}{}
 \\ &=&
 (\putemm{\self}{r,C}.\ome)\;;\;\getemm{r'}\left(
 (\putell{\self}{q,B}.w)\;;\;\getell{q'}(\putekk{\self}{p,A}.\putell{\req}{q,B}.w)
  \right)  \byeqnaa{by}{self}{}
  \\ [2.5pt] &=& 
  \left( 
  (\putemm{\self}{r,C}.\ome)
  \;;\;
  \getemm{r'}(\putell{\self}{q,B}.w)
  \right)
  \;;\; 
  \getelem{q'r'}(\putekk{\self}{p,A}.\putell{\req}{q,B}.w) 
  \begin{array}{l}
  \mbox{ by funct. of $\getemm{r'}$}
  \\ [-2.5pt]\mbox{ and assoc. of ;}
  \end{array}
  \\  [2.5pt] &=&
   (\putelem{\self}{qr,B}.\ome)
 \;;\; 
 \getelem{q'r'}(\putekk{\self}{p,A}.\putelem{\req}{qr,B}.\ome)
 \byeqnaa{by def. of $w$ and}{reqreq}{applied twice }
 \\ &=&
\putl^{\ekk\lmbr.\self}_{p(qr)}.\omega \byeqnaa{again by}{self}{}
\end{array}
\end{equation}


	\dend

	
			\clearpage
			\noindent{\LARGE Aux and OLd sections} 
				\section{Introduction. 
	(An old one but may  work for fossacs)
}

The paper presents a new version of {\em delta lenses} designed for applications in two practically important 
domains mentioned in the title of the paper. Evolving views are a version of a classical {\em view update} problem (VUP) studied in databases since the 70s \cite{} but still generating research \cite{?}.  Machine learning, and specifically supervised learning, is a promising endeavour of our days (which is also a popular buzzword but a good part of its glory is well-deserved). Several remarks on the history of ideas are in order. 

{\em State-based lenses (s-lenses)} as a mathematical model and an approach to VUP 
were proposed by Pierce \etal\  \cite{lenstr-023,nate-popl05} and categorified by Diskin \etal\ as {\em delta-lanes (d-lenses)} \cite{me-jot11,me-models11}. For -slenses, an update of a system is a pair $(A, A')$ of the original and the updated state, while for a delta lens, an update is an arrow \frar{u}{A}{A'} specifying what was changed, which makes the collection of system's states a category \spA\ rather than a set.
For s-lenses, consistency between states of different systems \spA, \spB, is a relation, say, $R\in\spA\timm\spB$, so that saying $A\in\spA$ and $B\in\spB$ are {\em consistent} amounts to stating $(A,B)\in R$. For d-lenses, we first define a set of {\corrce\ specifications} or just {\em corrs} $\Corr(A,B)$ for any two states $A\in |\spA|$, $B\in|\spB|$, and then consistency as a unary predicate $\Kcorr_{A,B}\subset \Corr(A,B)$ so that a triple $(A, r\inn\Corr(A, B), B)$ is {\em consisten} iff $r\in \Kcorr(A,B)$. 
\footnote{For a category \spC, we will denote its set of objects by $|\spC|$ and consider it as a discrete category where needed. 
} 
Thus, consistency is a property of $A,B)$ {\em together with a corr between them} or, in fact, a property of $r$ as $A,B$ are recovered from the type of $r$. Sometimes we will refer to updates as {\em vertical} and corrs as {\em horizontal} deltas.%
\footnote{Vert deltas were easy while horiz were hard.\zd{to be compl}}
In fact, corrs should be considered as (horis) arrows between states of different systems so that we come to a double category setting. We have a universe of systems to be kept in sync. ...

 $r\in\Corr(A, B)$ or, in fact, just $r$ as $A,B$ are recovered from $r$'s type. 

The present paper makes the next step and introduces delta lenses wtih indexed corrs to capture VUP in the {\em evolving} views setting -- the latter adds to delta lenses a new important ingredient called in the paper {\em indexed correspondences}. 

A compositional approach to supervised learning was proposed by Fong \etal\ in a seminal paper \citefst, and its tight connections to lenses was noticed and elaborated by Fong and Johnson in \citefjbx. The present paper continues this line and proposes to model supervised learning by delta lenses with indexed corrs and amendment. In contrast to \citefjbx, which used symmetric lenses as couterparts of learner, we model learners by {\em asymmetric} lenses: not only composition of asymmetric lenses is technically easier, but we argue that a learner {is} an asymmetric rather than symmetric construct -- indeed, after all, a learner produces a function rather than a relation and hence learners should be composable like functions  rather than relations (spans). This setting greatly simplifies the connection between learners and lenses. Indeed, in \citell, this connection is realized by a symmetric monoidal functor between categories of \lner s and symmetric lenses resp., \frar{\ltol_0}{\lnercat_0}{\slenscat_0}, and subindex $_0$ indicates that we live in the co-discrete setting. However, in the delta setting, a (delta-)learner {\em is} just an aa-lens with indexed corrs. In other words, categories $\lnercat_\Delta$ and $\aalenscat_\icorr$ coincide (up to the names of the ingredients, so that functor $\ltol_\Delta$ is identity).%
\footnote{According to our nomenclature, subindex $_\Delta$ could be omitted, but in \citell, name \lnercat\ is used for the discrete setting $\lnercat_0$. Also, the first \catname{A} in \aalenscat\ could be omitted as asymmetry is assumed by default, but then confusion arises about the meaning of the second \catname{A}, so, it's better to keep both.}
Here the contribution of the paper is a definition of the (horizontal) composition of aa-lenses and a (straightforward) demonstration that they form a symmetric monoidal category, plus a motivation of the delta setting for \lner s (more accurately, an attempt to guess about its possible benefits). This material is covered in Sect.3. 

Section 4 is about another dimension of the composition story:....

				\section{Overview of delta lenses with amendment and their compositionality}
\label{sec:backgr}

\subsection{Background: \bx\ world of model spaces and delta lenses}
In the (delta-based) \bx, the counterpart of values in Euclidean spaces dealt with in ML are complex artifacts called \bfem{models}. They can be seen as models of logical theories (of the first- or even higher-order) called \bfem{metamodels}. A typical way of specifying metamodels is to use graphs and constraints; in fact, metamodels can be seen as Makkai's generalized sketches \citemakkai, see also \cite{me-entcs08} for definitions and discussion in the MDE context. For our current discussion, it's enough to think about models as graphs with some (unspecified) extra structure, which behaves well enough to ensure the required categorical properties (the reader can think about {\em attributed typed graphs} as defined in \cite{ehrig-book06}); we will call them 'graphs'. In MDE applications, it normally does not make sense to distinguish isomorphic 'graphs' (indeed, objects identifiers are invisible) and thus we identify models with isomorphisms  classes of 'graphs' and assume categories of models to be strongly skeletal: the only isomorphisms are identity loops.  

Let \spG\ be a category of 'graphs'/models, $G, G'\in |\spG|$ are two objects, and \wrar{\tilde u}{G}{G'} is some informal notion of a model update (delta, change). We can formalize it by a span of monics $u=(G\lTo^i O\rTo^{i'} G')$ whose head $O$ specifies what part of $G$ is preserved in $G'$ so that the elements of $G$ beyond the image $i(O)$ are deleted, and the elements of $G'$ beyond $i'(O)$ are added during the update. 
We assume that \spG\ has pullbacks so that spans are composable (and produce spans of monics), and we obtain a strongly skeletal category $\Span_{\mathsf{mon}}(\spG)$, 
whose arrows are called \bfem{updates} or \bfem{deltas}. 
The delta lens \fwk\ abstracts away these details and simply identifies model spaces with strongly skeletal categories and forward \trafon s between model spaces with functors; the latter are usually called \bfem{get functors}  or just \bfem{get}s (read `` get the view''): a detailed discussion and examples can be found in \cite{me-jot11}.%
\footnote{Not all useful \trafon s are functors, but many are, and it is a standard assumption in the delta lens \fwk.} 

Now we will schematically discuss how \alalens es work (and then will relate them to the ML setting in \citefst). 
While an ordinary asymmetric  lens ``inverts'' a given functor \frar{\get}{\spA}{\spB} between model spaces, an \alalens\ ``inverts''  {\bf\em and} ``learns'' over a family of get functors \frar{\get_p}{\spA}{\spB} with $p$ being a parameter from a set $P$. We will actually consider a more general case with the parameter space also being a small category \spP, and the get family being a functor \frar{\get}{\spP}{[\spA, \spB]} into the category of functors from \spA\ to \spB\ and their natural transformations. 
This implies that for \frar{e}{p}{p'} in \spP, and any two objects $A,A' \in \spA$, there is a mapping \frar{\get_e}{\spA(A,A')}{\spB(A.\get_p, A'.\get_{p'})} provided by the naturality of transformation $\get(e)$ (in the next section we will consider this in detail). 
We describe the data above in the top row of diagram \figri a, which provides the  ``types'' of the story, while the story itself is unravelled on the level of instances shown by the rectangular diagram below.   

Suppose given a model $A\in\spA$ and its view, model $A.\get_p= B \in\spB$, so that the state $(A,p,B)$ is consistent. 
Note our notation for the fact $A.\get_p=B$: bulleted arrows $x \rMapsto y$ don't have an internal structure and just visualize pairs $(x,y)$ --- the UML calls them {\em links}. The label over link $(A,B)$ in \figri(a) says that the link is an instance of function $\get_p$ (note the colon symbol), in more detail, this instance is fully determined by object $A$: it is the result of executing function  $\get_p$ at object $A$.%
\footnote{\label{fnote:storywithdefs} In a fuller version of the story with the type side explicitly given by a category \defcat\ of metamodels and \trafon\ definitions, over which a category of models is fibred,  link $(A,B)$ would be the Cartesian lifting of \trafon\ definition \flar{\niget_p}{\nim}{\nin} at object $A$ living in fibre \spA\ over metamodel \nim, and arriving at $B$ in the fibre \spB\ over \nin. Although the syntactical side is important for applications, lenses traditionally ignore it and we follow this style in the paper. 
}


\newcommand\arI{
	\begin{array}{c}
		a'= \\ [-5pt]
		r_I(p,a,b)	
	\end{array}
}
\newcommand\bxdiag{
	\begin{diagram}[small]
		\nia &\rTo^{\nivv_p}_{p\in\spP}&\nib&&
		\\ &&&&
	\\	
		\spA &\rTo^{
			\get_{p}}_{p\in\spP}&\spB&&
		\\ 
		&& &&
			\\ 
		\dbox{A}& \rDashmapsto^{{:}\get_p}& {B}&
		\\ 
		&& & \rdTo>{v}&
		\\ 
		\dDotto<{u^{}} &\swssseTilearrowABx{{:}\putl_{p,A}}{e}&\dDashto>{u.\get_{e}}&&\dbox{B'}
		\\ 
		&&&\ldDotto>{v^{@}}&
		\\ 
		A' &\rDermapsto^{{:}\get_{p'}}&B'^@_{}&&	
	\end{diagram}
}
\newcommand\mldiagnew{
	\begin{diagram}
		\rcat^m &\rTo^{I(p,\_)}_{p\in\rcat^k}&\rcat^n&&
		\\ 
		&& &&
		\\ 
		\dbox{a}& \rDashmapsto^{I(p,-)}& {I(p,a)}&
		\\ 
		&& & \rdTo>{\del{=}\del_{I}(p,a,b)} 
		&
		\\ 
		\dDotto<{\della} &\swssseTilearrowAB{{:}\putl }{}
		&\dDashto>{\della_{pp'}}&&\dbox{b}
		\\ 
		&&&\ldDotto>{\del^@}
		&
		\\ 
		\arI &\rDermapsto^{I(p',-)}_{p'=\Uml_I(p,a,b)}
		&I(p',a')&&	
	\end{diagram}
}
%

\begin{wrapfigure}{R}{0.5\columnwidth}
\vspace{-0.5cm}
\centering
\begin{tabular}{c}
	\\  [15ex]
\end{tabular}
\caption{Change propagation in \alalens es. \spXX\ stands for $\get(\nix)$, $\nix\in \{\nia, \nib\}$, and $\get_p$ stands for $\get(\nivv_p)$. Arrow 	\frar{e}{p}{p'} is not shown. Derived/computed elements are specially visualized: nodes are not framed, arrows are dashed. 
}
\label{fig:bxmlPut}
\end{wrapfigure}

Now suppose that $B$ is updated to $B'$ with delta (update) $\vii\in\spB(B,B')$ as shown. 
In the asymmetric lens setting, any change of view $B$ destroys consistency (it's not so for the symmetric version of lenses), and an \alalens\ restores it by calling an operation \putl\ which a) changes the parameter with delta/update \frar{e}{p}{p'} (where $e$ refers to the {\em evolution} of the get), 
b) changes the source model $A$ with update \uiii,  and c) amends the update-violator $v$ with an {\em amendment} update $v^@$  so that the result of these changes is a consistent system $A'.\get_{p'}=B'^@$ as shown in the diagram (note the dashed arrows and index $p'$ near the bottom arrow). Thus, \putl\ is a triple of operations $(\putl^\upd, \putl^\req, \putl^\self)$, which for a given pair $(A,v)$ return a corresponding triple of updates \frar{e=\putl^\upd(A,v)}{p}{p'} (not shown in the diagram), \frar{u=\putl^\req(A,v)}{A}{A'}, \frar{v^@=\putl^\self(A,v)}{B'}{B'^@}; the superscripts \upd, \req\ abbreviate the corresponding names used in \citefst.   

Moreover, to ensure that operations \putl\ really invert functors \get, we require a much stronger consistency condition to hold:  
for any $v$ as above, we require the equality $u.\get_{e}= v;v^@$ between updates---because it is update/delta $v;v^@$ that gives model $B'^@$ its real meaning in a typical MDE scenario.  
In the lens parlance, the last equation and its versions are called the Putget law: if we substitute $\putl(v)$ for $u$, we obtain equality $(\putl.v).\get_{p'} =v;v^@$ to hold for any $v$, hence, the name of the law. Another important law is {\em Stability}: if $v=\id_B$, then all three deltas, $e,u,v^@$ are also identities. A lens satisfying Stability and Putget laws is called an {\em SP-alaLens}. 

Thus, an (SP) ala-lens is a pair $\ell=(\get,\putl)$ specified above, and we will write it as an arrow $\ell=(\get,\putl){:}\, \spA\stackrel{\spP}{\longrightarrow}\spB$ or just \frar{\ell=(\get,\putl)}{\spA}{\spB} keeping the parameter space \spP\ implicit. In the next section, we will show that SP-lenses can be composed sequentially and in parallel, and form an sm-category $\alalenscat$. 

An ala-lens \frar{\ell}{\spA}{\spB} is called {\em (elementarily) codiscrete} if all categories $\spA,\spP, \spB$ are codiscrete (below we will also consider non-elementarily codiscrete lenses). The full subcategory of (SP) codiscrete lenses will be denoted by $\alalenscat_0$ (and we have a full embedding $\alalenscat_0\rFEmb\alalenscat$).  

\subsection{Learners in the delta lens setting, I}

Diagram \figri(b) shows a specialization of the general ala-lens scenario in \figri(a) for ML-learners as defined in \citefst. Euclidean spaces are turned into categories whose arrows are differences: for any pair of tuples $a,a'\in\reals^m$, we have their componentwise difference $\del=(\del_i)_{i\le m}$, $\del_i=a'_i-a_i$, considered as an arrow \frar{\del}{a}{a'}.  Composition and identity are obvious, and any function \frar{f}{\reals^m}{\reals^n} gives rise to a functor \frar{\pmb f}{\rcat^m}{\rcat^n}.%
\footnote{A more refined model would be to consider monoids of differences as acting on values, say, monoid $\reals^m$ acts on $\reals^m$ so that given a value $a\in\reals^m$ and a difference (\textit{edit}, in the lens parlance) $\del\in\reals^m$, we compute $a'=a\odot\del$ where action $\odot$ is given by summation: $a\odot\del :=a+\del$. Thus, we have a monoid $\DDel^m=(\reals^m,+,0)$ that acts on $\reals^m$ by  \frar{\odot}{\DDel^m\timm\reals^m}{\reals^m}, and if function\frar{f}{\reals^m}{\reals^n} is differentiable, then its Jacobian matrix at point  $a$ can be seen as a monoid morphism \frar{\jacobf_a}{\DDel^m}{\DDel^n} so that for sufficiently small deltas, $f(a\odot\del)\approx f(a)\odot\jacobf_a(\del)$. Then functoriality of $f$ is based on its differentiability. We consider this model in more detail in \sectref{?}.
}
 The notation in diagram \figri(b) is borrowed from \citefst\ to ease seeing the parallelism (but note a discrepancy: $B'$ in \figri(a) corresponds to $b$  without prime in \figri(b)). Also note that for defining \putl, a special metrics is used: first, a differentiable error function \frar{\err}{\reals\timm\reals}{\reals} is defined, and then the total error is given by summation, $E_I(p,a,b):=\sum_{j\le n}\err(I_j(p,a), b_j)$ (see \citefst) so that $E_I(p,a,b)$ is a function of $\del_I(p,a,b)$ seen as an arrow. Finally, we can define update arrows using the gradient descent idea: \frar{\dellp}{p}{p'} with $\dellp=-\eps\nabla_p E_I(p,a,b)$ and \frar{\della}{a}{a'} with $\della=-\eps\nabla_a E_I(p,a,b)$ (of course, we can use different step sizes \eps\ for \dellp\ and \della).%
 \footnote{In \citefst, they define \della\ differently. First, function \frar{\prt_x\err(x_0,\_)}{\reals}{\reals} is considered invertible for all $x_0$, and let \flar{\rre_{x_0}}{\reals}{\reals} be its inverse (symbol \rre\ is the inversely read \err). Then for an $a\in\reals^m$, we have \flar{\rre_a}{\reals^m}{\reals^m}. Now  $\della$ can be defined as $\rre_a(\nabla_a E_I(p,a,b))$.} 
 Amendment is defined by differencing $\del^@=I(p',a')-b$. It easy to see that if $\del=0=\id_{I(p,a)}$, then $E_I(p,a,b)=0$ achieves its minimum and thus  $\dellp=\della=\del^@=0$ as well so that Stability holds. The Putget law trivially holds due to the definition of $\del^@$. Thus, learning based on the gradient descent gives us an SP-alaLens.

 \subsection{Functoriality of update policies}
 
\begin{wrapfigure}{R}{0.5\textwidth}
\vspace{-1cm}
	\centering
\[
\begin{diagram}[h=2.5em]
\\  
\nia_1&\rTo_{\nivv^1_{p_1}}^{{p_1\in\spP_1}}
&\nia_2&\rTo_{\nivv^2_{p_2}}^{{p_2\in\spP_2}} 
&\nia_3&\rTo_{\nivv^3_{p_3}}^{{p_3\in\spP_3}} 
& \nia_4
\\ &&& &&&
\\  
  \spA_1&\rTo_{\get^1_{p_1}}^{{p_1\in\spP_1}}
&\spA_2&\rTo_{\get^2_{p_2}}^{{p_2\in\spP_2}} 
&\spA_3&\rTo_{\get^3_{p_3}}^{{p_3\in\spP_3}} 
& \spA_4
\\ 
&&& &&&
\\ 
   \dbox{A_1}&\rDermapsto^{{:}\get^1_{p_1}}
&A_2&\rDermapsto^{{:}\get^2_{p_2}}
&A_3&\rDermapsto^{{:}\get^3_{p_3}}
&A_4
\\ 
&&&&&
&\dTo>{u_4}
\\ 
&&&&\dDotto<{u_3}&
\swssseTilearrowABx{1{:}\putl^3_{A_3,p_3}}{e_3}
&\dbox{A'_4}
\\ 
&&\dDotto<{u_2}&
&&&\dDotto<{u_4^@}
\\ 
&&&
\swssseTilearrowABx{2{:}\putl^2_{A2, p_2} }{e_2}
&A'_3
&\rDashmapsto^{{:}\get^3_{p'_3}}&{A'}_4^@
\\ 
\dDotto<{u_1}&&&&\dDotto<{u_3^@}&\stackrel{4{:}\get^3_{e_3}}{\Rightarrow}&\dDashto
\\ 
&\swssseTilearrowABx{3{:}\putl^1_{A_1,p_1}}{e_1}&A'_2&\rDashmapsto^{{:}\get^2_{p'_2}}&{A'}_3^@&\rDashmapsto^{{:}\get^3_{p'_3}}&\circ
\\ 
&&\dDotto<{u_2^@}&\stackrel{5_1{:}\get^2_{e_2}}{\Rightarrow}
&\dDashto&\stackrel{5_2{:}\get^3_{e_3}}{\Rightarrow}&\dDashto
\\ 
A'_1&\rDermapsto^{{:}\get^1_{p'_1}}&{A'}_2^@&\rDashmapsto^{{:}\get^2_{p'_2}}
&\circ
	&\rDashmapsto^{{:}\get^3_{p'_3}}&\circ
\end{diagram}
\]
\caption{Functoriality of update propagation policies: Triple for assoc}
\label{fig:policyfun-1}
\end{wrapfigure}
 
 The diagram in \figrii\ illustrates compositionality of operations \get\ (trivial) and \putl\ (non-trivial) (nodes $\circ$ denote objects without names). Given a sequence of parameterized functors \frar{\get^i_{p_i}}{\spA_i}{\spA_{i+1}}, $p_i\in\spP_i$, $i\le n$ ($n=3$ in the diagram), it is easy to define their composition \frar{\get_p}{\spA_1}{\spA_n} for $p=(p_1...p_n)\in\spP{=}\spP_1\timm\ldots\timm\spP_n$ as $$A.\get_p=(...(A.\get^1_{p_1}).\;\ldots\;).\get^n_{p_n},$$ and show that it is associative (when functors are considered up to a suitable equivalence). Composition of $\putl^i$, $i\le n$ is more interesting: the diagram shows a simple projection of story, in which  arrows \frar{e_i{=}\putl^{i\,\upd}}{p_i}{p'_i} are not shown. Triple-arrow symbols $j{:}\putl^i$ denote three consecutive applications ($j=1,2,3$) of operations $\putl^i$ appropriately indexed. Double-arrows labelled $j{:}\get^i_{e_i}$ denote application of the corresponding \get-functors to arrows to demonstrate that the Putget law holds.

 A pair $\ell_i=(\get^i,\putl^i)$ is an ala-lens \frar{\ell_i}{\spA_i}{\spA_{i+1}}, and the construction shown in the figure defines lens composition \frar{\ell_1;\ell_2;\ell_3}{A_1}{A_4} (it will be formally defined in \sectref{alalenscat}). 
 However, it's not the end of the story developed in \citefst. 
 So far, in the lens formalism as such, two components of a lens $\ell=(\get, \putl)$ are considered given independently although their interaction is regulated by lens laws ensuring a weaker or stronger sense of \putl\ being an inverse to \get. However, in practice of both BX and ML, the \putl-component is built based on the \get-component. As there are many ways of inverting a {\bf general} functor, finding reasonable single-valued operation \putl\ (a so called {\em update policy})  is normally based on the {\bf specific} form of \get\ so that \putl\ is carefully ``tuned'' for a given specific \get. In other words, for a given specific domain context \nix, model spaces $\spA_i$ are special categories and functors \get s are special functors for which a reasonable single-valued operations \putl\ can be defined. For example, ML provides a special context, in which model spaces are Euclidean spaces (considered as categories $\rcat^m$), \get s are differentiable mappings \frar{f}{\rcat^k\timm\rcat^m}{\rcat^n}, and the gradient descent idea Gd provides a reasonable definition of \putl s. We will return for this discussion in \sectref{policyfun-2}

	\section{Introduction}\label{sec:intro}
	Modelling normally results in a set of inter-related models presenting different views of a single system at different stages of development. The former differentiation is usually referred to as {\em horizontal} (different views on the same abstraction level) and the latter as {\em vertical} (different abstraction levels beginning from the most general requirements down to design and further on to implementation). A typical modelling environment in a complex project is thus a collection of models inter-related and inter-dependant along and across the horizontal and the vertical dimensions of the network. We will call the entire collection a {\em multimodel}, and refer to its component as to {\em local} models.  


If one of the local models changes and their joint consistency is violated, the related models should also be changed to restore consistency. This task of model \syncon\ 
is obviously of paramount importance for Model-Driven Engineering (MDE), but its theoretical underpinning is inherently difficult and reliable automatic synchronization solutions are rare in practice. 
Much theoretical work partially supported by implementation has been done for the binary case (synchronizing two models) by the bidirectional transformation community (bx) in the \fwk\ of so called {\em lenses} -- algebraic structures specifying sync of two models via change propagation  
\footnote{
	Term {\em \Bx} abbreviates ``bidirectional $x$ (something)'' and refers to bidirectional change propagation in different contexts in different domains: file \syncon\ in versioning, data exchange in databases, model \syncon\ and model transformation in 
	MDE,  
	see \cite{bx-cross} for some of these contexts. 
}   

\zd{below is the act19 version}

		\newcommand\figrexi{\figref{ex1-backProp}}
\actfoss{
\section{Overview of delta lenses with amendment and their compositionality}
}{
\section{Background: \bx\ world of model spaces and delta lenses}
}
\label{sec:backgr}

\zd{ a good piece is to be written to explain we need this sect: formalization of metamod is not the goal of the paper but we need to show the worls of metamodelling.....}

\subsection{Metamodels, models, and updates}
\input{ll-figures/pdfs/ex1-backProp}

Figure \ref{fig:ex1-backProp} presents a toy example of the view mechanism. The source metamodel \nia\ (note the grey roundtangle in the upper left corner and ignore blue in- and out-arrows for a while) is a graph whose nodes denote classes of objects and arrows are relations between them. Formally, a {\em (pre)model} of this graph is a functor \frar{\semm{..}}{\nia}{\relcat} into the category of sets and relations (by the abuse of notation, \nim\ will also denote the category freely generated by \nim). We will denote the class of all such functors by $|\relcat^\nia|$ or $Mod^\pre(\nia)$.  Node \days\ is a constant and its semantics is predefined and fixed for all models: $\semm{\days}=\{1,..,7\}$ with numbers  referring to week days. Other sets vary depending on the model, \eg, a model  $X$ is given by sets \semmX{\car}, \semmX{\boat}, and relations $\semmX{\usedon}\subset\semmX{\car}\timm\semm{\days}$ and
$\semmX{\amph}\subset\semmX{\car}\timm\semmX{\boat}$. Expressions \textcolor{red}{0..1} (in red) are constraints (called {\em multiplicities} in UML). The lower one is satisfied by model $X$ iff relation \semmX{\amph} is a partially defined injection, and the upper one requires the inverse of \semmX{\amph} to be a partially defined injection as well. Thus, a {\em legal} model $A$ of the metamodel \nia\ must ensure that \semmA{\amph} is a partially defined bijection. A simple example with $\semmA{\car}=\{c1, c2\}$, $\semmA{\boat}=\{b\}$, $\semm{\amph}=\varnothing$ is shown in the figure (relation \semmA{\usedon} is omitted to save space).%
\footnote{Any functor \frar{\semmA{..}}{\nia}{\relcat} can equivalently be presented in the (Grothendieck) fibrational way as a functor \frar{t_A}{\int \semmA{..}}{\nia} called the {\em typing mapping}. In \figrexi, the graph/category $\int \semmA{..}$ is denoted by $A$, and mapping $t_A$ is specified by colons, \eg, $c1{:}\car$ means that $t_A(c1)=\car$.
}

Thus, a general metamodel is a pair $\nim=(|\nim|, C_\nim)$ with $C_\nim$ a set of constraints interpretable in \relcat, which determines 
the class of all legal models $Mod(\nia)=\compr{A\in\relcat^A}{A\models C}$, where $A\models C$ means $A\models c$ for all $c\in C_\nim$. For example, for the metamodel \nia\ shown in the figure, $C_\nia=\{\days, [0..1]_{bot}, [0..1]_{up} \}$ 
where we consider constant \days\ as a constraints that demands $\semmx{X}{\days}=\{1..7\}$ for a legal model $X$. Morphisms between models are natural transformation that are partial injections: morphism \frar{u}{X}{X'} is a family of partial injections $u_n\subset\semmX{n}\timm\semmXp{n}$ indexed by \nim-nodes such that the square
$u_n;\semmXp{a}=\semmX{a};u_{n'}$ commutes for any arrow \frar{a}{n}{n'} in \nim\ (we will often denote a metamodel and its carrier graph by the same letter omitting the vertical bars). This gives us a category \modcat(\nim) of \nim-models and their updates, which is faithfully (but not full)  embedded  $\modcat(\nim)\subset \relcat^\nim$ into the category of all functors and all natural transforamtions. %
\footnote{An accurate formalization of this construction needs Makkai's generalized sketches or graphs with diagram predicates described in \cite{me-entcs08} in the equivalent Grothendieck's fibrations setting.}
To shorten formulas, we will denote categories of legals models over metamodels \nia, \nib, \etc by the respective bold letters, \spA, \spB, etc. 

\subsection{The view mechanism}

\subsubsection{Queries and their execution.}
Suppose that an insurance company needs some data about \nia-models for proper insuring. These data are specified by metamodel \nib\ as shown in \figrexi\ upper right. Whereas metamodel \nia\ does not have such classes immediately, they can be recovered by queries against \nia, \ie, specifications of operations that produce required classes and arrows from the given classes and arrows. For example, for any given model $A$ and relation 
\relrar{\semmA{\usedon}}{\semmA{\car}}{\semm{\days}}, we can define subset on usedOn-business days cars 
\begin{myeq}{bcar}
\semmA{\bcar}=\compr{c\in\semmA{\car}}{\semmA{\usedon}(c)\subset\{1..5\}}\subset\semmA{\car}
\end{myeq}
Similarly, we can derive a set of cars used on weekends, \semmA{\wcar} together with  relation \same\ between them that relates a b-car and an w-car iff they are two different roles of the same car (\eg, in model $A$ in the figure, the car $c2\in\semmA{\car}$ is such). Then we can take the coproduct of sets \semmA{\wcar} and \semmA{\boat}, and compose arrows $s1{:}\same$ and $i1{:}\inzz$. In more detail, we fix  some signature \Q\ of diagrammatic operations over graphs, which gives rise to a monad \frar{\QQ}{\graphcat}{\graphcat} over the category of graphs so that any graph, \eg,  $\nia$ can be freely extended with derived elements obtained by applying operations from \Q\ to \nia\ (see \cite{me-efest} for details). Category \relcat---as a graph--- can also be freely extended, but our operations are to be interpretable in \relcat\ so that the latter becomes a \QQ-algebra \frar{\alpha_\relcat}{\QQ(\relcat)}{\relcat} (of course, the size issues are to be cleanly managed to make it accurate). Then any functor \frar{\semmA{..}}{\nia}{\relcat} is uniquely extended to the functor  \frar{\semmQA{..}}{\QQ(\nia)}{\relcat} or, fibrationally, we have an extended typing mapping \frar{\QQ(t_A)}{\QQ(A)}{\nia} (with $\QQ(A)$ denoting $\int \semmQA{..}$); \figrexi\ shows an appropriate portion of $\QQ(A)$ and $\QQ(t_A)$ (via colons). 

\subsubsection{Views and their execution.}
Now we can define the view we need by specifying a (Kleisli) mapping \flar{v}{\QQ(\nia)}{\nib} as shown in the figure. Then we take the pullback of $\QQ(t_A)$ and $v$ in \graphcat\ and thus obtain the view \frar{t_B}{B}{\nib} together with mapping \flar{\ovrx{v}{A}}{\QQ(A)}{B} called {\em traceability} in the MDE parlance. (Note that although \ovr{v} is a Kleisli mapping, the pullback is taken in \graphcat\ rather than the Kleisli category $\graphcat_\QQ$.) Speaking fibrationally, we have a codomain fibration and hence a change of base functor 
\frar{v^*}{\relcat^\nia}{\relcat^\nib} 
between the the respective categories of premodels (defined by precomposition). To ensure that the restriction of functor $v^*$ to category $\spA=\modcat(\nia)$ of legal \nia-models is actually targeted at $\spB=\modcat(\nib)$, we need to require that definition mappings be compatible with constraints: if a subgraph $D$ of graph \nib\ is subjected to a constraint $c$, then its image $v(D)\subset\nia$ has to bear the same (basic or derived) constraint --- this is a standard requirement for being a correct sketch morphism.%
\footnote{For an accurate formulation, we need to define a constraint declaration $c$ over metamodel \nim\ as a pair $(\phi_c,\delta_c)$ where $\phi_c$ is a predicate over graphs (interpretable in \relcat), $G_{\phi_c}$ is the arity of the predicate (\eg, the arity of multiplicity is a singleton arrow, and the arity of, say, being a jointly monic pair of arrows is a span of arrows), and  \frar{\delta_c}{G_{\phi_c}}{\nim}{} is a diagram over which this predicate is declared (substitution of \nim-elements to arity placeholders).  A generalized sketch is a pair $\nib=(|\nib|, C_\nib)$ and a sketch morphism \flar{v}{\nia}{\nib} is a graph morphism \flar{v}{|\nia|}{|\nib|} such that $v(c) \eqdef (\phi_c, \delta_c;|v|)\in C_\nia$ for any $c\in C_\nib$; more generally, $A\models v(c)$ for any model $A\models C_\nia$, \ie, $C_\nia\models v(c)$.
}
Following the lens jargon, we denote the change of base functor \frar{v^*}{\spA}{\spB} by $\get_v$ and call it ``get the view functor'' or just ``get''. 

\subsection{Update propagation}
Suppose that model $B$ is updated tio model $B'$ as shown in \figrexi\ bottom right (ignore for while the block arrow $b$ and all links it covers). As the view changed, \eg, now it has only one commuting vehicle, we need to change the source model $A$ to $A'$ so that $B'=\get_v(A')$.  Importantly, having just a pair of models $(B, B')$ does not say much what actually happened. The two leisure vehicles in $B'$ can be new ones or old ones, and vehicle $c3$ csan be one of \{cv1, cv2\} or a new one. For example, cv3 can be cv2, and the reason that the same-link disappeared is that car c2 in model $A$ (as shown by the respective traceability link from cv2 to c2) is stopped being used on weekends and thus vehicle lv2 is not present in $B'$. Thus, understanding the update and hence its propagation does need knowing a relation between models $B$ and $B'$, formally a span from graph $B$ to graph $B'$ whose both legs are monic as we don't want objects to merge and fork. A simple possible case \relrar{B}{B}{B'} is shown in the figure: it says that both leisure vehicles in $B$ are kept in $B'$ while vehicles cv1 and cv2 were deleted, and vehicle cv3 added. 

Having this information, we can approach finding the necessary change for model $A$.  A simple reasoning over the traceability links shows that car c2 is not now used on business days but is still used on weekends (as it is kept as a leisure vehikle). This gives us preservation links (c2,c2') and (c2'',c2''') in the update $a$ we are building. As for car c1, its disappearance can be caused by either a) its deletion from $A'$ or b) stopping its usage on business days; the latter means that c1 is not used at all which is not prohibited by constraints as \usedon(c1) is allowed being empty. Indeed, uncertainty and non-determinism in update propagation is a fundamental feature of the problem. Often, different heuristics are used, \eg, as keeping a non-used car under insurance is relatively rare, our propagation policy can mandate to delete car c1 and thus no link from c1 occurs into $a$. Vehicle c3 is new according to $b$, and the view definition mapping says that a new car c3' is to be added to class \semmAp{\bcar}. This new \bcar\ cannot be car c2' as we just found that c2' was removed from \semmA{\bcar}, but if model $A$ would have one or several unused cars, it would be a possible solution to make one of them 
a commuting car.  Moreover, a new commuting vehicle can simply mean a new car in the system, and this policy is used in the version shown in the figure. We don't know the exact value of the set \usedon(c3) but we do know that $\usedon(c3)\cap\{6,7\}=\varnothing$ (otherwise car c3 would have a {:}\same\ link into \semmBp{LV}) and hence $\usedon(c3)\subset \{1...5\}$ to ensure appearance of c3' in \semmAp{\bcar}. Thus, for industrial size models we have a significant uncertainty to be somehow managed with heuristics, AI, users input and other means encompassed by the notion of update propagation policy. If the latter is complete, it allows us for any given model $A$ and update \relrar{b}{\get_v(A)}{B'} to compute update \relrar{a}{A}{A'} and updated traceability mapping \flar{r'}{\QQ(A')}{B'}. We thus have a family of operations $\putl_A^v$ such that $a=\putl_A^v(b)$ is an update from $A$. A fundamental requirement to a reasonable policy is the {\em correctness} of the propagation: the following equality called the Putget law is to hold for all $A$ and $b$:
\begin{myeq}{putgetlaw}
(\putl_A^v(b)).\get_v = b
\end{myeq}
  
\subsection{Delta lenses and compositonal view systems}

\begin{defin}[Lenses]\label{def:alens} 
	Let \spS, \spT\ be two categories. An {\em (asymmetric delta) lens} from \spS\ to \spT, written as \frar{\ell}{\spS}{\spT}, is a pair $\ell=(\get^\ell, \putl^\ell)$, where \frar{\get^\ell}{\spS}{\spT} is a functor and $\putl^\ell _S$ is a family of operations indexed by objects of \spS, $S\in|\spS|$. Given $S$, operation $\putl^\ell_S$ maps any arrow \frar{t}{T}{T'} where $S.\get=T$ to an arrow \frar{s}{S}{S'}. We will often omit the script \elll, and write a lens as \frar{(\get,\putl)}{\spS}{\spT} 
	
	A lens is called {\em well-behaved (wb)} if the following two equational laws hold:
		\\[1ex]
	\lawgap=1 ex
	\noindent \begin{tabular}{l@{\quad}l}
		\lawnamebr{Stability}{
		}	& \mbox{ 
			 $\id_S = \putl_S(\id_{T})$ for all $S\in|\spS|$}  
		\\ [\lawgap] \lawnamebr{Putget}{} 
		& 
		\mbox{ 
			$(\putl_S.t).\get  = t$ for all $S\in|\spS|$ and all $t\in\spT(T,\_)$
			} 
	\end{tabular}  
	\end{defin} 
\begin{wrapfigure}{R}{0.3\textwidth}
	\renewcommand\grt{}
\vspace{-0.5cm}
	\centering
	\begin{tabular}{c}
\begin{diagram}[small]
&&\vdefcat_\nix
&&
\\ 
&\ldTo<{\get_\nix}
&
&\rdTo>{~~\polupdX}
& 
\\ 
\catcat
&&\lTo^{\minusput} 
&&\alenscat
\end{diagram}
\\ [20pt] (a) 
\\ [20pt]
\begin{diagram}[small]
	&&\vdefcat_\nix
	&&
	\\ 
	&\ldTo<{\ovr{\get_\nix}}
	&
	&\rdTo>{~~\ovr{\polupdX}}
	& 
	\\ 
	\getcat_\nix
	&&\lTo^{\minusput_\nix} 
	&&\alenscat_\nix
	\\ 
	\dEmbed<{{-}_\nix}&&&&\dEmbed<{{-}_\nix}
	\\ 
	\catcat&&\lTo^{\minusput} &&\alenscat
\end{diagram}
\\ [20pt] (b) 
\end{tabular}
\vspace{-0.55cm}
\label{fig:policyfun-ml1}
\end{wrapfigure}
Asymmetric lenses are sequentially associatively composable and can be organized into a category \alenscat\ (see \cite{me-jot11}) equipped with a ioo forgetful functor \frar{(\minusput)}{\alenscat}{\catcat} defined in the obvious way.
 \begin{defin}[View systems]\label{def:viewsys} 
 	A {\em (compositional) view system} is given by a cospan of funtors $\nix=(\get_\nix, \polupdX)$ such that the inset triangle diagram (a) below commutes. We will  also use the image factorizations of the functors (diagram (b)) and call lenses from the image of \polupdX, $\alenscat_\nix$,  {\em concrete}, and functors from the image of $\getcat_\nix$ {\em gets}. 
  \end{defin}
 \zd{Image fact; concrete vs abstract lenses. Cats $\getcat_\nix$  and $\alenscat_\nix$}
 
\begin{wrapfigure}{R}{0.29\textwidth}
\vspace{-1.25cm}
	\centering
\[
\begin{diagram}[small]
\\ 
\grt{\paracat} 
&&\rTo^{\grt{L_{\eps,\err}}} 
&&\learncat 
\\ 
&\rdTo<{\grt{{-}\reals}}
&
&\ldTo>{{-}\putl}
& 
\\ 
&&\psetcat
&&
\end{diagram}
\]
\vspace{-0.75cm}
\label{fig:policyfun-ml1}
\end{wrapfigure}

Figure~\ref{fig:green-vs-orange}(b) presents a different \trafon\ $\anmt_2$. Now a boat gives rise to a commuting and a leisure vehicle, whereas a car only produces a leisure vehicle (think about people living on an island). Clearly, being executed for the same source model $A$, \trafon\ $\anmt_2$ produces the same target model consisting of three objects and a \same-link. More accurately, models $\anmt_1(A)$ and $\anmt_2(A)$ are isomorphic rather than equal, but the same \trafon\ $\anmt$ executed twice for the same model $A$ would also produce isomorphic rather than equal models as \mt s are normally defined up to OIDs. We will always understand equality of models up to OID isomorphism, and thus can write $\anmt_1(A)=\anmt_2(A)$. It is easy to see that such an equality will hold for any source model containing equal numbers of cars and boats. 

	\clearpage
	\section{Bits and pieces}

{\em	Contribution: Specification of algebraic laws for learning lenses and studying their preservation under lens composition. Building an sm-category \alalenscat\ of al-lenses with amendment.} 

For \bx, this is a crucial aspect: it says that if several synchronization modules satisfy some laws (are {\em well-behaved} in the lens parlance), then their composition is automatically well-behaved and hence the (usually expensive) integration testing is not needed. However, a major asymmetric BX lens law that requires consistency of the system after change propagation (called the PutGet law in the lens parlance) does not always hold in practice and needs to be relaxed \cite{me-fase18}. The relaxed PutGet is based on the idea of self-propagation: the change \warrbb\ is amended by a consecutive change \warrxy{b'}{b'^@:=f_{p'}(a')} so that the resulting state $(a',p',b'^@)$ is consistent.   Thus, \bx\ needs lenses with amendment, and the main construct of the paper is the notion of an \bfem{asymmetric learning lens with amendment, \alalens}, for which we will formulate \shortlong{two}{three} basic laws. We will define sequential and parallel composition of \alalens es, study their compatibility with the laws, and build an sm-category of well-behaved \alalens es, \alalenscat. The role of algebraic laws in the ML context is an intriguing subject left for future work. 



We will reconsider a major result in \citefst---compositionality of the mapping from learning spaces to learning algorithms, in the light of {\em delta} rather than {\em codiscrete} lenses. The compositionality above is specified by the commutative triangle in \figro,  in which nodes are sm-categories and arrows are sm-functors (and below we will omit the prefix sm). 

Node $\paracatre$ denotes the category of (equivalence classes of) parameterized differentiable mappings $P\times\reals^m\rightarrow\reals^n$ (with the parameter space $P=\reals^k$), which is denoted by \paracat\ in \citefst\ but we add subindex \reals\ to recall the context, which will be important in our setting.
Arrow  \leefun\ is the functor built in \citefst\ with \eps\ and \err\ referring to parameters of the gradient descent learning (a step size and an error function $\reals\timm\reals\rightarrow\reals$ resp.), and ${-}\reals$ and ${-}\putl$ are two forgetful functors into the category of sets and (equivalence classes of) parameterized functors, \psetcat\ (
considering parameterized functions up to some equivalence is needed to ensure associativity of their composition -- see \sectref{pcatcat} for details). Here and below we will denote a functor forgetting a part $X$ of its source objects' or/and arrows' structure by symbol ${-}X$. Similarly, functors adding a structural piece $X$ are denoted by $\plusx{\xyz}X$ with index \xyz\ referring to the mechanism by which this addition is realized.
The following notion is fundamental for this paper. A {\em parameterized functor} (shortly, {\em p-functor}) from a category \spS\ (the source) to a category \spT\ (the target) is a functor  
\frar{f}{\spP}{[\spS, \spT]} 
where \spP\ a category of {\em parameters} and 
$[\spS, \spT]$ 
is the category of functors from \spS\ to \spT\ and their natural transformations (all categories in this defintion are assumed to be small; see \sectref{pcatcat} for precise definitions). 
We will write p-functors as labelled arrows \fprar{f}{\spS}{\spT}{\spP} and sometimes omit label $\spP$ over the arrow. 

\zd{----?}
 The main goal of the present paper is to introduce a \fwk\ in which similarities between ML and \Bx become apparent. We will see that by abstracting away some details and ignoring the contextual differences, both problems can be seen as two different instantiations of the same algebraic structure defined in the paper and called \bfem{asymmetric learning lens, al-lens}. Indeed, lenses are devices specifying change propagation, and learning can be seen as a special change propagation case.....
\zd{------}
Fong and Johnson also built an sm-functor 
$\learncat\rightarrow\slenscat$
which maps learning algorithms to so called symmetric lenses.

\zd{------}

Our main observation is that functor \leefun\ is actually composed from two components 
as shown by commutative square  (1) in \figroo, in which \alalensdifcat\ is the category of ala-lenses whose source and target model spaces are equipped with an operation \dif\ that given two objects $x, x'$ (\eg, two tuples in \realsx{n}), produces a delta between them, \ie, an arrow \frar{\dif(x,x')}{x}{x'} (\eg, for $x,x'\in\realsx{n}$, $\dif(x,x')=(x'_i-x_i)_{i\le n}$), and lens operations are compatible with \dif. Each delta lens with differencing gives rise (in fact, collapses) into a codiscrete lens, whose operation $\putlcodi$ works by, first, computing the required delta and then applying to it the delta lens \putl. We consider this construction in \sectref{codifun} and show that this construction amounts to a functor \frar{\codiscr}{\alalensdifcat}{\alalenscatcodi}.%
\footnote{Importantly, as functor \codiscr\ forgets structural information about its source lenses,  its composition  with the obvious  full forgetful embedding of \alalenscatcodi\ into \alalensdifcat\ (see triangle (4) in the diagram, where this embedding is denoted by $\plusx{\pairing}(\deltas,\dif)$ as it adds necessary deltas and differencing by employing pairing of objects) is not identity on \alalensdifcat\ (but the converse composition $(\plusx{\pairing}(\deltas,\dif));\codiscr$ is the identity functor $\id_\alalenscatcodi$).}

\begin{figure}[th]
\vspace{-1cm}
	\centering
\[
\begin{diagram}
\\ 
\grt{\paracat_\reals}
& 
& 
&&  \rDashto^{\grt{\leefun}}_{
	\begin{array}{c}
		{}^{(1)}
	\end{array}
} 
&  
&  \learncat
&\rEQ 
&& \alalenscatcodi
\\ 
&\rdFEmb^\cong%
              _{\grt{\plusdelx{\dif}}}   
   \rdDashto(5,2)^{\grt{\leefunovr}}_{_{(2)}}  
&    
&   
&   
 &\ruTo(1,2)^{\codiscr}
& \textcolor{red}{{}^{(4)}}
&
&\ldFEmb(4,2)>{\plusx{\pairing}(\deltas,\dif)}
\\ 
&   
&\grt{\paracatrcat}
&&\rTo_{\vspace{-5ex}
	\begin{array}{cc}
	\grt{\plusx{Gd(\eps,\err)}\putl\qquad} &\qquad
	\\[5pt]
	{}^{(3)}&
	\end{array}
}
&\alalensdifcat
&  
&   
\\ 
&&
&\rdTo(2,2)<{\grt{{-}\reals}}
&   \ldTo(1,2)>{{-}\putl}
& &&
\\ 
&&&&\pcatcat
&&&&
\end{diagram}
\]
\vspace{-0.5cm}
\caption{Functoriality of learning algorithms, 2}
\label{fig:policyfun-ml2}
\end{figure}
In comparison with functor \leefun\ considered in \citefst, the idea of gradient descent learning is more accurately described by functor \leefunovr\ because it does not include the forgetful functor \codifun\ into  the story (see triangle (1) in the diagram). Moreover, \leefunovr\ is itself composed as shown by triangle (2) of the diagram: the first component (the left edge) is an isomorphic representation of functions between Euclidean spaces-as-sets (denoted \realsx{n}) as functors between Euclidean spaces-as-categories (\rcatx{n}), whose arrows are componentwise differences, and the second component (the horizontal edge) is augmenting these functors with their special inverses called \putl s in the lens parlance. Building operations \putl s is essentially based on the gradient descent idea, hence, the notation 
$\plusx{Gd(\eps,\err)}\putl$ for this functor.  Adding \putl\ does not change the original functors \get s and triangle (3) formed with two obvious forgetful functors into \pcatcat\ commutes (category \pcatcat\ is the category of (equivalence classes of ) parametrized functors between small categories specified in \sectref{pcatcat}). Note that while commutativity of forgetting for functor \leefun\ is considered over \psetcat\ (see \figro), the context for \leefunovr\ is \pcatcat\ so that we have a consistent delta setting for the gradient descent learning. (The triangle from \figro\ can be restored in \figroo\ as well by a suitable combination of embeddings --- see \sectref{codifun}).


\renewcommand\actfoss[2]{\par=== ACT19====\par #1\par===Fossacs19==== \par#2}
\actfoss{
	In a seminal paper \citefst, Fong, Spivak and Tuy\'eras showed how to compose supervised machine learning (ML) algorithms so that the latter form a symmetric monoidal (sm) category \learncat, and built an sm-functor 
	\frar{L}{\paracat}{\learncat}, 
	which maps a parameterized differentiable function 
	\frar{f}{P\times\reals^m}{\reals^n} (with the parameter space $P=\reals^k$)
	to a learning algorithm 
	that improves an initially  given function \frar{f(p_0, \_)}{\reals^m}{\reals^n} 
	by learning from a set of training pairs $(a,b)\in \reals^m\timm \reals^n$. 
	Recently, Fong and Johnson noticed in \citefjbx\ (quoting them directly) ``surprising links between two apparently disparate areas'': 
	ML (treated compositionally as above) and bidirectional model \trafon s or \Bx (also treated compositionally in a \fwk\ of mathematical structures called {\em lenses} \cite{bpierce-popl05}), whereas ``naively at least, there seemed to be little reason to expect them to be closely related mathematically''.%
	\footnote{
		Term {\em \Bx} abbreviates ``bidirectional $X$ (something)'' and refers to bidirectional change propagation in different contexts in different domains: file \syncon\ in versioning, data exchange in databases, model \syncon\ and model transformation in Model-Driven software Engineering (MDE), 
		see \cite{bx-cross} for some of these contexts. What makes \Bx a special and united community amongst many circles studying \syncon\ and change propagation, is a foundational lens-based algebraic \fwk.  Nevertheless, within \Bx the application context can vary, sometimes significantly, and in the present paper, \Bx will mainly refer to \Bx in the MDE context --- the author's area of expertize in \Bx. 
	}   
	Fong and Johnson also built an sm-functor 
	$\learncat\rightarrow\slenscat$
	which maps learning algorithms to so called symmetric lenses.
}{
	The notion of view is fundamental for system engineering: to understand a complex system $X$, we normally build and analyze different view models (or just views) $V_i$ of $X$, \eg, a car is a system of its mechanical, electrical, embedded software etc. views, each of which encompasses its own view system, and so on. Views are also fundamental for database management and software, \eg, an interface is basically a view (often rather intricate) of the underlying software. View models of approximately  the same level of abstraction are often called {\em horizontal}, while decreasing abstraction (\ie, going from more to less abstract models) or refinement is termed {\em vertical}. The latter is  ubiquitous in software engineering, whose basic relationship ``X implements Y'' can often be formalized by considering Y as an abstract view of X so that implementation means building a source whose view is given. For example, a UML model M is a view to a Java program J to be created, which is in its turn is a view to the bytecode B generated from J. 
}

	\section{Arrow diagrams to use}
\newcommand\figr{\figref{fig:fbPpg-oper-laws}}

\subsection{Functoriality of update policies}

\begin{wrapfigure}{R}{0.55\textwidth}
\vspace{-1.cm}
	\centering
\begin{tabular}{l}
	$
   \begin{diagram}[w=1.1\cellW,h=0.9\cellH] 
	\dbox{A} 
	& \rMapsto^{{:}\getio}  & {B} 
	& \rMapsto^{{:}\getiio} &  {C}%
	\\
	\dDotto<u & \swTilearrow{\putl{1}_A}~~~ & \dDotto>{v} %
	& \swTilearrow{\putl2_B} ~~~& \dTo>{w} 
	\\
	{A'} & \rDermapsto_{{:}\getio}  & %
	B'            & \rDermapsto_{{:}\getiio} & \dbox{C'} %
\end{diagram}%
$
\\ [7.5ex]
(a) ordinary lens composition
\\ [10pt]
$
\begin{diagram}[h=2.em]
\dbox{A}&\rDermapsto^{{:}\get1_{0}}
&B&\rDermapsto^{{:}\get2_{0}}
&C
\\ 
&&&& \dTo>{w}
\\ 
&& 
&
\swseTilearrowABx{1{:}\putl2_{B}}{}
&\dbox{C'}
\\ 
&&
\dDotto>{v}
&&\dDotto<{w^@}
\\ 
&
\swseTilearrowABx{2{:}\putl1_{A} }{}
&B'
&\rDashmapsto^{{:}\get2_{0}}&{C'}^@
\\ 
\dDotto>{u} 
&&\dDotto<{v^@}&\stackrel{3{:}\get2}{\Rightarrow}&\dDashto>{v^@.\get2}
\\ 
A'&\rDashmapsto^{{:}\get1_{0}}&{B'}^@&\rDashmapsto^{{:}\get2_{0}}
&C'^{@@}
\end{diagram}
$
\\ [15ex]
(b)  lenses with amendment 
\end{tabular}
\caption{Composition of lenses}
\label{fig:aalenscompose}
\end{wrapfigure}

The diagram in \figref{policyfun-tripleLane} illustrates compositionality of operations \get\ (trivial) and \putl\ (non-trivial) (nodes $\circ$ denote objects without names). Given a sequence of parameterized functors \frar{\get^i_{p_i}}{\spA_i}{\spA_{i+1}}, $p_i\in\spP_i$, $i\le n$ ($n=3$ in the diagram), it is easy to define their composition \frar{\get_p}{\spA_1}{\spA_n} for $p=(p_1...p_n)\in\spP{=}\spP_1\timm\ldots\timm\spP_n$ as $$A.\get_p=(...(A.\get^1_{p_1}).\;\ldots\;).\get^n_{p_n},$$ and show that it is associative (when functors are considered up to a suitable equivalence). Composition of $\putl^i$, $i\le n$ is more interesting: the diagram shows a simple projection of story, in which  arrows \frar{e_i{=}\putl^{i\,\upd}}{p_i}{p'_i} are not shown. Triple-arrow symbols $j{:}\putl^i$ denote three consecutive applications ($j=1,2,3$) of operations $\putl^i$ appropriately indexed. Double-arrows labelled $j{:}\get^i_{e_i}$ denote application of the corresponding \get-functors to arrows to demonstrate that the Putget law holds.

A pair $\ell_i=(\get^i,\putl^i)$ is an ala-lens \frar{\ell_i}{\spA_i}{\spA_{i+1}}, and the construction shown in the figure defines lens composition \frar{\ell_1;\ell_2;\ell_3}{A_1}{A_4} (it will be formally defined in \sectref{alalenscat}). 
However, it's not the end of the story developed in \citefst. 
So far, in the lens formalism as such, two components of a lens $\ell=(\get, \putl)$ are considered given independently although their interaction is regulated by lens laws ensuring a weaker or stronger sense of \putl\ being an inverse to \get. However, in practice of both BX and ML, the \putl-component is built based on the \get-component. As there are many ways of inverting a {\bf general} functor, finding reasonable single-valued operation \putl\ (a so called {\em update policy})  is normally based on the {\bf specific} form of \get\ so that \putl\ is carefully ``tuned'' for a given specific \get. In other words, for a given specific domain context \nix, model spaces $\spA_i$ are special categories and functors \get s are special functors for which a reasonable single-valued operations \putl\ can be defined. For example, ML provides a special context, in which model spaces are Euclidean spaces (considered as categories $\rcat^m$), \get s are differentiable mappings \frar{f}{\rcat^k\timm\rcat^m}{\rcat^n}, and the gradient descent idea Gd provides a reasonable definition of \putl s. We will return for this discussion in \sectref{policyfun-2}


\newarrow{Updto}{<}---{>}
\begin{figure}
\centering
\begin{tabular}{c@{\qquad}c@{\qquad}c}
\begin{diagram}[w=2em,h=1.5em] 
\dbox{A}&& \rCorrto^r &&\dbox{B}\\%
&\seTilearrow{\falt} &&\ldDercorrto(4,4)~{a{*}r}&\\
\dUpdto<a & && &\\%
&&&&\\
\dbox{A'} & &&&
\end{diagram}%
&
\begin{diagram}[w=2em,h=1.5em] 
\dbox{A} && \rCorrto^r &&\dbox{B}\\%
&\rdDercorrto(4,4)~{r{*}b} &&\swTilearrow{\balt}\quad~~&\\%
& &&&\dUpdto>b \\%
&&&&\\
&&&&\dbox{B'}
\end{diagram}%
&
\begin{diagram} 
\dbox{A} & \rCorrto^r &\dbox{B}\\%
\dUpdto<a %
& %
\ruDercorrto^{a{*}r~~~~~~~}\luDercorrto^{~~~~~~r{*}b}%
& \dUpdto>b \\%
\dbox{A'} & \rDercorrto^{r'} &\dbox{B'}
\end{diagram}%
\\[50pt]
\multicolumn{2}{c}{%
(a)  arities %
} &
(b) (\falt-\balt) law
\end{tabular}
\caption{Realignment operations and their laws \label{fig:align-oper}}%
\end{figure}

\cellW=7.25ex%
\cellH=5.0ex
\begin{figure}
\centering
\begin{tabular}{c@{\quad}c@{\quad}c@{\quad}c}
\begin{diagram}[w=1.1\cellW,h=1.1\cellH] 
\dbox{A} & \rCorrto^r &\dbox{B}\\%
\dCorrto<a & \seTilearrow{\fppg}~~~
            & \dDercorrto>b \\%
\dbox{A'} & \rDercorrto^{r'} & B'
\end{diagram}%
& 
\begin{diagram}[w=\cellW,h=\cellH] 
\dbox{A} & \rCorrto^{r} & \dbox{B}\\%
\dCorrto<{\ide{}{A}} & \seTilearrow{\fppg}~~~
            & \dDercorrto>{\ide{}{B}} \\%
\dbox{A} & \rDercorrto^{r} & B
\end{diagram}%
 &   
 \begin{diagram}[w=\cellW,h=\cellH] 
\dbox{A} & \rCorrto^r &\dbox{B}\\%
\dTo<a & \seTilearrow{\fppg}~~~
            & \dDashto>b \\%
\dbox{A'} & \rDercorrto^{r'} & B'
\end{diagram}%
&     
\begin{diagram}[w=\cellW,h=\cellH] 
\dbox{A} & \rCorrto^r &\dbox{B}\\%
\uTo<a & \seTilearrow{\fppg}~~~
            & \uDashto>b \\%
\dbox{A'} & \rDercorrto^{r'} & B'
\end{diagram}%
\\[30pt]
\begin{diagram}[w=\cellW,h=\cellH] 
\dbox{A} & \rCorrto^r & \dbox{B}\\%
\dDercorrto>a & \swTilearrow{\bppg}~~~
            & \dCorrto>b \\%
A' & \rDercorrto^{r'} & \dbox{B'}
\end{diagram}%
& 
\begin{diagram}[w=\cellW,h=\cellH] 
\dbox{A} & \rCorrto^{r} & \dbox{B}\\%
\dDercorrto<{\ide{}{A}} & \swTilearrow{\bppg}~~~
            & \dCorrto>{\ide{}{B}} \\%
A & \rDercorrto^{r} & \dbox{B}
\end{diagram}%
&   
\begin{diagram}[w=\cellW,h=\cellH] 
\dbox{A} & \rCorrto^r &\dbox{B}\\%
\dDashto<a & \swTilearrow{\bppg}~~~
            & \dTo>b \\%
A' & \rDercorrto^{r'} & \dbox{B'}
\end{diagram}%
&   
   \begin{diagram}[w=\cellW,h=\cellH] 
\dbox{A} & \rCorrto^r &\dbox{B}\\%
\uDashto<a & \swTilearrow{\bppg}~~~
            & \uTo>b \\%
A' & \rDercorrto^{r'} & \dbox{B'}
\end{diagram}%
\\[35pt]
(a) arities &
(b) (\idppglaw)  laws & %
\multicolumn{2}{c}{(c) monotonicity laws}
\end{tabular}
\caption{Operations of update propagation
\label{fig:fbPpg-oper-laws}}%
\end{figure}

\begin{figure}
		\centering
{ 
\begin{tabular}{c}
\begin{diagram}[w=1.1\cellW,h=0.9\cellH] 
\dbox{A} 
& \rMapsto^{{:}\getio}  & {B} 
& \rMapsto^{{:}\getiio} &  {C}%
\\
\dDotto<u & \swTilearrow{\putl{1}_A}~~~ & \dDotto>{v} %
                 & \swTilearrow{\putl2_B} ~~~& \dTo>{w} 
\\
{A'} & \rDermapsto_{{:}\getio}  & %
 B'            & \rDermapsto_{{:}\getiio} & \dbox{C'} %
\end{diagram}%
\end{tabular}
} 
\caption{Lens composition} %
\label{fig:alenscompose}%
\end{figure}

\begin{figure}
\centering
\begin{tabular}{c@\quad c}
\begin{diagram}[w=1.05\cellW,h=1.1\cellH] 
\dbox{A} & \rCorrto^r     & \dbox{B} & \rCorrto^{r} & \dbox{A}%
               &  \rCorrto^{r} & \dbox{B}
\\%
\dCorrto<a & \seTilearrow{\fppg}~~~ & \dDercorrto>b %
                 & \seTilearrow{\bppg}~~~ & \dDercorrto>{a_1}
                  & \seTilearrow{\fppg}~~~ & \dDercorrto>b
\\%
\dbox{A'} & \rDercorrto^{r'}  & %
B'            & \rDercorrto^{r_1'} & A_1' %
               & \rDercorrto^{r_1'} & B'
\end{diagram}%
&    %
\begin{diagram}[w=1.1\cellW,h=1.1\cellH] 
\dbox{A} & \rCorrto^r     & \dbox{B} & \rCorrto^{r} & \dbox{A}%
               &  \rCorrto^{r} & \dbox{B}
\\%
\dDercorrto<a & \swTilearrow{\bppg}~~~ & \dDercorrto>{b_1} %
                 & \swTilearrow{\fppg}~~~ & \dDercorrto>{a}
                  & \swTilearrow{\bppg}~~~ & \dCorrto>b
\\%
{A'} & \rDercorrto^{r_1'}  & %
 B_1'            & \rDercorrto^{r_1'} & A' %
               & \rDercorrto^{r'} & \dbox{B'}
\end{diagram}%
\\[40pt]
(a) \fbfppglaw\ law  & (b) \bfbppglaw\ law
\end{tabular}
\caption{Round-tripping laws. (Scenario in diagram (b) ``runs'' from the right to the left.)
\label{fig:invert-laws}}%
\end{figure}

\begin{wrapfigure} {r}{0.5\textwidth}%
	\includegraphics[width=0.475\columnwidth]{ll-figures/pdfs/bxScenario}%
	\caption{BX scenario specified in (a) delta-based and (b) state-based way}%
	\label{fig:bx-scenario}%
	\vspace{-0.2cm}%
\end{wrapfigure}%

\begin{wrapfigure} {R}{0.326\textwidth}%
	\vspace{-0.5cm}%
	\includegraphics[width=0.325\columnwidth]{ll-figures/pdfs/bx-sttFwk-errors}%
	\caption{State-based BX: erroneous horizontal composition}%
	\label{fig:bx-sttFwk-errors}%
	\vspace{-0.2cm}%
\end{wrapfigure}%


	\section{To do for the TR}

\begin{itemize}
	\item mention on p.13 that we should, in general, consider lax learning functors but we leave it for future work. Some remarks on future work, including this?  
	\item  
	Discuss the case of object structures and deltas for ML (?) Well, not for this paper - let it go to a future ML oriented paper
	\item 
\end{itemize}
	
}{
	
	\newcommand\figro{\figref{policyfun-ml1}}
	\newcommand\figroo{\figref{policyfun-ml2}}
	\renewcommand\figrii{\figref{policyfun-brief}}

	\renewcommand\figri{\figref{bxmlPut}}
	\renewcommand\figrii{\figref{policyfun-brief}}

          \renewcommand\figriii{\figref{policyfun}}
 
	\newcommand\figari{\figref{apa-arity}} 
	\newcommand\defri{\defref{def:aalens-arity}}
	\renewcommand\ppgB{\ensuremath{\putl^B}}
	   \renewcommand\kl{\ensuremath{\ekk;\ell}} 
		\renewcommand\figrii{\figref{kl-lens}}


\appendix
\section{Appendices}

\newcommand\ideny{identity}
\renewcommand\lm{\enma{\ell\emm}}
\renewcommand\klbr{\enma{(\ekk\ell)}}
\renewcommand\lmbr{\enma{(\ell\emm)}}
\renewcommand\putekel[2]{\putUL{\kl.#1}{#2}}
\renewcommand\putelem[2]{\putUL{\lm.#1}{#2}}


\newcommand\figkl{\figref{kl-lens}}
\renewcommand\figrii{\figkl} 
\newcommand\defrii{\defref{def:aalens-hComp}}

\renewcommand\kl{\ensuremath{\ekk;\ell}} 
\renewcommand\lm{\enma{\ell;\emm}}
\renewcommand\lmbr{\enma{(\ell;\emm)}}
\renewcommand\kl{\enma{\ekk\ell}}
\renewcommand\lm{\enma{\ell\emm}}
\renewcommand\klbr{\enma{(\ekk\ell)}}
\renewcommand\lmbr{\enma{(\ell\emm)}}
\renewcommand\putekel[2]{\putUL{\kl.#1}{#2}}
\renewcommand\putelem[2]{\putUL{\lm.#1}{#2}}

\bibliographystyle{splncs04} 
\bibliography{refsGrand17}

\begin{thebibliography}{10}
\providecommand{\url}[1]{\texttt{#1}}
\providecommand{\urlprefix}{URL }
\providecommand{\doi}[1]{https://doi.org/#1}

\bibitem{viewmain-vldb98}
Abiteboul, S., McHugh, J., Rys, M., Vassalos, V., J.Wiener: {I}ncremental
  {M}aintenance for {M}aterialized {V}iews over {S}emistructured {D}ata. In:
  Gupta, A., Shmueli, O., Widom, J. (eds.) {VLDB}. Morgan Kaufmann (1998)

\bibitem{tony-bxbook1}
Anjorin, A.: An introduction to triple graph grammars as an implementation of
  the delta-lens framework. In: Gibbons, J., Stevens, P. (eds.) Bidirectional
  Transformations - International Summer School, Oxford, UK, July 25-29, 2016,
  Tutorial Lectures. Lecture Notes in Computer Science, vol.~9715, pp. 29--72.
  Springer (2016). \doi{10.1007/978-3-319-79108-1\_2},
  \url{https://doi.org/10.1007/978-3-319-79108-1}

\bibitem{tony-benchmarx1}
Anjorin, A., Diskin, Z., Jouault, F., Ko, H., Leblebici, E., Westfechtel, B.:
  Benchmarx reloaded: {A} practical benchmark framework for bidirectional
  transformations. In: Eramo and Johnson  \cite{DBLP:conf/etaps/2017bx}, pp.
  15--30, \url{http://ceur-ws.org/Vol-1827/paper6.pdf}

\bibitem{andy-20years}
Anjorin, A., Leblebici, E., Sch{\"{u}}rr, A.: 20 years of triple graph
  grammars: {A} roadmap for future research. {ECEASST}  \textbf{73} (2015).
  \doi{10.14279/tuj.eceasst.73.1031},
  \url{https://doi.org/10.14279/tuj.eceasst.73.1031}

\bibitem{tony-viewtgg-ecmfa14}
Anjorin, A., Rose, S., Deckwerth, F., Sch{\"{u}}rr, A.: Efficient model
  synchronization with view triple graph grammars. In: Cabot, J., Rubin, J.
  (eds.) Modelling Foundations and Applications - 10th European Conference,
  {ECMFA} 2014, Held as Part of {STAF} 2014, York, UK, July 21-25, 2014.
  Proceedings. Lecture Notes in Computer Science, vol.~8569, pp. 1--17.
  Springer (2014). \doi{10.1007/978-3-319-09195-2\_1},
  \url{https://doi.org/10.1007/978-3-319-09195-2\_1}

\bibitem{bryce-act19}
Clarke, B.: Internal lenses as functors and cofunctors. In: Pre-proceedings of
  {ACT'19}, Oxford, 2019.
  Http://www.cs.ox.ac.uk/ACT2019/preproceedings/Bryce

\bibitem{bx-cross}
Czarnecki, K., Foster, J.N., Hu, Z., L{\"a}mmel, R., Sch{\"u}rr, A.,
  Terwilliger, J.F.: Bidirectional transformations: A cross-discipline
  perspective. In: Theory and Practice of Model Transformations, pp. 260--283.
  Springer (2009)

\bibitem{me-bx17}
Diskin, Z.: Compositionality of update propagation: Lax putput. In: Eramo and
  Johnson  \cite{DBLP:conf/etaps/2017bx}, pp. 74--89,
  \url{http://ceur-ws.org/Vol-1827/paper12.pdf}

\bibitem{myX-learning19}
Diskin, Z.: General supervised categorical learning as change propagation with
  delta lenses. CoRR  \textbf{abs/1911.12904} (2019),
  \url{http://arxiv.org/abs/1911.12904}

\bibitem{me-jss15}
Diskin, Z., Gholizadeh, H., Wider, A., Czarnecki, K.: A three-dimensional
  taxonomy for bidirectional model synchronization. {Journal of System and
  Software}  \textbf{111},  298--322 (2016). \doi{10.1016/j.jss.2015.06.003},
  \url{https://doi.org/10.1016/j.jss.2015.06.003}

\bibitem{me-faoc19}
Diskin, Z., K{\"{o}}nig, H., Lawford, M.: Multiple model synchronization with
  multiary delta lenses with amendment and {K-Putput}. Formal Asp. Comput.
  \textbf{31}(5),  611--640 (2019). \doi{10.1007/s00165-019-00493-0},
  \url{https://doi.org/10.1007/s00165-019-00493-0}, (Sect.7.1 of the paper is
  unreadable and can be found in http://arxiv.org/abs/1911.11302)

\bibitem{me-entcs08}
Diskin, Z., Wolter, U.: {A} {D}iagrammatic {L}ogic for {O}bject-{O}riented
  {V}isual {M}odeling. Electr. Notes Theor. Comput. Sci.  \textbf{203}(6),
  19--41 (2008)

\bibitem{me-jot11}
Diskin, Z., Xiong, Y., Czarnecki, K.: {F}rom {S}tate- to {D}elta-{B}ased
  {B}idirectional {M}odel {T}ransformations: the {A}symmetric {C}ase. Journal
  of Object Technology  \textbf{10},  6: 1--25 (2011)

\bibitem{me-models11}
Diskin, Z., Xiong, Y., Czarnecki, K., Ehrig, H., Hermann, F., Orejas, F.: From
  state-to delta-based bidirectional model transformations: the symmetric case.
  In: MODELS, pp. 304--318. Springer (2011)

\bibitem{viewmain-elke06}
El-Sayed, M., Rundensteiner, E.A., Mani, M.: {I}ncremental {M}aintenance of
  {M}aterialized {XQuery} {V}iews. In: Liu, L., Reuter, A., Whang, K.Y., Zhang,
  J. (eds.) {ICDE}. p.~129. IEEE Computer Society (2006).
  \doi{10.1109/ICDE.2006.80}

\bibitem{DBLP:conf/etaps/2017bx}
Eramo, R., Johnson, M. (eds.): Proceedings of the 6th International Workshop on
  Bidirectional Transformations co-located with The European Joint Conferences
  on Theory and Practice of Software, Bx@ETAPS 2017, Uppsala, Sweden, April 29,
  2017, {CEUR} Workshop Proceedings, vol.~1827. CEUR-WS.org (2017),
  \url{http://ceur-ws.org/Vol-1827}

\bibitem{fj-bx19}
Fong, B., Johnson, M.: Lenses and learners. In: Cheney, J., Ko, H. (eds.)
  Proceedings of the 8th International Workshop on Bidirectional
  Transformations co-located with the Philadelphia Logic Week, Bx@PLW 2019,
  Philadelphia, PA, USA, June 4, 2019. {CEUR} Workshop Proceedings, vol.~2355,
  pp. 16--29. CEUR-WS.org (2019), \url{http://ceur-ws.org/Vol-2355/paper2.pdf}

\bibitem{fst-backprop}
Fong, B., Spivak, D.I., Tuy{\'{e}}ras, R.: Backprop as functor: {A}
  compositional perspective on supervised learning. In: 34th Annual {ACM/IEEE}
  Symposium on Logic in Computer Science, {LICS} 2019, Vancouver, BC, Canada,
  June 24-27, 2019. pp. 1--13. {IEEE} (2019). \doi{10.1109/LICS.2019.8785665},
  \url{https://doi.org/10.1109/LICS.2019.8785665}

\bibitem{jules-corr19}
Hedges, J.: From open learners to open games. CoRR  \textbf{abs/1902.08666}
  (2019), \url{http://arxiv.org/abs/1902.08666}

\bibitem{frank-sosym15}
Hermann, F., Ehrig, H., Orejas, F., Czarnecki, K., Diskin, Z., Xiong, Y.,
  Gottmann, S., Engel, T.: Model synchronization based on triple graph
  grammars: correctness, completeness and invertibility. Software and System
  Modeling  \textbf{14}(1),  241--269 (2015). \doi{10.1007/s10270-012-0309-1},
  \url{https://doi.org/10.1007/s10270-012-0309-1}

\bibitem{jr-unified}
Johnson, M., Rosebrugh, R.D.: Unifying set-based, delta-based and edit-based
  lenses. In: Proceedings of the 5th International Workshop on Bidirectional
  Transformations, Bx 2016. pp. 1--13 (2016),
  \url{http://ceur-ws.org/Vol-1571/paper_13.pdf}

\bibitem{mtbe-kappel12}
Kappel, G., Langer, P., Retschitzegger, W., Schwinger, W., Wimmer, M.: Model
  transformation by-example: {A} survey of the first wave. In: Conceptual
  Modelling and Its Theoretical Foundations - Essays Dedicated to Bernhard
  Thalheim on the Occasion of His 60th Birthday. pp. 197--215 (2012).
  \doi{10.1007/978-3-642-28279-9\_15},
  \url{https://doi.org/10.1007/978-3-642-28279-9\_15}

\bibitem{makkai-ske}
Makkai, M.: Generalized sketches as a framework for completeness theorems.
  Journal of Pure and Applied Algebra  \textbf{115},  49--79, 179--212,
  214--274 (1997)

\bibitem{groundtram1}
Sasano, I., Hu, Z., Hidaka, S., Inaba, K., Kato, H., Nakano, K.: Toward
  bidirectionalization of {ATL} with {GRoundTram}. In: Cabot, J., Visser, E.
  (eds.) Theory and Practice of Model Transformations - 4th International
  Conference, {ICMT} 2011, Zurich, Switzerland, June 27-28, 2011. Proceedings.
  Lecture Notes in Computer Science, vol.~6707, pp. 138--151. Springer (2011).
  \doi{10.1007/978-3-642-21732-6\_10},
  \url{https://doi.org/10.1007/978-3-642-21732-6\_10}

\bibitem{tony-emoflon-bx19}
Weidmann, N., Anjorin, A., Fritsche, L., Varr{\'{o}}, G., Sch{\"{u}}rr, A.,
  Leblebici, E.: Incremental bidirectional model transformation with emoflon: :
  Ibex. In: Cheney, J., Ko, H. (eds.) Proceedings of the 8th International
  Workshop on Bidirectional Transformations co-located with the Philadelphia
  Logic Week, Bx@PLW 2019, Philadelphia, PA, USA, June 4, 2019. {CEUR} Workshop
  Proceedings, vol.~2355, pp. 45--55. CEUR-WS.org (2019),
  \url{http://ceur-ws.org/Vol-2355/paper4.pdf}

\end{thebibliography}
\dend




\clearpage
\noindent{\LARGE Aux and OLd sections}

\clearpage

\section{To do for the TR}

\begin{itemize}
	\item mention on p.13 that we should, in general, consider lax learning functors but we leave it for future work. Some remarks on future work, including this?  
	\item  
	Discuss the case of object structures and deltas for ML (?) Well, not for this paper - let it go to a future ML oriented paper
	\item 
\end{itemize}


\vfill

{\small\medskip\noindent{\bf Open Access} This chapter is licensed under the terms of the Creative Commons\break Attribution 4.0 International License (\url{http://creativecommons.org/licenses/by/4.0/}), which permits use, sharing, adaptation, distribution and reproduction in any medium or format, as long as you give appropriate credit to the original author(s) and the source, provide a link to the Creative Commons license and indicate if changes were made.}

{\small \spaceskip .28em plus .1em minus .1em The images or other third party material in this chapter are included in the chapter's Creative Commons license, unless indicated otherwise in a credit line to the material.~If material is not included in the chapter's Creative Commons license and your intended\break use is not permitted by statutory regulation or exceeds the permitted use, you will need to obtain permission directly from the copyright holder.}

\medskip\noindent\includegraphics{cc_by_4-0.eps}

\dend
\clearpage

\dend

\clearpage
\medskip
\noindent{\Large\sc Appendices} 
%

	
}
\end{document}